\documentclass[letterpaper,11pt,fleqn]{article}
\pdfoutput=1
\usepackage{jheppub}
\usepackage{graphicx}
\usepackage{bm,amsmath,amssymb}
\usepackage[mathscr]{eucal}

\long\def\comment#1{ }
\newcommand{\eqn}[1]{Eq.~(\ref{#1})}
\newcommand{\beq}{\begin{equation}}
\newcommand{\eeq}{\end{equation}}
\newcommand{\nn}{\nonumber\\}

\newcommand{\rmd}{{\rm d}}
\newcommand{\rme}{{\rm e}}

\newcommand{\del}{\partial}

\newcommand{\order}[1]{\mcal{O}{(#1)}}
\newcommand{\mcal}{\mathcal}

\newcommand{\bk}{\bm{k}}

\newcommand{\abar}{\bar{\alpha}}

\newcommand{\F}{\mcal{F}}
\newcommand{\xc}{{x_c}}
\newcommand{\xt}{{x_{\rm th}}}

\title{\Large Medium--induced jet evolution:  wave turbulence and energy 
loss}

\author{Leonard Fister}

\author{and Edmond Iancu}

\affiliation{Institut de Physique Th\'{e}orique de Saclay,
F-91191 Gif-sur-Yvette, France}

\emailAdd{leonard.fister@cea.fr}
\emailAdd{edmond.iancu@cea.fr}

\abstract{
We study the gluon cascade generated via successive medium-induced branchings by an 
energetic parton propagating through a dense QCD medium.  We focus on
the high-energy regime where the energy $E$ of the leading particle is much larger than the
characteristic medium scale $\omega_c=\hat q L^2/2$, with $\hat q$ the jet quenching parameter
and $L$ the distance travelled through the medium.  In this regime 
the leading particle loses only a small fraction $\sim\alpha_s(\omega_c/E)$ of its energy
and can be treated as a steady source of radiation for gluons with energies 
$\omega \le \omega_c$. For this effective problem with a source, we obtain exact analytic
solutions for the gluon spectrum and the energy flux. These solutions exhibit
wave turbulence: the basic physical process is a continuing fragmentation which is
`quasi-democratic' (i.e. quasi-local in energy) and which
provides an energy transfer from the source to the medium at a rate (the energy flux
 $\F$) which is quasi-independent of $\omega$.
The locality of the branching process implies a spectrum of the Kolmogorov-Obukhov
type,  i.e. a power-law spectrum which is a fixed point of the branching process and
whose strength is proportional to the energy flux: $D(\omega)\sim \F/\sqrt{\omega}$
for $\omega \ll \omega_c$. Via this turbulent flow, the gluon cascade loses towards the medium
an energy  $\Delta E\sim\alpha_s^2 \omega_c$, which is independent of the initial energy 
$E$ of the leading particle and of the details of the thermalization mechanism
at the low-energy end of the cascade. This energy is carried away by very soft gluons, which
propagate at very large angles with respect to the jet axis. Our predictions for the value
of $\Delta E$ and for its angular distribution appear to agree quite well, 
qualitatively and even semi-quantitatively, with the phenomenology 
of di-jet asymmetry in nucleus-nucleus collisions at the LHC.
}

\begin{document}
\maketitle
%\flushbottom

\section{Introduction}
\label{sec:intro}

The experimental observation of the phenomenon known as `di--jet asymmetry' in Pb+Pb collisions
at the LHC \cite{Aad:2010bu,Chatrchyan:2011sx,Chatrchyan:2012nia,Aad:2012vca,Chatrchyan:2013kwa,Chatrchyan:2014ava,Aad:2014wha,Gulhan:2014} 
has triggered intense theoretical efforts \cite{MehtarTani:2010ma,MehtarTani:2011tz,CasalderreySolana:2010eh,Qin:2010mn,CasalderreySolana:2011rz,Blaizot:2012fh,CasalderreySolana:2012ef,Blaizot:2013hx,Blaizot:2013vha,Iancu:2014aza,Kurkela:2014tla,Blaizot:2014ula,Apolinario:2014csa} aiming at 
understanding the evolution of an energetic jet propagating through a dense QCD medium,
such as a quark--gluon plasma. The crucial observation is that the part of the jet fragmentation 
which is triggered by interactions inside the medium is controlled by relatively soft 
gluon emissions, with energies $\omega$ well below the characteristic medium scale 
$\omega_c=\hat q L^2/2$ and a formation time $t_{\rm br}(\omega)$ much smaller than $L$. 
(Here, $\hat q$ is the jet quenching parameter, $L$ is the distance travelled by the
`leading particle' --- the energetic parton which has initiated the jet --- 
through the medium, and the `formation time' $t_{\rm br}(\omega)
\sim\sqrt{\omega/\hat q}$ is the typical duration of the branching process.)  
This observation has far reaching consequences:

The soft gluons can be easily deviated towards large angles by rescattering in the medium,
so their abundant production via jet fragmentation may explain the significant transport of
energy at large angles with respect to the jet axis --- the hallmark of di--jet asymmetry.
Also, the subsequent emissions of soft gluons can be viewed as independent
from each other and hence described as a classical, probabilistic, branching process. Indeed, 
the quantum coherence effects and the associated interference phenomena are efficiently
washed out by rescattering in the medium \cite{MehtarTani:2010ma,MehtarTani:2011tz,CasalderreySolana:2010eh}:  the loss of color coherence occurs on a time scale 
comparable to that of the branching process, so that gluons that emerge from a splitting
propagate independently of each other \cite{Blaizot:2012fh}. 

Based on such considerations, one has been able to derive a classical effective theory 
for the gluon cascade generated via successive medium--induced gluon branchings 
\cite{Blaizot:2013hx,Blaizot:2013vha} (see also 
Refs.~\cite{Baier:2000sb,Jeon:2003gi} for earlier, related, studies).
This is a stochastic theory for a Markovien process in which the branching rate is given by the
BDMPSZ spectrum \cite{Baier:1996kr,Baier:1996sk,Zakharov:1996fv,Zakharov:1997uu,Baier:1998kq} 
for a single, medium--induced, gluon emission. The branching probability corresponding to
a distance $L$ is parametrically of order $\abar[L/t_{\rm br}(\omega)]$,
with $\abar\equiv\alpha_s N_c/\pi$. This probability becomes
of order one (meaning that the branching dynamics becomes non--perturbative) when
$\omega\lesssim \omega_s\equiv \abar^2\omega_c$. As we shall see, this `soft' scale $\omega_s$
is truly semi--hard (in the ballpark of a few GeV), meaning that there
is a significant region in phase--space where perturbation theory breaks down. The effective
theory put forward in Refs.~\cite{Blaizot:2013hx,Blaizot:2013vha} allows one to deal with such
non--perturbative aspects, by resuming soft multiple branchings to all orders.

The original analysis in \cite{Blaizot:2013hx} demonstrated that the non--perturbative
dynamics associated with multiple branchings has a remarkable consequence: it leads to
{\em wave turbulence}  \cite{KST,Nazarenko}. The leading particle, whose initial energy $E$
is typically much larger than the non--perturbative scale $\omega_s$, promptly and abundantly 
radiates soft gluons with energies $\omega\lesssim \omega_s$ and thus loses an
amount of energy of order $\Delta E\sim \omega_s$ {\em event by event} 
(that is, with probability of order  one). After being emitted, these soft primary gluons keep
on branching into even softer gluons, and their subsequent branchings are {\em quasi--democratic}~:
the two daughter gluons produced by a typical splitting have comparable 
energies\footnote{It is interesting to notice that a similar branching process occurs in a different physical
context, namely the thermalization of the quark--gluon plasma produced in the intermediate
stages of a ultrarelativistic heavy ion collision: during the late stages of the  `bottom--up' 
scenario \cite{Baier:2000sb}, the hard particles lose energy towards the surrounding  thermal
bath via soft radiation giving rise to quasi--democratic cascades \cite{Kurkela:2011ti}.}. 
The locality of the branchings in $\omega$ is the key ingredient for turbulence. 
It leads to a power--law spectrum $D(\omega)\propto 1/\sqrt{\omega}$,  which emerges
as the Kolmogorov--Zakharov (KZ) fixed point \cite{KST,Nazarenko} 
of the branching process (this KZ spectrum is formally similar to  the BDMPSZ spectrum),
and to an energy flux which is independent of $\omega$ --- the {\em turbulent flow}.
An energy flux which is uniform in $\omega$ means that the energy flows from the high--energy end
to the low--energy end of the cascade, without accumulating at any intermediate value of $\omega$. 
For an ideal cascade, where the branching law remains unchanged down to arbitrary
small values of $\omega$, the energy carried by the flow accumulates into a 
condensate at $\omega=0$. In practice, we expect the branching process to be modified 
when the gluon energies become comparable to the medium `temperature' $T$ 
(the typical energy of the medium constituents): the soft gluons with $\omega\sim T$
`thermalize', meaning that they transfer their energy towards the medium. Assuming
the medium to act as a perfect sink  at $\omega\simeq T$, we conclude that the rate for 
energy loss is fixed by the turbulent flow and thus independent of the details of the 
thermalization mechanism (`universality').

%In that context too, wave turbulence has been observed in numerical simulations
%\cite{Berges:2013eia}.
%The rate for this energy transfer is fixed by the
%branching dynamics and hence is independent of the details of the thermalization mechanism %(`universality').

An essential property of the turbulent flow is 
the fact that it allows for the  {\em transfer of a significant fraction of the total energy 
towards arbitrarily soft quanta}. To better appreciate how non--trivial this situation is, let us compare 
it with a more traditional parton cascade in perturbative QCD: the DGLAP cascade, as
driven by bremsstrahlung in the vacuum. In that case, the typical splittings are very asymmetric,
due to the `infrared' ($\omega\to 0$) singularity of bremsstrahlung, and lead to a rapid rise in
the number of gluons with small values of the energy fraction $x\equiv\omega/E$. 
Yet, the total energy
carried by these `wee' gluons with $x\ll 1$ is very small: the energy fraction contained in the 
region of the spectrum at $x< x_0$ vanishes as a power of $x_0$ when $x_0\to 0$. Most of the 
original energy remains in the few partons with larger values of $x$. This is due to the fact that,
after a very asymmetric splitting, the parent parton preserves most of its original energy.

By contrast, for the medium--induced cascade, the energy contained
in the bins of the spectrum at $x< x_0$
is only a {\em part} of the total energy associated with modes 
softer than $x_0$. The other part is the energy carried by the turbulent flow, 
which ends up at arbitrarily low values of $x$ (at least, for an ideal cascade) and hence
is independent of $x_0$. Depending upon the size $L$ of the medium, this flow
energy can be as large as the original energy $E$ of the leading particle (see the discussion
in Sect.~\ref{sec:phys}). In the presence of a thermalization mechanism at $\omega\sim T$,
the above argument remains valid so long as $x_0\ge x_{\rm th}$, with  $x_{\rm th}\equiv T/E$.
In practice, this `thermal' value $x_{\rm th}\sim 10^{-2}$ is quite small, so most of the energy lost
by the gluon cascade towards the medium is associated with the turbulent flow, and {\em not}
with the (BDMPSZ--like) 
gluon spectrum\footnote{Incidentally, this explains why earlier studies of the energy
distribution based on the BDMPSZ spectrum alone, which have not included 
the effects of multiple branchings, 
concluded that there should be very little energy in the gluon cascade 
at small $x$ and large angles \cite{Salgado:2003rv}, 
and thus failed to predict the phenomenon of di--jet asymmetry.}. Without this flow, there
would be no significant energy transfer towards very small $x\sim x_{\rm th}$.

Soft gluons propagate at large angles $\theta$ with respect to the jet axis: 
$\theta\sim k_\perp/\omega$, where $ k_\perp$ is the typical transverse momentum acquired
by the gluon via rescattering in the medium, and is at most weakly dependent upon $\omega$.
So, the ability of the medium--induced cascade to abundantly produce soft gluons provides a
natural explanation for the main feature of di--jet asymmetry: the fact that the energy difference
between the trigger jet and the away jet is carried by many soft ($p_T\lesssim 2$~GeV)
hadrons propagating at large angles ($\theta \gtrsim 0.8$) with respect to the axis
of the away jet \cite{Chatrchyan:2011sx}. This qualitative explanation has been originally
proposed in \cite{Blaizot:2013hx} and further developed in 
Refs.~\cite{Blaizot:2013vha,Iancu:2014aza,Kurkela:2014tla,Blaizot:2014ula}. 
However, these previous studies were not fully conclusive, 
as they did not explicitly consider the kinematical regime which is pertinent 
for di--jet asymmetry.
Namely, they focused on the `low--energy' regime where
the energy $E$ of the leading particle (LP) is smaller than the medium scale $\omega_c$.
Albeit the value of $\omega_c$ is not precisely known from first principles,
its current phenomenological estimates are well below the energy $E\gtrsim 100$~GeV
of the trigger jet in the experimental measurements of di--jet asymmetry (see the discussion 
in Sect.~\ref{sec:phys}). It is our main objective in this paper to provide 
a thorough analysis of the high--energy regime at $E\gg\omega_c$,
including its implications for the phenomenology.

In order to describe our results below, it is useful to recall the physical meaning of
the medium scale $\omega_c=\hat q L^2/2$: this is the highest possible
energy of a medium--induced emission by a parton with energy $E>\omega_c$
which crosses the medium over a distance $L$. The emission of a gluon with energy
$\omega_c$ has a formation time $t_{\rm br}(\omega_c)=L$ and hence a small probability 
$\abar[L/t_{\rm br}(\omega_c)]\sim \abar$: this is a {\em rare} event. Still, such rare
but hard emissions dominate the {\em average} energy loss by the LP, 
estimated as $\langle\Delta E\rangle\sim \abar \omega_c$
 \cite{Baier:1996kr,Baier:1996sk,Zakharov:1996fv,Zakharov:1997uu}. Hence, a very energetic
particle with $E\gg\omega_c$ loses only a small fraction $\abar(\omega_c/E)\ll 1$ of its
original energy and thus emerges from the medium with an energy  $E'\sim E$, which is
much larger than the maximal energy $\omega_c$ of its radiation. Accordingly, 
the spectrum shows a {\em gap} between a peak at $\omega\sim E$, which represents
the LP,  and a continuum at $\omega\le\omega_c$, which describes the radiation. 
The detailed structure of the peak is irrelevant for studies of the di--jet asymmetry: 
the energy carried by the LP is very closely collimated around the jet axis,
within a small angle\footnote{Here, 
$Q^2_L\equiv \hat q L$ is the transverse momentum broadening acquired by the LP while
crossing the medium over a distance $L$. Some typical
values are $Q_L=2$~GeV, $E=100$~GeV, and hence $\theta_{\rm LP}\sim 0.02$.}
 $\theta_{\rm LP}\sim Q_L/E\ll 1$, which is much 
smaller than the angular opening of the experimental `jet'. This is in agreement with the
experimental observation  \cite{Aad:2010bu,Chatrchyan:2011sx} that the azimuthal distribution 
of di--jets in Pb+Pb collisions is as narrowly peaked at $\Delta\phi=\pi$ as the 
corresponding distribution in p+p collisions.

In view of the above, our subsequent analysis will focus on the radiation part of the spectrum
at $x\le x_c$, where $x= \omega/E$ and $x_c= \omega_c/E\ll 1$. This part includes the
essential physics of multiple branching leading to energy loss via many
soft particles propagating at large angles. For the purposes
of this analysis, the LP can be treated as a steady source of radiation for gluons
with energy fractions $x\le x_c$. For this effective problem with a source, we will be able to
construct exact solutions for the gluon spectrum $D(x,t)$ at any time $t\le L$, and also for the
energy flux $\F(x,t)$ (the rate for energy transfer through the cascade; 
see Sect.~\ref{sec:low} for a precise definition). This energy flux,
and more precisely its `flow' limit $ \F_{\rm flow}(t)\equiv \F(x=0,t)$, is
the most interesting quantity in the present context, since it controls the energy transfer
by the gluon cascade to the medium.

A non--zero `flow' component in the energy flux is the main 
signature of turbulence  \cite{KST,Nazarenko} (e.g., there is no such a component
for the DGLAP cascade).
An important property of turbulence, which follows from the locality of the branchings,
is the fact that, within the `inertial range'  deeply between
the `source' and the `sink', the spectrum is fully determined by the energy flux together
with the KZ scaling law. For the standard turbulence in 3+1 dimensions, this relation is
known as the `Kolmogorov--Obukhov spectrum'. For our present problem in 1+1 dimensions 
(energy and time),   the `source' is the leading 
particle, the `sink' is the thermal bath, and the `inertial range' correspond to $x_{\rm th} \ll x\ll 1$.
{\em A priori}, our problem differs from the familiar turbulence set--up via its explicit
time--dependence: the source acts only up to a finite time $t_{\rm max}=L$, which moreover is quite
small, in the sense that $\hat q L^2\ll E$. 
%(In that sense, our problem corresponds to transient phenomena
%in the traditional turbulence problems, for which very little is known in general.) 
Notwithstanding, we shall demonstrate that a time--dependent generalization of 
the Kolmogorov--Obukhov relation holds for the problem at hand:
%\footnote{The validity of this relation is not
%restricted to the `high--energy' regime $E\gg\omega_c$: a similar relation holds
%when $E\lesssim\omega_c$, but it has not been noticed in Ref.~\cite{Blaizot:2013hx}
%(see the discussion in Sect.~\ref{sec:lowflux} below).}: 
the gluon spectrum at $x\ll x_c$ is fully determined
by the flow component %(the $x\to 0$ limit) 
of the energy flux, together with the characteristic scaling behavior of the BDMPSZ spectrum
(the KZ scaling for the present problem). Namely, we shall find $D(x,t)\propto \F_{\rm flow}(t)/\sqrt{x}$,
where the proportionality constant is under control. 

%\bigskip
 \begin{figure}[t]
	\centering
	\includegraphics[width=0.6\textwidth]{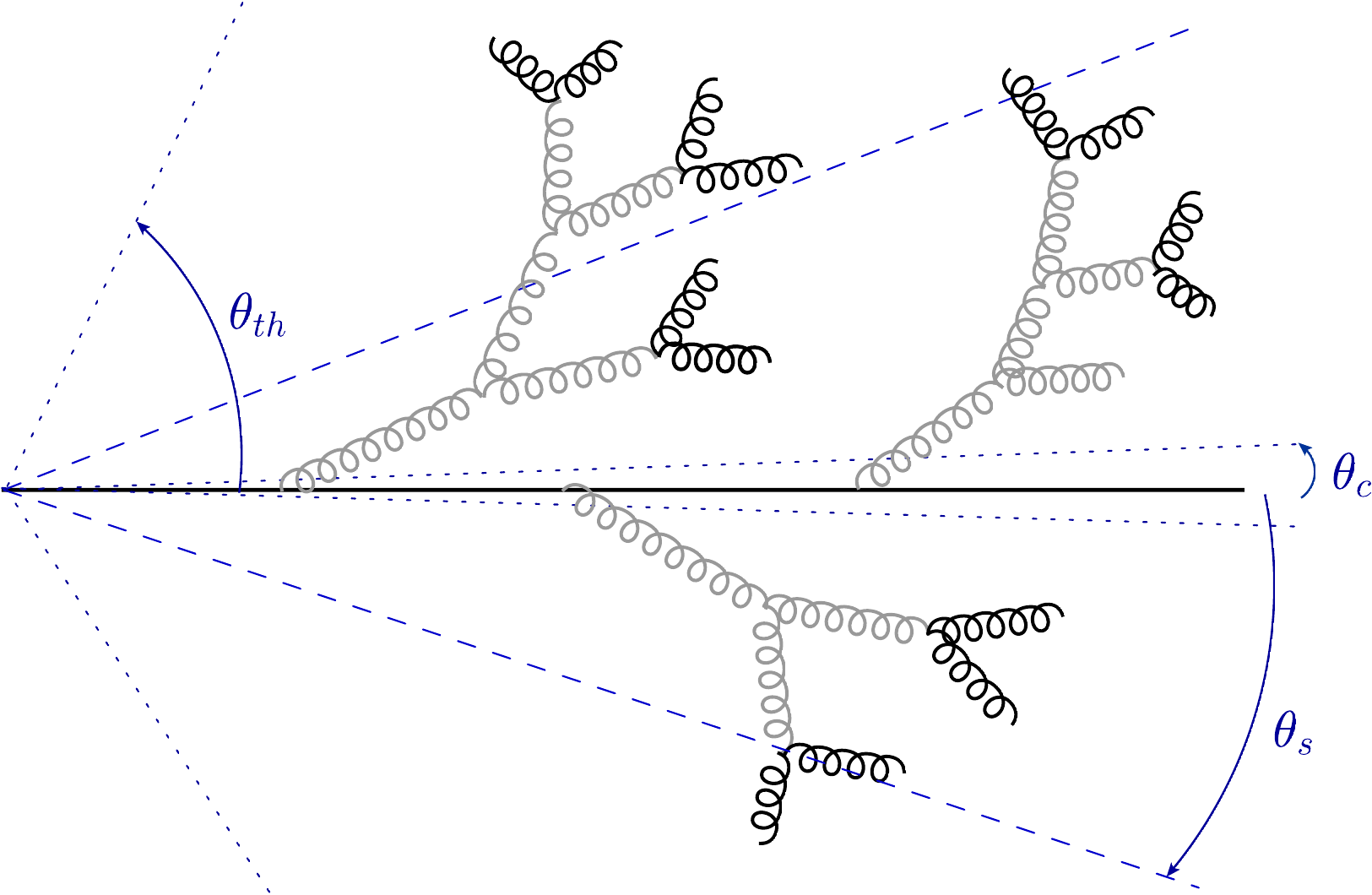}
		\caption{\sl  A typical gluon cascade as generated via medium--induced gluon
		branchings. The small angle $\theta_c\simeq Q_L/\omega_c$ is the propagation angle 
		for a relatively hard gluon with energy $\omega\sim\omega_c$. Such a hard 
		emission is a rare event and hence is not included in our typical event.
		All the shown gluons (besides the LP) have soft energies
		$\omega\lesssim\omega_s=\abar^2\omega_c$, hence their emissions
		occur with probability of $\order{1}$. The primary gluons are emitted (by the LP) 
		at a typical angle $\theta_s=\theta_c/\abar^2$ and subsequently disappear via
		branching into even 
		softer gluons. The opaque lines refer to gluons which exist at intermediate
		stages of the cascade, while the black lines refer to the `final' gluons, which
		thermalize and propagate at even larger angles, $\theta\sim 
		\theta_{\rm th}\gg \theta_s$  (see Sect.~\ref{sec:eloss} for details).}
		\label{fig:cascade}
\end{figure}
%\bigskip

The energy transferred by the gluon cascade to the medium can be identified with the energy
$\Delta E_{\rm flow}$ carried away by the flow, i.e. the time integral of $ \F_{\rm flow}(t)$ 
between $t=0$ and $t=L$.
For the high--energy regime under consideration, this quantity turns out to be independent
of the original energy $E$ of the LP and to have a transparent physical
interpretation\footnote{This estimate for $\Delta E_{\rm flow}$ 
holds to leading order in $\abar$~; see \eqn{DEflow} and
the plots in Sect.~\ref{sec:eloss} for more accurate results.}:
$\Delta E_{\rm flow}\simeq {\upsilon}\,\omega_s$, where $\omega_s=\bar{\alpha}^2 \omega_c$ 
and ${\upsilon}\simeq 4.96$ is a pure number which can be interpreted as
the average number of soft primary emissions with energies $\omega\sim \omega_s$.
Such soft gluons are radiated by the LP with probability of order one and 
they subsequently transfer their energy towards the medium via successive, quasi--democratic, branchings. A typical gluon cascade is illustrated in Fig.~\ref{fig:cascade}.
Using phenomenologically motivated values for $\hat q$ and $L$, we find
$\Delta E_{\rm flow}\simeq 10\div 20$~GeV  (see Sect.~\ref{sec:eloss}). Since carried by
very soft gluons, with energies $\omega\sim T\ll \omega_s$, this energy propagates at
very large angles with respect to the jet axis, at least as large as $\theta_s\equiv Q_L/\omega_s
\sim 0.5$. ($\theta_s$ is the typical propagation angle of the soft primary gluons, and its above
estimate will be discussed in Sect.~\ref{sec:eloss}.)
By progressively increasing the jet opening angle $\theta_0$ within a rather wide range,
say from  $\theta_0\sim \theta_s$  up to $\theta_0\sim 1$, we can recover part of the missing energy,
{\em but only very slowly} : most of this energy lies at even larger angles, 
$\theta\sim \theta_{\rm th} \gg \theta_s$ (see Fig.~\ref{fig:cascade} and Sect.~\ref{sec:eloss} for details).
The above predictions --- the numerical estimate for the energy loss at large angles 
$\Delta E_{\rm flow}$ and its extremely weak dependence upon the jet opening angle $\theta_0$ ---
are in good agreement, qualitative and even semi--quantitative, with the phenomenology of
di--jet asymmetry at the LHC \cite{Chatrchyan:2011sx,Aad:2012vca,Chatrchyan:2013kwa,Gulhan:2014}. 
Vice versa, we believe that these particular LHC data  could not be understood in a scenario 
which neglects
multiple branchings, nor in one which uses a vacuum--like model for the in--medium gluon 
fragmentation, that is, a model which ignores the quasi--democratic nature of the
soft branchings and the associated turbulent flow.

Our paper is organized as follows. In Sect.~\ref{sec:phys} we shall introduce, via qualitative
considerations and parametric estimates, the main physical scales which control the 
medium--induced gluon branching and allow one to separate between various physical regimes.
In Sect.~\ref{sec:low}, we shall consider the low--energy regime at $E\lesssim \omega_c$
as a warm--up. Besides a succinct review of the main results obtained in 
Ref.~\cite{Blaizot:2013hx}, this section will also contain some new material, like the
explicit calculation of the energy flux and a first discussion of the Kolmogorov--Obukhov relation.
Sects.~\ref{sec:highD} and \ref{sec:eloss} will be devoted to the main new problem of interest
for us here: the high--energy regime at $E\gg\omega_c$. Sect.~\ref{sec:highD} will present
the main theoretical developments: the justification of the effective problem with a source,
the exact, analytic and numerical, solutions for the radiation spectrum at $\omega\le\omega_c$
and for the turbulent flow, the democratic nature of the branchings and its physical
implications, and the proof of
the (time--dependent version of the) Kolmogorov--Obukhov relation for the branching 
dynamics at hand. Finally, in Sect.~\ref{sec:eloss} we shall discuss some phenomenological
consequences of this dynamics for the energy lost by the jet
via soft gluons propagating at large angles.

%\newpage

%%%%%%%%%%%%%%%%%%%%%%%%%%%%%%%%%
\section{Typical scales and physical regimes}
\label{sec:phys}
%%%%%%%%%%%%%%%%%%%%%%%%%%%%%%%%%

We would like to study the gluon cascade generated via successive medium--induced gluon
branchings by an original gluon --- the `leading particle'  (LP) --- with energy $E$ which propagates
through a dense QCD medium along a distance $L$. For the present purposes, the medium is 
solely characterized by a transport coefficient $\hat q$, known as the `jet quenching parameter',
which measures the dispersion in transverse momentum acquired by a parton propagating 
through this medium per unit length (or time).
Depending upon its energy, the leading particle can either escape the medium, or disappear 
inside it (in the sense of not being distinguishable from its products of fragmentation).
The actual scenario depends upon the ratio between $E$ and a characteristic medium
scale  $\omega_c\equiv \hat q L^2/2$,
which is the maximal energy of a gluon whose emission can be triggered by multiple
scattering in the medium: gluons with an energy $\omega\sim \omega_c$ have a formation
time of order $L$ and an emission probability of order $\abar\equiv\alpha_s N_c/\pi$.
Another energy scale that will play an important role in what follows is the {\em soft} scale
$\omega_s\equiv\abar^2\omega_c$~: gluons with $\omega\sim \omega_s$ have 
a relatively short formation time $t_{\rm br}(\omega)
\sim \abar L$ and an emission probability of order 1. This scale is `soft' since  
$\omega_s\ll \omega_c$ at weak coupling and since one generally has $E\gg \omega_s$
in the applications to phenomenology (see below).

More generally, the elementary
probability $\Delta \mcal{P}$ for a gluon with energy $\omega$ to be radiated (via the BDMPSZ
mechanism) during a time interval $\Delta t$ can be parametrically estimated as
\beq\label{DeltaP}
\Delta \mcal{P}\,\sim\,\abar\,\frac{\Delta t}{t_{\rm br}(\omega)}
 \,\sim\,\abar\,\sqrt{\frac{\hat q}{2\omega}}\,\Delta t\,,\eeq
where  $t_{\rm br}(\omega)\simeq\sqrt{2\omega/\hat q}$ is the `gluon formation time'
--- more precisely, the typical duration of a branching process in which the softest of the two 
daughter gluons has an energy $\omega \ll \omega_c$. \eqn{DeltaP} holds so long as 
$\Delta \mcal{P}\ll 1$. When $\Delta \mcal{P}\sim \order{1}$, the multiple branchings become important
and the evolution of the gluon cascade becomes non--perturbative (in the sense that the effects
of multiple branchings must be resumed to all orders). As clear from \eqn{DeltaP}, for any
$\Delta t < L$, there exists a sufficiently soft sector where the branching
dynamics is non--perturbative:
this occurs at $\omega\lesssim \omega_s(\Delta t) \equiv  \abar^2 \hat q \Delta t^2/2$. In particular,
for $\Delta t=L$, this yields back the `soft' scale aforementioned:
$\omega_s(L)= \omega_s$.

The above discussion in particular implies that the quantity $\omega_s$ sets the scale for the 
energy lost by the LP in a {\em typical event}~: with a probability of $\order{1}$, the LP particle
emits primary gluons with energies of $\order{\omega_s}$, and thus loses an energy $\Delta E
\sim \omega_s$. Accordingly, the {\em typical} energy loss,
as measured event--by--event, is sensitive to multiple
branchings. On the other hand, the {\em average} energy loss 
$\langle \Delta E\rangle$ is dominated by rare but hard emissions, with energies 
$\omega \gg \omega_s$, for which the effects of multiple branchings are negligible.
One finds indeed
\begin{eqnarray}\label{Eav}
\langle \Delta E\rangle\simeq \int^{\omega_{_{\rm max}}}{\rm d} \omega\,
  \omega\,\frac{{\rm d} N}{{\rm d} \omega}\,\simeq \abar
  \int^{\omega_{_{\rm max}}}{\rm d} \omega\, \sqrt{\frac{\omega_c}{\omega}}
 \sim\, \bar{\alpha}\sqrt{\omega_c\, \omega_{_{\rm max}}}\,,\end{eqnarray}
where the gluon spectrum $ \omega({\rm d} N/{\rm d} \omega)$ is essentially
the elementary probability for a single branching, \eqn{DeltaP}, 
evaluated for $\Delta t=L$ and the upper limit 
 $\omega_{_{\rm max}}\equiv {\rm min}(\omega_c, E)$ is typically much
larger than $\omega_s$. The integral in \eqn{Eav} is dominated
by its upper limit, i.e. by energies $\omega \sim \omega_{_{\rm max}} \gg  \omega_s$.

The global features of the medium--induced gluon
cascade depend upon the relative values of these three scales $E$,
$\omega_c$, and $\omega_s$.  Namely,  for a given medium scale $\omega_c$,
one can distinguish between three interesting physical regimes, depending
upon the energy $E$ of the leading particle : \texttt{(i)} {\em high energy} $E \gg \omega_c$, 
 \texttt{(ii)} {\em intermediate energy} $\omega_c  \gtrsim E \gg \omega_s$,  and
 \texttt{(iii)} {\em low energy} $E \lesssim \omega_s$. Recalling that 
$\omega_c= \hat q L^2/2$, we see that the `high energy' regime can also be
viewed as the limit where the in--medium path $L$ is relatively small, whereas the `low energy'
case corresponds to relatively large values of $L$.

In case  \texttt{(i)}, both the average energy loss $\langle \Delta E\rangle\sim \abar\omega_c$
and its typical value $\Delta E\sim\omega_s$ are much smaller than $E$, and the
probability to find the LP outside the (already narrow) energy interval $(E-\omega_c, E)$ is
negligibly small. Accordingly, in this case there is a {\em gap in the spectrum} between a
`peak' at $\omega\simeq E$ representing the leading particle  
and a `continuum' at $\omega \lesssim \omega_c$ representing the radiated gluons.

In case  \texttt{(ii)}, the typical
energy loss is still much smaller than $E$, so the leading particle survives in most of the
events, yet there is a sizable fraction of the events, of $\order{\abar}$ or larger, where both
fragmentation products carry similar energies. Accordingly, the LP peak is visible
in the spectrum, but there is no gap anymore. The average energy loss 
 $\langle \Delta E\rangle\sim \abar\sqrt{\omega_c E}$ is still  smaller than the original
 energy $E$, but it represents a relatively large fraction of it, of order $\gtrsim{\abar}$.

In case  \texttt{(iii)}, both the typical and the average energy loss are of order $E$,
meaning that the LP undergoes strong fragmentation and `disappears' in most, if not all, of the events.
Of course, this should be also the faith of the very soft $(\omega \lesssim\omega_s$) gluons
produced via radiation in cases \texttt{(i)} and \texttt{(ii)}. So,
in this third case, the spectrum contains no peak or other structure suggestive of the LP.

To summarize, the first two cases have in common the
fact that the LP survives after crossing the medium, but they differ in the actual shape of the spectrum
(with or without a gap). The last two cases are both characterized by the absence of a gap,
but they differ in the fact that the LP peak is still visible in  case  \texttt{(ii)}, whereas it is
totally washed out in  case  \texttt{(iii)}.

To make contact with the phenomenology, we chose $\hat q=1$~GeV$^2$/fm (a
reasonable estimate for a weakly coupled quark--gluon plasma \cite{Baier:1996kr} which moreover
appears to be consistent with  recent analyses of data  \cite{Burke:2013yra}), 
$\abar=0.3$, and let $L$ vary from 2 to 6~fm. For the three 
particular values $L=(2,\,4,\,6)$~fm, we deduce $\omega_c\simeq (10,\,40,\,90)$~GeV
and  $\omega_s\simeq (1,\,4,\,9)$~GeV. Hence,  when one is interested in the phenomenology
of high--energy jets with $E\ge 100$~GeV, as in the studies of di--jet asymmetry at the LHC,
one should mainly consider the case  \texttt{(i)} above.
On the other hand, for studies of the nuclear modification factor $R_{AA}$, where the energies
of the measured hadrons vary from 1~GeV to about 20~GeV, one is mostly in the situations
covered by cases  \texttt{(ii)} and \texttt{(iii)}.  These last two cases have been thoroughly 
discussed  in the recent literature, in particular in relation with the disappearance of the leading
particle and the energy transport at large angles 
\cite{Blaizot:2013hx,Blaizot:2013vha,Iancu:2014aza,Kurkela:2014tla,Blaizot:2014ula}, 
but to our knowledge the first case has not been studied in detail so far. 
From the previous discussion, is should be clear that this
is the most relevant case for a study of di--jet asymmetry in Pb+Pb collisions at the LHC.
This is the main problem that we would like to address in what follows.

%%%%%%%%%%%%%%%%%%%%%%%%%%%%%%%%%
\section{The low--energy regime}
\label{sec:low}
%%%%%%%%%%%%%%%%%%%%%%%%%%%%%%%%%

In preparation for the discussion of the high--energy regime at $E \gg \omega_c$,
it is useful to first review some known results concerning the low and intermediate
regimes at $E \lesssim \omega_c$ \cite{Blaizot:2013hx,Blaizot:2013vha} (see also 
Refs.~\cite{Baier:2000sb,Jeon:2003gi,Schenke:2009gb} for earlier, related, studies).
These two regimes can be simultaneously discussed, as they refer to different limits
of a same theoretical description. 

\subsection{The rate equation}

Throughout this paper we shall focus on the gluon
spectrum integrated over transverse momenta, i.e.
\beq\label{spectrum}
D(\omega,t)\,\equiv\,\omega\,\frac{\rmd N}{\rmd \omega}\,=\int \rmd^2\bk\ \omega\,
\frac{\rmd N}{\rmd \omega \rmd^2\bk}\,,\eeq
where $\omega\le E$ and $\bk$ denote the energy and respectively transverse momentum
of a gluon in the cascade, $N$ is the number of gluons,  and it is understood that the evolution time obeys
$0\le t\le L$. The function $D(\omega,t)$ describes the energy distribution within the cascade and
its evolution with time. For sufficiently soft gluons at least, namely so long as $\omega\ll
\omega_c$, and to leading order\footnote{A class of particularly large radiative corrections,
which are enhanced by the double--logarithm $\ln^2(LT)$, can be
effectively resummed into the effective dynamics by replacing the `bare' value of
the jet quenching parameter $\hat q$ by its renormalized value, as recently computed in 
Refs.~\cite{Liou:2013qya,Iancu:2014kga,Blaizot:2014bha,Iancu:2014sha,Wu:2014nca}.}
 in $\alpha_s$, this evolution can be described as a classical 
stochastic branching process \cite{Blaizot:2012fh,Blaizot:2013hx,Blaizot:2013vha}, 
with the elementary splitting rate determined by the BDMPSZ spectrum
\cite{Baier:1996kr,Baier:1996sk,Zakharov:1996fv,Zakharov:1997uu,Baier:1998kq}. 
Specifically, the differential probability per unit time 
and per unit $z$ for a gluon with energy $\omega$ to split into two gluons with 
energy fractions respectively $z$ and  $1-z$ is
%reads
 \beq\label{Pdef}
 \frac{\rmd^2 \mcal{P}_{\rm br}}{\rmd z\,\rmd t}\,=\,\frac{\alpha_s}{2\pi}\,
 \frac{P_{g\to g}(z)}{t_{\rm br}(z, \omega)},
  \eeq
where $P_{g\to g}(z)=N_c [1-z(1-z)]^2/z(1-z)$, with $0< z <1$, 
is the leading order gluon--gluon splitting function,
$N_c$ is the number of colors, and  $t_{\rm br}(z, \omega)$
is the typical duration of the branching process:
 \beq\label{taubr}
 t_{\rm br}(z, \omega)\equiv \sqrt{\frac{z(1-z)\omega}
 {\hat q_{\text{eff}}(z)}}\,,\qquad 
 \hat q_{\text{eff}}(z)\equiv \hat q \left[1-z(1-z)\right]\,.\eeq
Note that this branching time depends upon both the energy $\omega$
of the parent gluon and the splitting fraction $z$, and that it is much smaller than $L$
whenever at least one of the two daughter particles,
with energies $z\omega$ and respectively $(1-z)\omega$,  is soft compared to $\omega_c$. 
%with energies $z\omega$ and respectively $(1-z)\omega$,  are .

The elementary splitting rate \eqref{Pdef} together with the requirement of probability
conservation completely specifies the structure of the stochastic branching process and,
in particular, the evolution equation obeyed by the gluon spectrum. So long as $E < \omega_c$,
this equation reads
 \begin{align}\label{eqDf}
   \frac{\del D(x,\tau)}{\del\tau}\,
  =\,\abar\int \rmd z \,{\cal K}(z)  
  \left[\sqrt\frac{z}{x}
  D\Big(\frac{x}{z}, \tau\Big)-\frac{z}{\sqrt{x}}\,
  D\big({x},\tau\big)\right],
  \end{align}
in convenient notations where $D(x,\tau)\equiv D(\omega,t)$,  $x\equiv \omega/E \le 1$ is
the energy fraction with respect to the leading particle, and 
 \beq\label{tau}
\tau\,\equiv\,\sqrt{\frac{\hat q}{E}}\,t
 =\sqrt{2\,x_c}\,\frac{t}{L}\,,\qquad x_c\equiv \frac{\omega_c}{E}\,,
 \eeq
is the reduced time (the evolution time in dimensionless units). Notice that $x_c>1$
for the physical problems discussed in this section. The splitting kernel ${\cal K}(z)$ is defined as
 \beq\label{Kdef}
 \mcal{K}(z)\,\equiv\,\frac{f(z)}{[z(1-z)]^{3/2}}=\mcal{K}(1-z)\,,\qquad
 f(z)\equiv\big[ 1-z(1-z)\big]^{5/2}\,.\eeq
It depends only upon the splitting fraction $z$ since the corresponding dependence
upon the energy (fraction) $x$ of the leading particle, cf. \eqn{Pdef}, 
has been explicitly factored out in writing \eqn{eqDf}.
 
We shall  refer to the r.h.s. of Eq.~(\ref{eqDf}) as the `branching term' and denote it as $\abar{\cal I}[D]$.
This is the sum of two terms, which can be recognized as the familiar `gain' and 
`loss' terms characteristic of a branching process. The first term,  which is positive
and nonlocal in $x$, is the {\em gain term} : it describes the rise in the number of gluons
at  $x$ due to emissions from gluons at larger $x'=x/z$. The respective integral over $z$
is restricted to  $x< z < 1$ by the support of $D({x}/{z}, \tau)$.
The second, negative, term, which is local
in $x$, represents the {\em loss term} and describes the reduction in the
number of gluons at $x$ due to their decay into gluons with smaller  $x'=zx$.
Taken separately, the gain term and the loss term
in \eqn{eqDf} have endpoint singularities at $z=1$, but these singularities exactly 
cancel between the two terms and the overall equation is well defined.

As anticipated,  \eqn{eqDf} encompasses the two regimes at `low' and `intermediate' energies
introduced in Sect.~\ref{sec:phys}. In fact, there is no fundamental difference between the dynamics
in these two regimes, rather they differ only in the maximal value for the reduced time $\tau$
which is allowed in practice. This maximal value, namely $\tau_L\equiv \sqrt{2\,x_c}=
\sqrt{{\hat q}/{E}}\,L$, increases with the medium size $L$, but decreases with the energy $E$
of the leading particle. In the `intermediate energy' regime, the evolution is limited to relatively small times,
$1\lesssim \tau_L \ll 1/\abar$, whereas in the `low energy' one, it can extend
up to much larger values: $\tau_L \gtrsim 1/\abar$. This explains the qualitative differences between
the two regimes that were anticipated in Sect.~\ref{sec:phys} and will be now demonstrated via
explicit solutions to  \eqn{eqDf}.

\subsection{The spectrum and the flow energy}
\label{sec:Dlow}

To study the effects of multiple branchings, one needs a non--perturbative solution 
to Eq.~\eqref{eqDf}. Whereas it is straightforward to solve this equation via numerical methods,
for the purpose of demonstrating subtle physical phenomena, it is much more convenient 
to dispose of an analytic solution. Such a solution has been obtained in Ref.~\cite{Blaizot:2013hx},
but for the simplified kernel ${\cal K}_0(z)\equiv 1/{[z(1-z)]^{3/2}}$, which is obtained 
from \eqn{Kdef} after replacing the slowly varying factor $f(z)$ in  the numerator by 1.
This simplified kernel has the same singularities at $z=0$ and $z=1$ as the original
kernel ${\cal K}(z)$, hence it is expected to have similar physical implications,
at least qualitatively. (This will also  be checked via numerical simulations later on; see
e.g. Fig.~\ref{fig:fullvsmodel}.)

For the simplified kernel ${\cal K}_0(z)$ and the initial condition ${D}(x,\tau=0)=\delta(x-1)$, 
corresponding to a single gluon (the `leading particle' ) 
carrying all the energy at $\tau=0$, the exact solution reads \cite{Blaizot:2013hx}
\beq\label{Dexact}
  D(x,\tau)\,=\,\frac{\abar\tau}{\sqrt{x}(1-x)^{3/2}}\ \exp\left\{-\frac{\pi\abar^2\tau^2}{1-x}\right\}\,.\eeq  
This is recognized as the product between the BDMPSZ spectrum \cite{Baier:1996kr,Baier:1996sk,Zakharov:1996fv,Zakharov:1997uu,Baier:1998kq} (which is the same as the
result of the first iteration of  \eqn{eqDf}),
 \beq\label{Dzero}
  D_0(x,\tau)\,=\,\frac{\abar\tau}{\sqrt{x}(1-x)^{3/2}}\,,
  \eeq
and a Gaussian factor describing, at early times, the broadening of the peak associated with the LP \cite{Baier:2001yt} and, at late times, the suppression of the spectrum as whole.

To be more specific, consider increasing the time from $\tau=0$
up to the maximal value $\tau_L=\sqrt{2\,x_c}$, where we recall that $x_c>1$.
When $\tau\to 0$, the r.h.s. of \eqn{Dexact} approaches $\delta(x-1)$, as it should.
So long as $\tau$ is small enough for $\pi\abar^2\tau^2\ll 1$, the spectrum 
exhibits a pronounced peak in the vicinity of $x=1$, which describes the leading particle:
the maximum of this peak lies at $x_p$ with $1-x_p \simeq (2\pi/3)\abar^2\tau^2$ 
and its width $\Delta x$ around $x_p$ is of order $\pi\abar^2\tau^2$. 
The fact that the peak gets displaced below 1
is a consequence of the Gaussian factor in \eqn{Dexact}, which strongly suppresses the spectrum 
for $x$ close to 1, within a window
\beq\label{suppress}
1-x \,\lesssim\, \pi\abar^2\tau^2\,\ll\,1\,.\eeq
The physical origin of this suppression should be clear in view of the discussion in Sect.~\ref{sec:phys}:
for $x$ close to 1, the quantity $\epsilon\equiv (1-x)E$ is the energy lost by the leading particle 
via radiation. \eqn{suppress} shows that the typical value of this energy is
$\epsilon(t)\simeq 2\pi \omega_s(t)$, with $\omega_s(t) = \abar^2 \hat q t^2/2$ the
non--perturbative scale for the onset of multiple branching, as introduced in Sect.~\ref{sec:phys}.
That is, the LP copiously radiates very soft gluons, for which the emission probability is of $\order{1}$,
and thus loses an energy of order $\omega_s(t)$. Interestingly, this energy loss
is enhanced by the relatively large numerical factor $2\pi$, which can be interpreted as the average
number of gluons with energy $\omega\sim\omega_s(t)$ that are emitted by the LP during a time
interval $t$. This interpretation will be supported by other findings below.

%Similar %, purely numerical, 
%enhancement factors will often appear in the subsequent analysis; they are potentially important
%for the phenomenology, as they amplify some physical effects 
%beyond the respective parametric estimates.

Let us now increase $\tau$ towards larger values $\pi\abar^2\tau^2\gtrsim1$. This is of course
possible only in the `low energy' regime where $\tau_L \gtrsim 1/\abar$. Then the Gaussian
suppression extends to all values of $x$, the LP peak gets washed out  ---
it broadens, it moves towards smaller values of $x$, and its height is decreasing --- and eventually
disappears from the spectrum. One can say that a LP with energy
$E\lesssim \abar^2 \omega_c$ has a finite `lifetime' inside the 
medium, of order $\Delta\tau\sim 1/\abar$ or, in physical units (cf. \eqn{tau}),
\beq\label{Deltat}
\Delta t\,\sim\,\frac{1}{\abar}\,
\sqrt{\frac{E}{\hat q}}\ .\eeq
More precisely, this means that the LP has fragmented into gluons
which carry a sizable fraction of its original energy $E$. 
Via successive branchings, the energy
gets degraded to lower and lower values of $x$, and it is interesting to understand this evolution
in more detail. A priori, one might expect this energy to accumulate in the small--$x$ part
of the spectrum, and notably at $x\lesssim x_s(\tau) \equiv \abar^2\tau^2$
(corresponding to  $\omega\lesssim\omega_s(t)$),
but \eqn{Dexact} shows that this is actually {\em not} the case: for $x\ll 1$,
 \eqn{Dexact} reduces to
\beq\label{Dsmallx}
  D(x,\tau)\,\simeq\,\frac{\abar\tau}{\sqrt{x}}\ \rme^{-\pi\abar^2\tau^2}\,,\eeq
which has exactly the same shape in $x$ as the small--$x$ limit of the BDMPSZ spectrum, \eqn{Dzero}. 
In fact, \eqn{Dsmallx}  formally looks like the BDMPSZ spectrum produced via a single emission 
by the LP, times a Gaussian factor describing the decay of the LP with increasing time. This
interpretation seems to imply that multiple branchings are not important at 
small $x$, but from the discussion in Sect.~\ref{sec:phys} we know that this cannot be true:
after a time $t$, the single--branching probability becomes of order one (meaning that 
multiple branching becomes important) for all the soft modes obeying 
$x < x_s(\tau)$. This last condition can also be inferred from
\eqn{Dzero}: when $x \sim x_s(\tau)\ll 1$, the BDMPSZ spectrum becomes of $\order{1}$.

We are thus facing an apparent paradox --- in spite of the importance of multiple branching,
the energy does not get accumulated in the bins of the spectrum at small $x$ --- which finds
its solution in the
phenomenon of {\em wave turbulence} \cite{Blaizot:2013hx}. The BDMPSZ spectrum at
small $x$ is not modified by the fragmentation because this represents a {\em fixed point} of
the rate equation \eqref{eqDf} at small $x\ll 1$: the branching term vanishes 
(meaning that the `gain' and `loss' terms compensate each other) when evaluated 
with the `scaling' spectrum $D_{\rm sc}(x)\equiv 1/{\sqrt{x}}$. This can be recognized
as the Kolmogorov--Zakharov (KZ) spectrum \cite{KST,Nazarenko} for the branching process 
at hand. In turn, the existence of this
fixed point  implies that, via successive branchings, the energy gets transmitted from large 
$x$ to small $x$, without accumulating at any intermediate value of $x$ : it rather 
flows throughout the spectrum and accumulates into a condensate at $x=0$.

 \begin{figure}[t]
	\centering
	\includegraphics[width=0.7\textwidth]{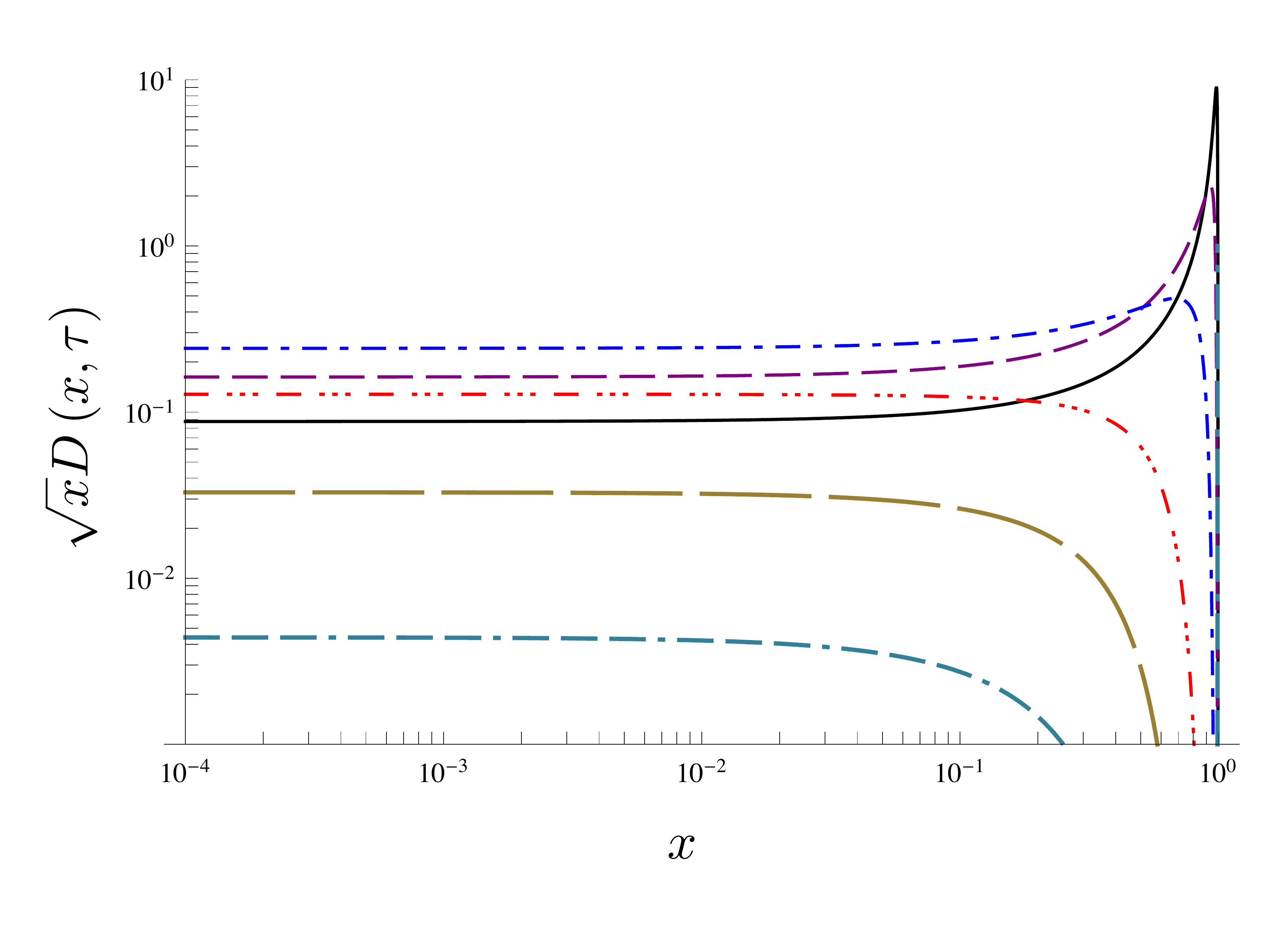}
		\caption{\sl Plot (in log-log scale) of $\sqrt{x} D(x,\tau)$, 
		with $D(x,\tau)$ given by Eq.~(\ref{Dexact}), as a function of $x$ for various values of $\tau$:
		solid (black): $\tau=0.3$; dashed (purple): $\tau=0.6$;
 dashed--dotted (blue): $\tau=1.3$; dashed--triple dotted (red): $\tau=2.5$;
  long--dashed (brown): $\tau=3.5$; triple dashed--dotted (green): $\tau=4.5$.
		We use $\bar{\alpha}=0.3$.
}
		\label{fig:Dlargexc}
\end{figure}

This is illustrated in Fig.~\ref{fig:Dlargexc}, where the exact solution \eqref{Dexact} 
is represented as a function of $x$ for several values of $\tau$, up to relatively large
values, such that $\pi\abar^2\tau^2\gtrsim1$. The early--time set of curves, at $\tau\lesssim 1$,
where the LP peak is still visible, is representative for the `intermediate energy' regime, 
where the late--time 
curves, from which the LP has disappeared and where the spectrum is seen to be
suppressed as a whole, correspond to the `low energy' case.

The energy flow can also be studied analytically, on the basis of \eqn{Dexact}. To that
aim, consider the energy balance between spectrum and flow. The energy fraction contained 
in the spectrum after a time $\tau$ is computed as
 \cite{Blaizot:2013hx} 
\begin{eqnarray}\label{Espec} {\cal E}(\tau)\,=\,
\int_0^1 {\rm d} x \,D(x,\tau)\,=\,{\rm e}^{-\pi\abar^2\tau^2}\,,\end{eqnarray}
and decreases with time. The difference 
\begin{eqnarray}\label{Eflow} 
  {\cal E}_{\rm flow}(\tau)\,\equiv \,1-{\cal E}(\tau)\,=\, 1- {\rm e}^{-\pi\abar^2\tau^2}\,,\end{eqnarray}  
is the energy fraction carried by the flow, i.e. by the multiple branchings, and which formally ends up
in a condensate at $x=0$.  For sufficiently large times $\abar\tau\gtrsim1$ (corresponding to
the low--energy regime),
this can be as large as the total initial energy of the LP. 

It is also interesting to consider the small time limit of \eqn{Eflow}, that is
\beq
\label{Eflowtau}
 {\cal E}_{\rm flow}(\tau)\,\simeq\,\pi\abar^2\tau^2\,=\,2\pi x_s(\tau)\quad\mbox{for}\quad
 \pi\abar^2\tau^2\ll 1\,.
 \eeq
This result can be interpreted as follows: $\upsilon_0\equiv 2\pi$ is the average number of
primary gluons with energies of the order of $\omega_s(t) = \abar^2 \hat q t^2/2$
that are emitted by the leading particle during a time $t$. This number is
independent of $t$ or $\abar$, since such gluon emissions occur with probability of order one.
Stated differently, the typical time interval between two successive such emissions
is of order $t$. [This interval can be estimated from the condition that $\Delta \mcal{P}\sim \order{1}$,
with $\Delta \mcal{P}$ given by \eqn{DeltaP} with $\omega\sim\omega_s(t)$~;
this implies $\Delta t\sim (1/\abar)
t_{\rm br}(\omega_s(t))\simeq t$.] After being emitted,
these soft primary gluons rapidly cascade into even softer gluons 
and thus eventually transmit (after a time $\Delta t\sim t$ estimated as above)
their whole energy to the arbitrarily soft quanta which compose the flow.
This argument also shows that the gluons with energies $\omega\sim\omega_s(t)$
not only are emitted with a probability of $\order{1}$ during an interval of order ${t}$,
but also have a `lifetime' $\Delta t\sim t$ before they branch again with probability of $\order{1}$.
 
\subsection{Energy flux, turbulence, and thermalization}
\label{sec:lowflux}

\begin{figure}[t]
	\centering
	\includegraphics[width=0.7\textwidth]{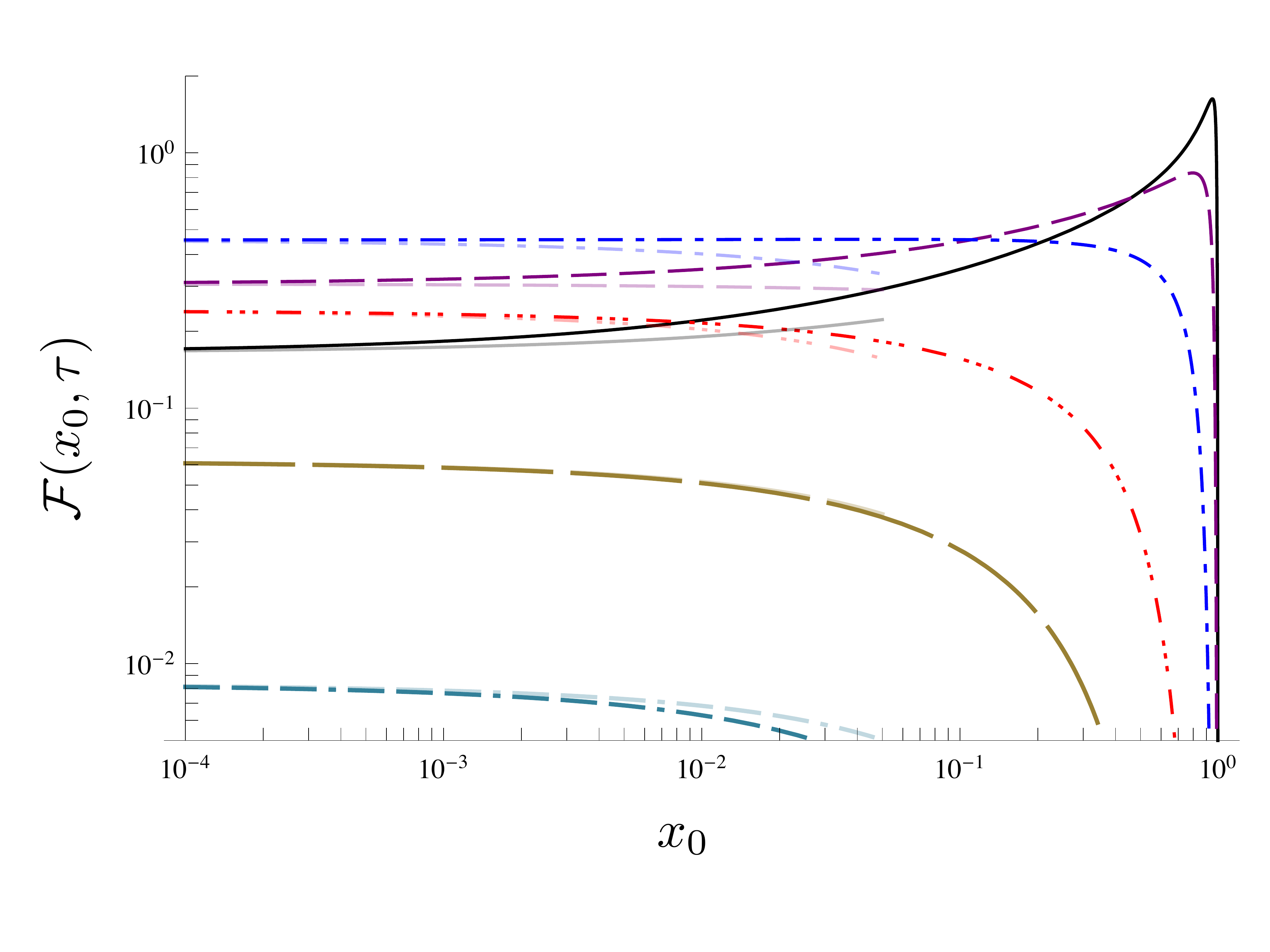}
		\caption{\sl Plot (in log-log scale) of the energy flux $\F(x_0,\tau)$, cf. Eq.~(\ref{Fx0}), as a function of $x_0$ for various values of $\tau$. We use the same conventions  as in Fig.~\ref{fig:Dlargexc}. 
		The thin curves, which are drawn for $x_0\le 0.05$, represent the approximation in \eqn{Fsmallx0}, which is valid at small $x_0$.}
		\label{fig:Plargexc}
\end{figure}

The physical interpretation of \eqn{Eflow} in terms of multiple branchings and, in particular,
its relation to turbulent flow become more transparent if one studies a more differential
quantity,  the {\em energy flux} $ \F(x_0,\tau)$. This is defined as the 
rate for energy transfer from the region $x>x_0$ to the region $x < x_0$. 
Since the energy in the region $x>x_0$ is decreasing with time, via branchings,
it is natural to define the flux as the following, positive, quantity 
 \begin{align}\label{Fx0}
 \F(x_0,\tau)&\,\equiv\,-\frac{\partial {\cal E}^{\,>}(x_0,\tau)}{\partial\tau}\,=\,
 \frac{\partial {\cal E}^{\,<}(x_0,\tau)}{\partial\tau}\,,
 %2\pi \abar^2\tau {\cal E}^{\,>}(x_0,\tau)+2\abar\sqrt{\frac{x_0}{1-x_0}}\
% \rme^{-\frac{\pi\abar^2\tau^2} {1-x_0}}\,.
 % &\simeq \Big(2\pi \abar^2\tau +2\abar\sqrt{x_0}\Big)
 % {\rm e}^{-\pi\abar^2\tau^2}\,,
 \end{align}
where ${\cal E}^{\,>}(x_0,\tau)$ is the energy fraction contained %at time $\tau$
in the bins of the spectrum with $x>x_0$, that is,
%(the second equality below follows after using  \eqn{Dexact}),
\beq\label{Dexint}
 {\cal E}^{\,>}(x_0,\tau)=\int_{x_0}^1\rmd x\, D(x,\tau)
 %\,=\,\rme^{-\pi \abar^2\tau^2}\,\frac{2}{\sqrt{\pi}}
% \int_{\eta_0}^\infty\rmd\eta \,\rme^{-\eta^2}\,,\qquad \eta_0\equiv \abar\tau\sqrt{\frac{\pi x_0}{1-x_0}}
 \,,\eeq
whereas the complementary quantity ${\cal E}^{\,<}(x_0,\tau)$ 
is the energy fraction carried by the modes with $x < x_0$. In turn, ${\cal E}^{\,<}(x_0,\tau)$ 
is the sum of two contributions : the flow energy \eqref{Eflow} and the energy contained
in the bins of the spectrum at $x < x_0$~; that is,
 \begin{align}\label{Elow}
% {\cal E}^{\,>}(x_0,\tau)&\,=\int_{x_0}^1\rmd x\, D(x,\tau) \,,\nn
 {\cal E}^{\,<}(x_0,\tau)&\,=\, 1 - {\cal E}^{\,>}(x_0,\tau)\,=\,{\cal E}_{\rm flow}(\tau)
 \,+\,\int_0^{x_0}\rmd x\, D(x,\tau)\,.
 \end{align}
 Using the above definitions together with \eqn{Dexact} for $D(x,\tau)$, it
is straightforward to numerically compute the energy flux $ \F(x_0,\tau)$, 
with the results displayed in Fig.~\ref{fig:Plargexc}. For a 
physical discussion, it is convenient to focus on the behavior 
at small $x_0\ll 1$. In that region, one can use \eqn{Elow} together with the small--$x$ approximation 
to the spectrum, \eqn{Dsmallx}, to deduce the analytic estimate
\begin{align}\label{Fsmallx0}
  \F(x_0,\tau)&\,\simeq \Big[2\pi \abar^2\tau +2\abar\sqrt{x_0} \big(1- 2\pi\abar^2\tau^2\big)\Big]
  {\rm e}^{-\pi\abar^2\tau^2}\,.
 \end{align}
The first term within the square brackets, which is independent of $x_0$,  is the flow contribution, 
 \begin{align}\label{Fflow}
  \F_{\rm flow}(\tau)\,\equiv \,
  \frac{\partial {\cal E}_{\rm flow}(\tau)}{\partial\tau}\,=\,2\pi \abar^2\tau \,
    {\rm e}^{-\pi\abar^2\tau^2}\,,
 \end{align}
while the second term, proportional to  $\sqrt{x_0}$, is the rate at which the energy 
changes in the region of the spectrum at $x\le x_0$. Clearly, the flow component
in \eqn{Fflow} dominates over the non--flow one at sufficiently small values of $x_0$, such that
$x_0\lesssim x_s(\tau)= \abar^2\tau^2$.  This is also visible in Fig.~\ref{fig:Plargexc}, where
the various curves become indeed flat at sufficiently small $x_0$.

\comment{ \begin{align}\label{Fx0}
 \F(x_0,\tau)&\,=\,-\frac{\partial {\cal E}^{\,>}}{\partial\tau}\,=\,
 2\pi \abar^2\tau {\cal E}^{\,>}(x_0,\tau)+2\abar\sqrt{\frac{x_0}{1-x_0}}\
 \rme^{-\frac{\pi\abar^2\tau^2}
  {1-x_0}}\,.
 % &\simeq \Big(2\pi \abar^2\tau +2\abar\sqrt{x_0}\Big)
 % {\rm e}^{-\pi\abar^2\tau^2}\,,
 \end{align}
This last expression becomes more transparent when specialized to small $x_0\ll 1$, where it can be approximated as}

We thus see that the small--$x$ behavior of the flux, and
unlike the corresponding behavior of the spectrum, does reveal the non--perturbative nature
of the multiple branchings and of the associated scale $x_s$ : in the soft region at 
$x_0\lesssim x_s(\tau)$, the flux $ \F(x_0,\tau)$ is controlled by its `flow' component
and is quasi--independent of $x_0$. A {\em uniform} energy flux is the 
distinguished signature of (wave) turbulence \cite{KST,Nazarenko}. It physically means that the 
energy flows through the spectrum without accumulating at intermediate values of $x$.
To see this, let us compute the rate of change for ${\cal E}(x_1,x_2,\tau)$  ---
the energy fraction contained within the interval $x_1<x<x_2$ :
\beq
{\cal E}(x_1,x_2,\tau)\,=\int_{x_1}^{x_2}\rmd x\, D(x,\tau)\ \ \Longrightarrow\ \
\frac{\partial {\cal E}(x_1,x_2,\tau)}{\partial\tau}\,=\, \F(x_2,\tau)- \F(x_1,\tau)\,.\eeq
This rate vanishes if the flux is independent of $x$. In our case, the flux is not strictly uniform,
not even at very small values of $x$ (see \eqn{Fsmallx0}). Yet, the energy flux which crosses a bin at 
$x \lesssim x_s(\tau)$ is much larger than the rate for energy change in that bin: 
the energy flows through the bin, without accumulating there.

It is intuitively clear that a quasi--uniform flux requires the branchings to be {\em quasi--local} in $x$
(or `quasi--democratic'). Since, if the typical branchings were strongly asymmetric, then after each
branching most of the energy would remain in the parent gluon
and the energy would accumulate in the bins at large $x$. 
It is also quite clear, in view of the general arguments in Sect.~\ref{sec:phys}, that
in the non--perturbative region at $x \lesssim x_s(\tau)$ the branchings are  indeed quasi--local:
a gluon with energy $x\sim x_s(\tau)$ splits with probability of $\order{1}$ during a time
interval $\tau$ {\em irrespective of the value $z$ of the splitting fraction}. 
Hence, there is no reason why special
values like $z\ll 1$ or $1-z\ll 1$ should be favored. A more elaborate argument in favor of
democratic branchings will be presented in Sect.~\ref{sec:flow2}.

The locality of the interactions is a fundamental property of turbulence   \cite{KST,Nazarenko}.
In the traditional turbulence problem, where the energy is injected by a time--independent source 
which is localized in energy and produces a steady spectrum, this property ensures that the 
energy spectrum in the `inertial range'  (i.e. sufficiently far away from the source) can
be expressed in terms of the (steady) flux $\F$ and a special power--like spectrum,
the `Kolmogorov--Zakharov spectrum', which is a fixed--point of the `collision term'. 
In the case of hydrodynamic turbulence in 3+1 dimensions, this relation between the energy
spectrum and the flux is known as the `Kolmogorov--Obukhov spectrum'.
 
For the problem at hand, where the `source' is the leading particle originally localized at $x=1$,  
the `inertial region'  corresponds to $x\ll 1$, the  `collision term' term is the branching 
term $\abar\mcal{I}[D]$,  and the fixed--point solution is the scaling spectrum 
$D_{\rm sc}(x)= 1/{\sqrt{x}}$. But unlike for the more conventional set--up, 
our current problem is clearly {\em not} stationary: 
the `source' (the LP) loses energy and can even disappear at large times, 
so both the spectrum and the energy flux have non--trivial time dependencies.
%It is possible, albeit physically less interesting, to associate a steady problem to our
%branching dynamics as well. Examples in that sense will be given in Sect.~\ref{sec:highD}
%(see also  \cite{Blaizot:2013hx}) and shown to follow the expected pattern for turbulence.
Notwithstanding, it turns out that the fundamental relation alluded to above, 
between the energy spectrum and the flux, also 
holds for the time--dependent physical problem at hand. Namely, by
inspection of Eqs.~\eqref{Dsmallx} and \eqref{Fflow}, it is clear than one can write
 \beq\label{Kolmog}
  D(x,\tau)\,\simeq\,\frac{1}{2\pi\abar}\,
  \frac{\F_{\rm flow}(\tau)}{\sqrt{x}}\qquad\mbox{for}\quad x\,\ll\,1\,.\eeq
This relation can be recognized as a version of the celebrated Kolmogorov--Obukhov scaling 
adapted to the current problem and generalized to
a time--dependent situation. Note that \eqn{Kolmog} involves only the flow contribution to the
flux, albeit this relation holds for any $x\ll 1$ and not only in the `non--perturbative' sector
at $x\lesssim x_s(\tau)$. At this level, the
relation  \eqref{Kolmog} might look fortuitous, but in Sect.~\ref{sec:flow2}
we shall present a general argument showing that it has a deep physical motivation.

So far, we have implicitly assumed that the branching dynamics as described   
by \eqn{eqDf}  extends all the way down to $x=0$, that is, it includes arbitrarily soft gluons.
In reality, the dynamics should change at sufficiently low energies, for various reasons.
First, when the gluons in the cascade become as soft as the medium constituents
--- that is, their energies become comparable to the 
temperature $T$ --- they rapidly thermalize via collisions in the medium and thus
`disappear' from the cascade.  Second, the BDMPSZ branching law \eqref{Pdef}
assumes the dominance of multiple soft scattering and hence it ceases to be valid when the
branching time $t_{\rm br}(z, \omega)$ becomes as low as the mean free path
$\ell$ between successive collisions in the medium. This condition restricts the gluon energies
to values $\omega\gtrsim \omega_{_{\rm BH}}\equiv\hat q\ell^2/2$. 
For a weakly coupled quark--gluon plasma, the `Beithe--Heitler' scale $\omega_{_{\rm BH}}$
is comparable to the temperature $T$.  (Indeed, in this case, one has
$\hat q\sim \abar^2 T^3$ and $\ell\sim (\abar T)^{-1}$ to parametric accuracy.)
With this example in mind, we shall not distinguish
between these two scales anymore, but simply assume that
the dynamics described by \eqn{eqDf} applies for all the energies
$\omega\gtrsim T$, i.e., for all $x\gtrsim x_{\rm th}\equiv T/E$.
In all the interesting problems, the thermal scale $x_{\rm th}$ is 
small enough to allow for multiple branchings: $x_{\rm th} \ll x_s=\abar^2 x_c$.  
For instance, in the case of a weakly coupled plasma, the above condition
is tantamount to % $\omega_{_{\rm BH}}\ll \abar^2 \omega_c$, or 
$L\gg \ell/\abar \sim (\abar^2 T)^{-1}$, which is indeed satisfied since the interesting values for $L$ are
much larger than the typical relaxation time $\lambda_{\rm rel}\sim (\abar^2 T)^{-1}$ of the plasma.

Notice that we implicitly assume here that the thermalization mechanism acts as a `perfect
sink' at $x\sim x_{\rm th}$. (A similar
assumption was made e.g. in the `bottom--up' scenario for thermalization \cite{Baier:2000sb}.)
That is, the surrounding medium 
absorbs the energy from the cascade at a rate equal
to the relevant flux $\F(x_{\rm th},\tau)$, without modifying the branching dynamics
at higher values $x\gg x_{\rm th}$.  This is a rather standard assumption in the context of
turbulence and is well motivated for the problem at 
hand, as we argue now.
To that aim, one should compare the relaxation time $\lambda_{\rm rel}\sim (\abar^2 T)^{-1}$
aforementioned, which represents the characteristic  thermalization time
at weak coupling, with the lifetime $\Delta t (\omega)$  of a gluon generation 
(the time interval between two successive branchings)
for gluons with energy $\omega\sim T$, which is the characteristic time scale for
the turbulent flow. This $\Delta t (\omega)$ can be estimated 
as explained at the end of Sect.~\ref{sec:Dlow}, and reads (to parametric accuracy)
 \beq\label{lifetime}
 \Delta t(\omega) \sim \frac{1}{\abar}\,
t_{\rm br}(\omega)\,\sim\, \frac{1}{\abar}\sqrt{\frac{\omega}{\hat q}}\,.
\eeq
Using $\omega\sim T$ and the perturbative estimate
$\hat q\sim \abar^2 T^3$, one deduces
$\Delta t (T)\sim (\abar^2 T)^{-1}\sim  \lambda_{\rm rel}$. We thus conclude
that the physics of thermalization is as efficient  in dissipating the energy
as the turbulent flow. This implies that there should be no
energy pile--up towards the low--energy end of the cascade.

Under these assumptions, 
it is interesting to compute the total energy lost by the cascade towards
the medium, i.e. `the energy which thermalizes'. 
This is the same as the energy which has the crossed the bin $x_{\rm th}$
during the overall time $\tau_L$, namely (cf. \eqn{Elow})
 \beq\label{Eth}
{\cal E}_{{\rm th}} \equiv {\cal E}^{\,<}(x_{\rm th},\tau_L)\,\simeq\,1-
\rme^{-\pi \abar^2\tau_L^2}+2
 \abar\tau_L\sqrt{x_{\rm th}}\,\rme^{-\pi \abar^2\tau_L^2}\,,
\eeq
where the approximate equality holds since $x_{\rm th}\ll 1$.
\eqn{Eth} is recognized as the sum of the flow energy, \eqn{Eflow}, and of the energy that
would be contained in the spectrum at $x\le x_{\rm th}$, cf. \eqn{Dsmallx}.
Using $\tau_L=\sqrt{2x_c}$ and $x_{\rm th} \ll x_s=\abar^2 x_c$, it is easy to check that 
the flow component dominates over the spectrum piece, and hence ${\cal E}_{{\rm th}} \simeq
{\cal E}_{\rm flow}(\tau_L)$.
This implies that the energy lost by the gluon cascade
towards the medium is independent of the details of the thermalization process, like the
precise value of $x_{\rm th}$.
This universality too is a well known feature of a turbulent process  \cite{KST,Nazarenko}.

%This is obvious in the low--energy regime, where $\pi\abar^2\tau^2_L\gtrsim1$ and
%${\cal E}_{\rm th}\sim 1$, but
%it is also true for higher energies, where $\pi\abar^2\tau^2_L <1$, since
%the thermalization scale is small enough: $x_{\rm th} \ll x_s\sim \abar^2\tau_L^2$.

\comment{In the formal limit $\pi\abar^2\tau^2_L \ll 1$, \eqn{Eth} would predict that the total 
energy $ \Delta E_{\rm th}$ transferred to the medium
is independent of the original energy $E$ of the LP:
\beq \label{Eth0}
{\cal E}_{{\rm th}}\,\simeq \,\pi \abar^2\tau_L^2\,=\,2\pi \abar^2 x_c
\ \Longrightarrow\ \Delta E_{\rm th}\,\equiv\,E\,{\cal E}_{{\rm th}}
\,=\,2\pi \abar^2 \omega_c\,.\eeq
Recalling that $\tau^2_L=2 x_c$ and $\abar\simeq 0.3$, it is however quite clear
that this particular limit $\pi\abar^2\tau^2_L \ll 1$ is not truly accessible in the present regime
where $x_c > 1$. Rather, this becomes interesting in the high--energy regime at $x_c\ll 1$, to be
discussed starting with the next section.}

%%%%%%%%%%%%%%%%%%%%%%%%%%%%%%%%%
\section{The high--energy regime}
\label{sec:highD}
%%%%%%%%%%%%%%%%%%%%%%%%%%%%%%%%%

With this section, we begin the study of the main physical problem of interest for us in this paper,
namely the gluon cascade produced in the medium by a very energetic leading particle,
with original energy $E\gg \omega_c$.
The main new ingredient as compared to the previous discussion is a kinematical restriction
on the primary gluon emissions that can be triggered by interactions in the 
medium: the energy $\omega$ of the gluons emitted by the LP
cannot exceed a value $\omega_c$ in order for the respective formation times to
remain smaller than $L$. When $x_c\equiv \omega_c/E\ll 1$, this restriction has important consequences: 
it implies that the LP loses only a small fraction of its total energy, of order $\abar x_c\ll 1$.
Our main focus in what follows will not be on this average energy lost by the LP
(this is well understood within the original BDMPSZ formalism, including multiple soft
emissions of primary gluons \cite{Baier:2001yt}), but rather on the further evolution of
this radiation via multiple branchings and the associated flow of energy
towards small values of $x$ %(down to the thermalization scale) 
and large angles.

\subsection{The coupled rate equations}

Since the radiation is restricted to relatively low energies $\omega\le\omega_c\ll E$, 
or $x\le x_c\ll 1$, it is clear that the part of the spectrum at higher energies $x_c<x<1$
has to be associated with the LP. This makes it natural to decompose
the overall spectrum as
 \beq\label{Dtwo}
 D(x,\tau)\,=\,\big[\Theta(x-x_c)+\Theta(x_c-x)\big] D(x,\tau)\,\equiv\,
  D_{\rm LP}(x,\tau)\,+\,  D_{\rm rad}(x,\tau)\,.\eeq
In reality, the LP piece $D_{\rm LP}(x,\tau)$ is a rather narrow peak located in the
vicinity of $x=1$ (see below), 
so there is a large gap between the two components of the spectrum.

The evolution of the radiation via successive branchings involves no special constraint, so 
the respective rate equation can be obtained simply by replacing $ D(x,\tau)$ according to 
\eqn{Dtwo} in the r.h.s. of the general equation \eqn{eqDf} (restricted to $x<x_c$, of course). This
yields
 \begin{align}\label{Drad}
   \frac{\del D_{\rm rad}(x,\tau)}{\del\tau}=
  \mcal{S}(x,\tau)+\,  \abar\int \rmd z \,{\cal K}(z)  
  \bigg\{%\Theta\bigg(z-\frac{x}{x_c}\bigg)
  \sqrt\frac{z}{x}\, D_{\rm rad}\bigg(\frac{x}{z}, \tau\bigg)\,-\,\frac{z}{\sqrt{x}}\,
  D_{\rm rad}\big({x},\tau\big)\bigg\},
  %\nn &\equiv \mcal{S}(x,\tau) \,+\,\mcal{I}[D_{\rm rad}](x,\tau)
  \end{align}
 where the {\em source}  $\mcal{S}(x,\tau)$ is the energy per unit time and per unit $x$ radiated
by the LP:
   \begin{align}\label{Source}
   \mcal{S}(x,\tau)\equiv %\Theta(x-x_c)
   \abar \int \rmd z \,{\cal K}(z)  %\,\Theta\bigg(\frac{x}{x_c}-z\bigg)
  \,\sqrt\frac{z}{x} \,  D_{\rm LP}\bigg(\frac{x}{z}, \tau\bigg).
  \end{align}
It is here implicitly understood that this source has support at $x\le x_c$ and that it acts
over a limited interval in time, at $0\le \tau\le \tau_L\equiv \sqrt{2\,x_c}$, which is moreover
small, $\tau_L\ll 1$, in the high--energy regime of interest.
The integral over $z$ in the gain term of \eqn{Drad} is 
restricted to ${x}/{x_c}< z <1$, where the lower limit is introduced by the support of the function 
$D_{\rm rad}({x}/{z}, \tau)$. 

In the rate equation for the leading particle, one needs to enforce the condition that
the radiated gluons have energy fractions smaller than $x_c$. The ensuing equation
reads (with $x>x_c$)
  \begin{align}\label{DLP}
   \frac{\del D_{\rm LP}(x,\tau)}{\del\tau}\
  &=\abar\int \rmd z \,{\cal K}(z) \bigg\{\Theta\bigg(z-\frac{x}{x+x_c}\bigg)
  \sqrt\frac{z}{x}\, D_{\rm LP}\bigg(\frac{x}{z}, \tau\bigg)\nn
  &\qquad\,-\,\frac{z}{\sqrt{x}}\,
  D_{\rm LP}\big({x},\tau\big)\bigg[\Theta\bigg(z-1+\frac{x_c}{x}\bigg)+
  \Theta\bigg(\frac{x_c}{x}-z\bigg)\bigg]\bigg\}
  \end{align}
where the various $\Theta$--functions enforce the kinematical constraint: In the gain term,
one requires that the unmeasured gluon emitted (with splitting fraction $1-z$) by the LP (with 
initial energy fraction $x/z$) be softer than $x_c$ : $(1-z)(x/z)<x_c \Longrightarrow z > x/(x+x_c)$. 
In the loss term, one requires that one of the daughter gluons be soft: either 
$zx < x_c$, or $(1-z)x < x_c$. 

As it should be clear from the previous discussion, the functions $D_{\rm LP}(x,\tau)$ and
$D_{\rm rad}({x},\tau)$ at any time $\tau<\tau_L$ also depend upon $x_c$,
hence upon the overall size $L$ of the medium, via the kinematical constraints on
the gluon emissions. This shows that the dynamics in
this high energy regime is  {\em non--local in time}~; e.g., the branching rate
in \eqn{DLP} `knows' about the maximal time $\tau_L$ via the various
$\Theta$--functions, which involve $x_c$. This property reflects
a true non--locality of the underlying quantum dynamics: it takes some time to emit a gluon
and this time cannot  be larger than $L$. Accordingly, at any $\tau<\tau_L$, one should
only initiate emissions whose energies are smaller than $\omega_c$~: gluon fluctuations with 
higher energies would have no time to become on--shell. The kinematical constraint 
$\omega\le\omega_c$ reflects only in a crude way the actual non--locality 
of the quantum emissions. The classical description at hand, as based on rate equations,
is truly appropriate only for the sufficiently soft emissions with
% with energies $\omega\ll \omega_c$ and 
small formation times $t_{\rm br}(\omega)\ll L$. Fortunately, these are
the most important emissions for the physics problems that we shall here address.

In the zeroth order approximation, which is strictly valid as $\tau\to 0$,
one can use $D_{\rm LP}(x,\tau)=\delta(1-x)$, and then the source in \eqn{Source} 
reduces to the BDMPSZ spectrum, as expected:
\beq\label{Source0}
   \mcal{S}_0(x)\,\equiv\,\abar x {\cal K}(x)\,\simeq\,\frac{\abar}{\sqrt{x}}\,.\eeq
In writing the second, approximate, equality we have used the fact  that $x$ is small,
$x\le x_c \ll 1$, to simplify the
expression of the splitting kernel (cf. \eqn{Kdef}): $ {\cal K}(x)\simeq x^{-3/2}$
for $x\ll 1$.

We shall now argue that the expression \eqref{Source0}, which is time--independent,
remains a good approximation for all the times $\tau$ of interest. 
Of course, the  spectrum $D_{\rm LP}(x,\tau)$ of the LP changes quite
fast with increasing $\tau$, notably due to the prompt radiation of very soft quanta with 
energy fractions $x\lesssim x_s(\tau)=\abar^2 \tau^2$. This leads to a broadening
of the LP peak on the scale $\Delta x\sim \abar^2
\tau^2\lesssim \abar^2 x_c \ll 1$, similar to that exhibited by \eqn{Dexact} at small times.
Yet, the probability to emit a
relatively hard gluon with $x\sim x_c$ is very small, of $\order{\abar}$.
Accordingly,  the support of the function $D_{\rm LP}(x,\tau)$ remains
limited to a narrow band at $1-x_c\lesssim x < 1$, which is well separated from
the radiation spectrum at $x < x_c$. Hence, the integration over $z$ 
in \eqn{Source} is effectively restricted to a narrow range close to $x$, namely
$x<z<x/(1-x_c)$, and the integral can be approximated as
 \begin{align}\label{Source1}
   \mcal{S}(x,\tau)\,\simeq\,
      \abar x{\cal K}(x)   \int \rmd x' \,   D_{\rm LP}(x', \tau)\,\simeq\,\frac{\abar}{\sqrt{x}}
      \Big[1 + \order{x, \,\abar x_c}\Big]\,.
  \end{align}
Here we have used the fact that   the
overall strength of the function $D_{\rm LP}(x,\tau)$, i.e. the energy fraction carried by the 
LP after a time $\tau$, can be estimated as
 \beq 
{\cal E}_{{\rm LP}}(\tau)\equiv \int\rmd x\,D_{\rm LP}(x,\tau)
\,\simeq \,1 - 2\abar\tau\sqrt{x_c}\,,\eeq
that is, the initial energy minus the energy lost via radiation of soft gluons with $x\le x_c$,
cf. \eqn{Source0}.

 \begin{figure}[t]
	\centering
	\includegraphics[width=0.85\textwidth]{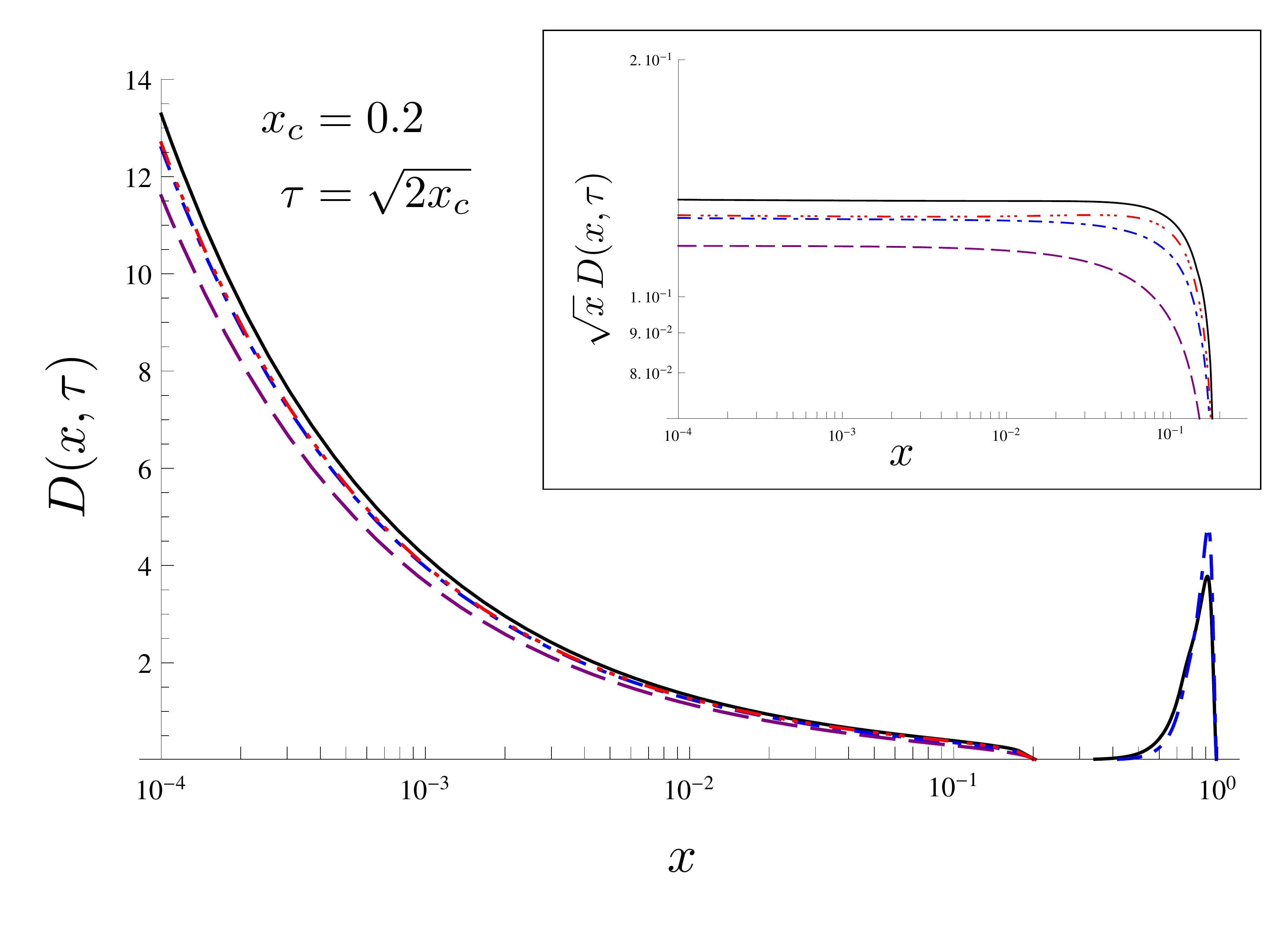}
		\caption{\sl
		The full spectrum
		$D(x,\tau)= D_{\rm LP}(x,\tau)+ D_{\rm rad}(x,\tau)$ obtained by numerically
		solving the coupled equations \eqref{Drad} and \eqref{DLP}, versus the 
		 radiation spectrum predicted by \eqn{Drad0} with a source.
		 We use both versions
		of the kernel, $\mcal{K}$ and $\mcal{K}_0$, together with  $x_c=0.2$
		and  $\tau=\sqrt{2x_c}\simeq 0.63$.
		(i) Simplified kernel $\mcal{K}_0$: black curve: Eqs.~\eqref{Drad}--\eqref{DLP}; purple,
		 dashed: \eqn{Drad0}. (ii) Full kernel $\mcal{K}$: blue, dashed--dotted:                                                     Eqs.~\eqref{Drad}--\eqref{DLP}; red, dashed--triple dotted: \eqn{Drad0}.
		 In the insert: the same plots  (for the radiation part only) in log--log scale.}
		\label{fig:fullvsmodel}
\end{figure}

To summarize, after an evolution time $\tau_L$, the energetic LP loses only a 
small fraction $\abar\sqrt{x_c}\tau_L \sim \abar x_c\ll 1$ of its total energy and its
spectral density remains peaked near $x=1$. Accordingly, it can be effectively
treated as a steady source $\mcal{S}_0(x)$
for the soft radiation at $x \ll 1$. This is verified in the plots
in Fig.~\ref{fig:fullvsmodel}, where we perform two types of comparisons:
\texttt{(i)} between the evolution with the exact kernel $\mcal{K}$ in \eqn{Kdef}
and that with the simplified kernel $\mcal{K}_0$, and \texttt{(ii)} between the
solution to the coupled system of equations \eqref{Drad} and \eqref{DLP}
and that to the effective equation with a source, i.e. \eqn{Drad} 
with $\mcal{S}(x,\tau)\to \mcal{S}_0(x)$. As one can see in this plot, the two choices
for the kernel lead indeed to results which are qualitatively similar and numerically 
very close to each other. Furthermore, the radiation spectrum at $x\le x_c$ produced
by the `model' equation with a source is indeed close to the respective prediction 
of the coupled rate equations. (In fact, for the exact kernel $\mcal{K}$, this similarity
looks even more striking --- the respective curves almost overlap
with each other at sufficiently small $x$ --- but in our opinion this is merely a coincidence.)
In the next subsection, we shall construct an exact analytic solution for the equation 
with the source, for the case of the simplified kernel $\mcal{K}_0$.

\subsection{The radiation spectrum}
\label{sec:rad}

In the remaining part of this section, we shall concentrate on the solution to the following equation
 \begin{align}\label{Drad0}
   \frac{\del D_{\rm rad}(x,\tau)}{\del\tau}\,&=\,
  \frac{\abar}{\sqrt{x}}\,+\,  \abar\int \rmd z \,{\cal K}(z)  
  \bigg\{%\Theta\bigg(z-\frac{x}{x_c}\bigg)
  \sqrt\frac{z}{x}\, D_{\rm rad}\bigg(\frac{x}{z}, \tau\bigg)\,-\,\frac{z}{\sqrt{x}}\,
  D_{\rm rad}\big({x},\tau\big)\bigg\}
  \nn &\equiv \,\mcal{S}_0(x) \,+\,\abar\mcal{I}[D_{\rm rad}](x,\tau)\,,
  \end{align}
which, as above argued, offers a good approximation for the dynamics of the medium--induced
radiation by a leading particle with high energy $E\gg \omega_c$. This is an inhomogeneous equation
with vanishing initial condition and can be solved with the help of the respective Green's function:
 \begin{align}\label{DradG}
   D_{\rm rad}(x,\tau)\,=\,\int_x^{x_c}\rmd x_1\int_0^\tau\rmd \tau_1\,
   G(x, x_1, \tau-\tau_1)\,\mcal{S}_0(x_1)\,. \end{align}
The Green's function $G(x, x_1, \tau)$ obeys the homogeneous version of \eqn{Drad0} with
initial condition $G(x, x_1, \tau)=\delta(x-x_1)$. 

From now on, we shall again restrict ourselves to the case of the simplified splitting kernel
${\cal K}_0(z)$, which we recall is
obtained by replacing $f(z)\to 1$ in \eqn{Kdef}. For this case, the Green's function $G(x, x_1, \tau)$
can be exactly computed, since it is closely related to the  function ${D}(x,\tau)$ in \eqn{Gexact}:
both functions obey Eq.~\eqref{eqDf}, but with slightly different
initial conditions. It is easy to check that the corresponding
solutions are related via an appropriate rescaling of the variables:
\beq\label{Gexact}G(x, x_1, \tau)\,=\,\frac{1}{x_1}
  D\bigg(\frac{x}{x_1},\frac{\tau}{\sqrt{x_1}}\bigg)
  \,=\,\sqrt\frac{x_1}{x}\,
  \frac{\abar\tau}{(x_1-x)^{3/2}}\ 
  \exp\left\{-\frac{\pi\abar^2\tau^2}{x_1-x}\right\}\,.\eeq  
Since the source $\mcal{S}_0(x_1)$ in \eqn{DradG} is independent of time, the integral
over $\tau_1$ involves only the  Green's function and can be readily computed:
\beq\label{intG}
\int_0^\tau\rmd \tau_1\,   G(x, x_1, \tau-\tau_1)\,=\,\frac{1}{2\pi\abar}\sqrt\frac{x_1}{x(x_1-x)}\,
 \bigg[1-
  \exp\left\{-\frac{\pi\abar^2\tau^2}{x_1-x}\right\}\bigg]\,.
  \eeq
To also compute the integral over $x_1$, it is convenient to change the integration variable
according to $u\equiv \pi\abar^2\tau^2/(x_1-x)$. One thus easily finds
\begin{align}\label{Dradex}
   D_{\rm rad}(x,\tau)&\,=\, \frac{\abar\tau}{\sqrt{x}}\,\frac{1}{2\sqrt{\pi}}\int_\zeta^\infty
   \frac{\rmd u}{u^{3/2}}\Big[1-\rme^{-u}\Big]
   \,=\, \frac{\abar\tau}{\sqrt{x}}\,\left\{ \frac{1}{\sqrt{\pi}}\,\Gamma\Big(\frac{1}{2},\zeta\Big)
   +\frac{1-\rme^{-\zeta}}{\sqrt{\pi\zeta}}\right\}
\,, \end{align}
where 
\beq\label{zeta}
\zeta\,\equiv\,\zeta(x_c-x,\tau)\,\equiv\,\frac{\pi\abar^2\tau^2}{x_c-x}\,,\eeq
and
\beq\label{Gamma}
\Gamma\Big(\frac{1}{2},\zeta\Big)\,\equiv\,
\int_\zeta^\infty
   \frac{\rmd z}{\sqrt{z}}\,\rme^{-z}
   \,=\,\sqrt{\pi}\,-\,
\int_0^\zeta\frac{\rmd z}{\sqrt{z}}\,\rme^{-z}
     \,=\,\sqrt{\pi}\,-\,\gamma\Big(\frac{1}{2},\zeta\Big)\,,
   \qquad \eeq
is the upper incomplete Gamma function (whereas $\gamma(1/2,\zeta)$ is the
respective lower function). Note that $D_{\rm rad}(x,\tau)$ is also a function of the
limiting energy fraction $x_c$, but in our notations this dependence is left implicit.
A similar observation applies to all formul{\ae} that appear in this section.

For what follows, it is also useful to single out the piece of the spectrum 
that would be produced by the source term alone, in the absence of branchings. Specifically,
using \eqn{Gamma}, we can write
\begin{align}\label{Dradex2}
   D_{\rm rad}(x,\tau)&\,=\, \frac{\abar\tau}{\sqrt{x}}\,-\,\delta D_{\rm br}(x,\tau)\,,\nn
   \delta D_{\rm br}(x,\tau)&\,\equiv
 \frac{\abar\tau}{\sqrt{x}}\,\left\{ \frac{1}{\sqrt{\pi}}\,\gamma\Big(\frac{1}{2},\zeta\Big)
   -\frac{1-\rme^{-\zeta}}{\sqrt{\pi\zeta}}\right\}
   \,\equiv
 \frac{\abar\tau}{\sqrt{x}}\,h(\zeta)
   \,,
    \end{align}
where the quantity $\delta D_{\rm br}(x,\tau)$ is the change in the spectrum due
to multiple branchings and it is positive semi--definite, as one can easily check
--- meaning that the effect of branchings is a {\em depletion} in the spectrum, 
at any $x\le x_c$. This depletion reflects the flow of energy from one parton 
generation to the next one, via parton branching --- a phenomenon to which we shall return
in the next subsection. 
But before doing that, let us discuss the radiation spectrum \eqref{Dradex} in more detail. 
 \begin{figure}[t]
	\centering
	\includegraphics[width=0.7\textwidth]{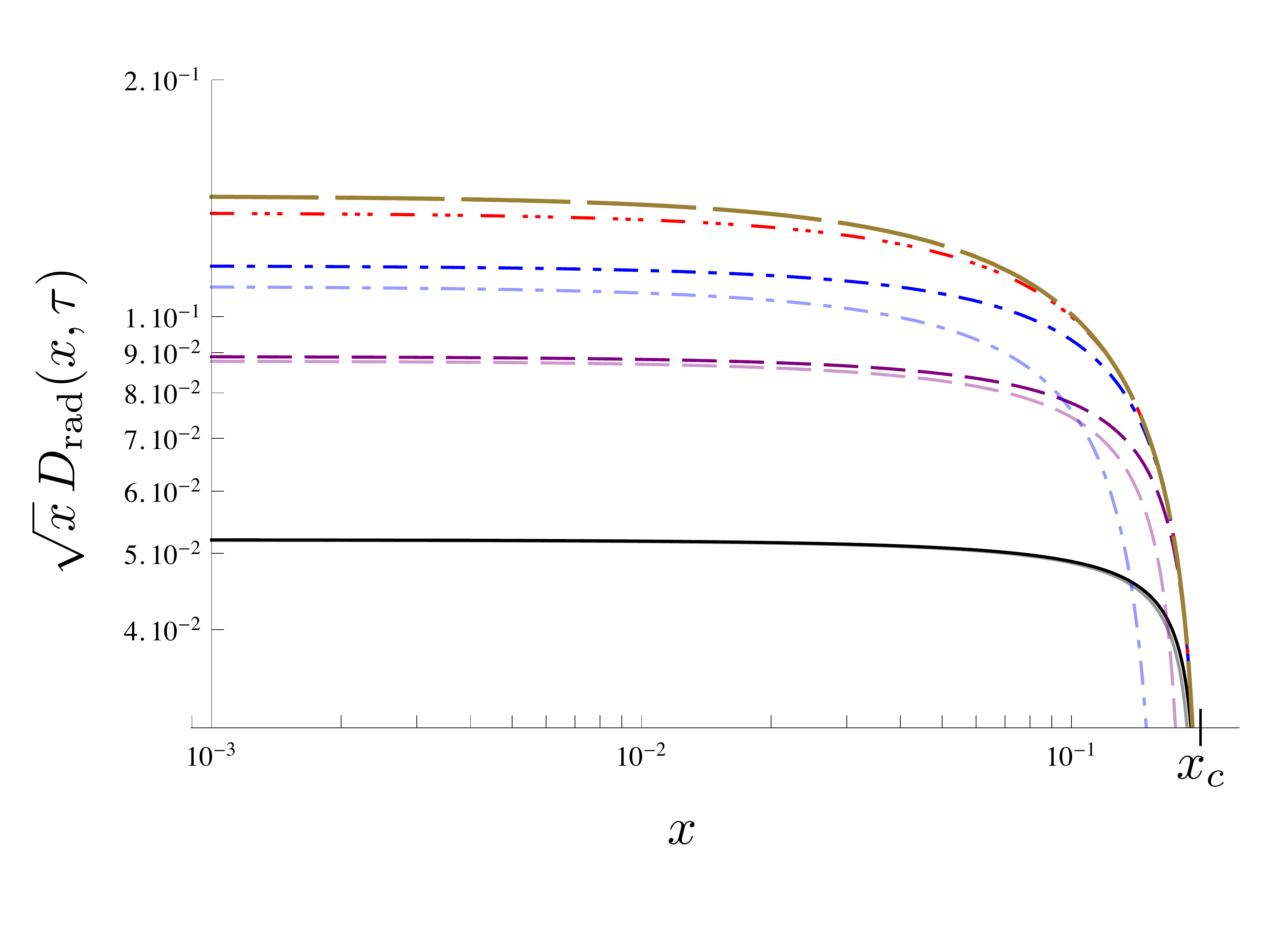}
		\caption{\sl Plot (in log-log scale) of $D_{\rm rad}(x,\tau)$, cf. Eq.~(\ref{Dradex}), as a function of 
		$x$ for $x_c=0.2$ and various values of $\tau$. The thick curves show the function $\sqrt{x}D_{\rm rad}(x,\tau)$ for $\tau=0.2$ (black, solid), $\tau=0.4$ (purple, dashed), $\tau=0.63$ (blue, dashed--dotted),
		and $\tau=1.0$ (red, dashed--triple dotted). Note that the maximal value
		for $\tau$ which is physically allowed is $\tau_L=\sqrt{0.4}\simeq 0.63$.
		The thin curves, shown for $\tau\le\tau_L$, 
		 represent the corresponding approximations at small $\zeta(x_c-x,\tau)$,
		as obtained by keeping only the first 2 terms in the Taylor expansion in \eqn{Dradexp1}.
		The enveloping curve (brown, long--dashed) is the limiting curve at large $\zeta$, 
		cf. \eqn{Dradas}.}
		\label{fig:Drad}
\end{figure}

This spectrum is depicted in Fig.~\ref{fig:Drad} as a function of $x$ for various values of $\tau$.
The different limiting behaviors can be also understood in analytic terms. 
To that aim, it is useful to notice a few properties of the function $h(\zeta)$. This function
is monotonously increasing and interpolates between $h=0$ at $\zeta=0$ and $h\to 1$ as
$\zeta\to \infty$. Furthermore, the ratio
$h(\zeta)/\sqrt{\zeta}$ is an analytic function of $\zeta$ with infinite radius of convergence 
and a rapidly converging Taylor expansion:
\beq\label{hexp}
\sqrt{\frac{\pi}{\zeta}}\,h(\zeta)\,=\,\int_0^1 \rmd u\, u^{-1/2}\,\rme^{-\zeta u}\,-\,
\frac{1-\rme^{-\zeta}}{\zeta}\,=\,1 - \frac{1}{6}\zeta +\frac{1}{30}\zeta^2 \,+\,\order{\zeta^3}\,.
\eeq
Finally, for large $\zeta$, one finds the asymptotic behavior
\beq\label{has}
1-h(\zeta)\,=\,\frac{1}{\sqrt{\pi\zeta}}\,+\cdots\,,\eeq
where the dots stand for terms which are exponentially suppressed.

Returning to the spectrum in \eqn{Dradex2}, we first observe that at small $x\ll x_c$ this reduces
to the scaling spectrum $D_{\rm sc}(x)= 1/{\sqrt{x}}$ --- the expected fixed point of the 
branching dynamics at small $x$. Indeed, when $x\ll x_c$, one can approximate 
$\zeta(x_c-x,\tau)\simeq \zeta(x_c,\tau)$ and therefore
 \begin{align}\label{Dradsc}
   D_{\rm rad}(x,\tau)\,\simeq\,\frac{\abar\tau}{\sqrt{x}}\,\big[1-h(\zeta_0)\big]\,,\qquad
   \zeta_0\,\equiv\,\zeta(x_c,\tau)\,=\,\frac{\pi\abar^2\tau^2}{x_c}
    \,.\end{align}
Interestingly, at the end of the evolution, i.e. for
$\tau=\tau_L=\sqrt{2x_c}$, \eqn{Dradsc} reduces to the BDMPSZ spectrum times a function
of the QCD coupling $\abar$, which is strictly smaller than 1 and which expresses the reduction
in the spectrum due to multiple branchings:
\begin{align}\label{DradtauL}
   D_{\rm rad}(x,\tau_L)\,\simeq\,\abar\sqrt{\frac{2x_c}{x}}\,\Big[1-h\big(2\pi\abar^2\big)\Big]
   \quad\mbox{for}\quad x \ll x_c
  \,.\end{align}

Consider now larger values of $x$, where the deviations from the scaling spectrum start to
be important. So long as $x$ is not too close to $x_c$,  such that $\zeta\lesssim 1$, 
the spectrum can be expanded in powers of $\zeta$, with the help of \eqn{hexp}. One
thus finds
\begin{align}\label{Dradexp1}
   D_{\rm rad}(x,\tau)&\,=\, \frac{\abar\tau}{\sqrt{x}}\left\{
   1\,-\,\frac{\abar\tau}{\sqrt{x_c-x}}\,+\,\frac{\pi}{6}
\left(   \frac{\abar\tau}{\sqrt{x_c-x}}\right)^3 \,+\,\cdots
   \right\}\quad\mbox{when}\quad \zeta(x_c-x,\tau)\,\lesssim\,1\,,
      \end{align}
where the dots stand for terms of $\order{\zeta^{5/2}}$ and higher.
This expansion is rapidly converging for any $\zeta\lesssim 1$. Given that
$\zeta(x_c,\tau_L) =2\pi\abar^2$ is a relatively small number
($2\pi\abar^2\simeq 0.6$ for $\abar=0.3$), we expect a limited expansion like
\eqn{Dradexp1} to be quite accurate for any $\tau\lesssim\tau_L$ and for $x$ values
in the bulk. And indeed, the curves obtained by keeping just the first 2 terms in this expansion 
provide an excellent approximation to the full curves in Fig.~\ref{fig:Drad}
for any  $\tau\le \tau_L$, except of course for $x$ very close to $x_c$. Notice that the inclusion
of the first correction in \eqn{Dradexp1}, which expresses the dominant effect of the
multiple branchings at small $\abar\tau$, is truly
essential in order to obtain such a good agreement.
Indeed, for $\tau\sim\tau_L$ and  $x$ values in the bulk, that correction is numerically
important, of relative order  $\abar\tau_L/\sqrt{x_c}=\sqrt{2}\abar\simeq 0.4$. 

\comment{To summarize, the effects of the multiple branchings are significantly large for $\tau\sim\tau_L$,
but they can be accurately computed by using just the first term in the `small--time' (i.e.
small $\abar\tau$) expansion of $\delta D_{\rm br}(x,\tau)$, except for values of $x$ which
are very close to the endpoint at $x=x_c$. This remark is interesting since, as we shall
shortly discuss, this dominant contribution to $\delta D_{\rm br}(x,\tau)$ can also be obtained
within perturbation theory, without an exact knowledge of the spectrum.

 This flow mechanism 
being very robust, we conclude that the shape of the spectrum near its upper endpoint  is well under
control, in spite of the fact that our effective theory becomes inaccurate for relatively hard emissions
with $\omega\sim\omega_c$.
}

The expansion in \eqn{Dradexp1} breaks down when the first correction becomes of $\order{1}$
or larger, namely for $x_c-x\lesssim \abar^2\tau^2$. This is in agreement with the fact that the
emission of very soft gluons, with energy fractions $x\lesssim x_s(\tau)=\abar^2\tau^2$, is
non--perturbative. To investigate the effect of such emissions via analytic approximations,
let us consider the behavior near the endpoint of the spectrum, at $x\to x_c$. In that limit,
one has $\zeta\gg 1$, so one can use \eqn{has} to deduce
\begin{align}\label{Dradas}
   D_{\rm rad}(x,\tau)&\,\simeq\,\frac{1}{\pi}\,\sqrt{\frac{x_c-x}{x}}
   \quad\mbox{when}\quad \zeta(x_c-x,\tau)\,\gg\,1\,.
      \end{align}
This result is time--independent and
shows that the spectrum vanishes when $x\to x_c$ at any time $\tau$. This demonstrates
the efficiency of the soft branchings in depleting the spectrum near its endpoint. 
The energy which is transferred in this way towards the bins at $x<x_c$  
cannot be compensated by a corresponding flow of energy coming 
from the bins at $x>x_c$, since the spectrum ends at $x_c$.

The steady spectrum in \eqn{Dradas} also represents the limiting curve
for the function $D_{\rm rad}(x,\tau)$ in the formal large--time limit at 
$\zeta(x_c,\tau)=\pi\abar^2\tau^2/x_c\gg 1$. That is, in this limit, the spectrum takes the
form  in \eqn{Dradas} for {\em any} $x\le x_c$. This large--time limit is merely 
formal, since, as already mentioned, the maximal value for $\zeta(x_c,\tau)$ which is physically allowed 
is $\zeta(x_c,\tau_L) =2\pi\abar^2$, which is not that large.
 Still, this limit is conceptually interesting, in that it corresponds to the more familiar
turbulence set--up: a steady situation in which the whole energy injected by the source flows
through the spectrum into the  `sink' at $x=0$ (see the discussion in the next
subsection).

It is finally interesting to clarify the suitability of perturbation theory (by which we mean
the iterative solution to \eqn{Drad0} in which the branching term  $\abar\mcal{I}[D_{\rm rad}]$
is treated as a small perturbation) for the problem at hand. Via successive
iterations, one can construct a perturbative solution for $D_{\rm rad}(x,\tau)$ in the form of a series
in powers of $\abar\tau$ and is interesting to compare this series to the small--$\zeta$ expansion
of the exact solution in \eqn{Dradexp1}. Clearly, we do not expect this perturbative approach to 
be reliable near the endpoint of the spectrum at $x_c$, but one may hope that it
becomes meaningful for $x$ well below $x_c$ and for small times $\abar\tau\ll 1$
--- that is, in the region where the expansion  \eqref{Dradexp1} 
can be viewed too as a series in powers of $\abar\tau$. But even this last expectation is naive,
as shown by following argument: a perturbative solution via iterations would generate both
odd and even powers of $\abar\tau$, whereas the corresponding expansion  in \eqn{Dradexp1}
contains only odd powers. 

To further clarify this mismatch, we shall construct in Appendix~\ref{sec:PT}
the perturbative solution to low orders: $D_{\rm rad}= D_{\rm rad}^{(0)}+ D_{\rm rad}^{(1)}
+ D_{\rm rad}^{(2)}+\cdots$. The zeroth order result is, clearly, 
$D_{\rm rad}^{(0)}={\abar\tau}/{\sqrt{x}}$, while the first iteration, as obtained
by evaluating the branching term  $\abar\mcal{I}[D_{\rm rad}]$
with the zeroth order result, yields precisely the correction of $\order{\abar\tau}$
shown in \eqn{Dradexp1}, that is,
 \beq\label{D1}
 D_{\rm rad}^{(1)}(x,\tau)=-
 \frac{\abar^2\tau^2}{\sqrt{x(x_c-x)}}\,.\eeq
But a subtle issue shows up starting with the second iteration: the first--order correction
$ D_{\rm rad}^{(1)}$ turns out to be an {\em exact fixed point} of the branching kernel:
 $\mcal{I}[D_{\rm rad}^{(1)}]=0$. Accordingly, the second--order correction is
 exactly zero, $ D_{\rm rad}^{(2)}=0$ (still in agreement with \eqn{Dradexp1}), but then 
 the same is true for all the subsequent iterations: $ D_{\rm rad}^{(n)}=0$ for any $n\ge 2$.
 That is, the perturbative expansion, as computed without any approximation, terminates 
after just one non--trivial  iteration and predicts
$D_{\rm rad}= D_{\rm rad}^{(0)}+ D_{\rm rad}^{(1)}$. This prediction is certainly incorrect
(except as an approximation at small times and small $x$):
it differs from the actual expansion \eqn{Dradexp1} of the exact result and, in particular, it 
becomes negative and divergent when $x\to x_c$. 

The mathematical origin of this failure
will be clarified in Appendix~\ref{sec:PT}. But its physical origin should be quite clear:
we have already noticed the non--perturbative nature of the dynamics associated 
with the emission of very soft quanta, with energy fractions $x\lesssim x_s=\abar^2\tau^2$.
For such emissions, the effects of multiple branchings must be resumed to all orders
and cannot be accurately studied via iterations. This non--perturbative dynamics is
responsible for the rapid broadening of the LP peak and also for the fact that the radiation
spectrum in Eq.~(\ref{Dradex}) exactly vanishes as $x\to x_c$ for any $\tau$.
%(in sharp contrast with the zeroth order spectrum $D_{\rm rad}^{(0)}$ 
%that would be produced by the source alone). 
Similar, non--perturbative aspects 
affect the spectrum at any value of $x$, including the intermediate bins at $x_s \ll x \ll x_c$,
since the occupation of any such a bin can change via the emission of very
soft gluons. Hence, not surprisingly, the spectrum $D(x,\tau)$ cannot be faithfully computed
within perturbation theory for generic values $(x,\,\tau)$, albeit interesting information can be 
obtained via this method in special cases, as we shall see.

\subsection{The energy flux}
\label{sec:Fsmallxc}

As in Sect.~\ref{sec:low}, the dissipative properties of the cascade, in particular, the rate
for energy loss towards the medium, can be best studied by computing the energy flux 
 associated with branchings. Let ${\cal E}(x_0,x_c,\tau)$ denote
the energy which at time $\tau$ is contained in the modes in the spectrum 
within the interval $x_0<x<x_c$ :
 \beq\label{Ex0} {\cal E}(x_0,x_c,\tau)\,=\,
\int_{x_0}^\xc {\rm d} x \,D_{\rm rad}(x,\tau)\,.
\eeq
When increasing $\tau$, this energy can change via two mechanisms: \texttt{(i)}
it increases due to additional radiation by the source, at a rate $\int_{x_0}^\xc {\rm d} x\,
\mcal{S}_0(x)$, and  \texttt{(ii)} it decreases due to the energy transfer towards the modes
at $x<x_0$ via gluon branching, at a rate which is by definition the energy flux $ \F(x_0,\tau)$
through the bin $x_0$. Hence, we can write
\beq\label{balance}
\frac{\partial {\cal E}(x_0,x_c,\tau)}{\partial\tau}\,=\,\int_{x_0}^\xc {\rm d} x\,
\mcal{S}_0(x)\,-\,
\F(x_0,\tau)\,,\eeq
which immediately implies
\beq\label{Fx0xc}
\F(x_0,\tau)\,=\,\int_{x_0}^\xc {\rm d} x\
\frac{\partial \,}{\partial\tau}\,\delta D_{\rm br}(x,\tau)\,=\,-
\abar\int_{x_0}^\xc {\rm d} x\,\mcal{I}[D_{\rm rad}](x,\tau)\,,
\eeq
where the first equality follows after recalling the definition \eqref{Dradex2} of
$\delta D_{\rm br}(x,\tau)$, and the second one after also using the rate equation
\eqref{Drad0}. Each of the two integral representations for $\F(x_0,\tau)$ in the
equation above has its own virtues. When combined with the explicit result for $\delta D_{\rm br}(x,\tau)$ 
shown in  \eqn{Dradex2}, the first representation allows for efficient numerical
calculations, with results that we shall shortly  describe. On the other hand,
this formula is not well suited for analytic studies, as we shall see. The second integral
representation, which involves the branching term $\mcal{I}[D_{\rm rad}]$, is more directly 
connected to the dynamics of branchings and admits a transparent physical interpretation, 
to be discussed in Sect.~\ref{sec:flow2}. A priori, this representation seems to be mathematically 
more involved, in that it involves a double convolution over the spectrum. 
Yet, as we shall see, this representation allows for more accurate analytic studies.
In particular, it will permit us to deduce an exact analytic result in the important limit $x_0\to 0$.

\begin{figure}[t]
	\centering
	\includegraphics[width=0.7\textwidth]{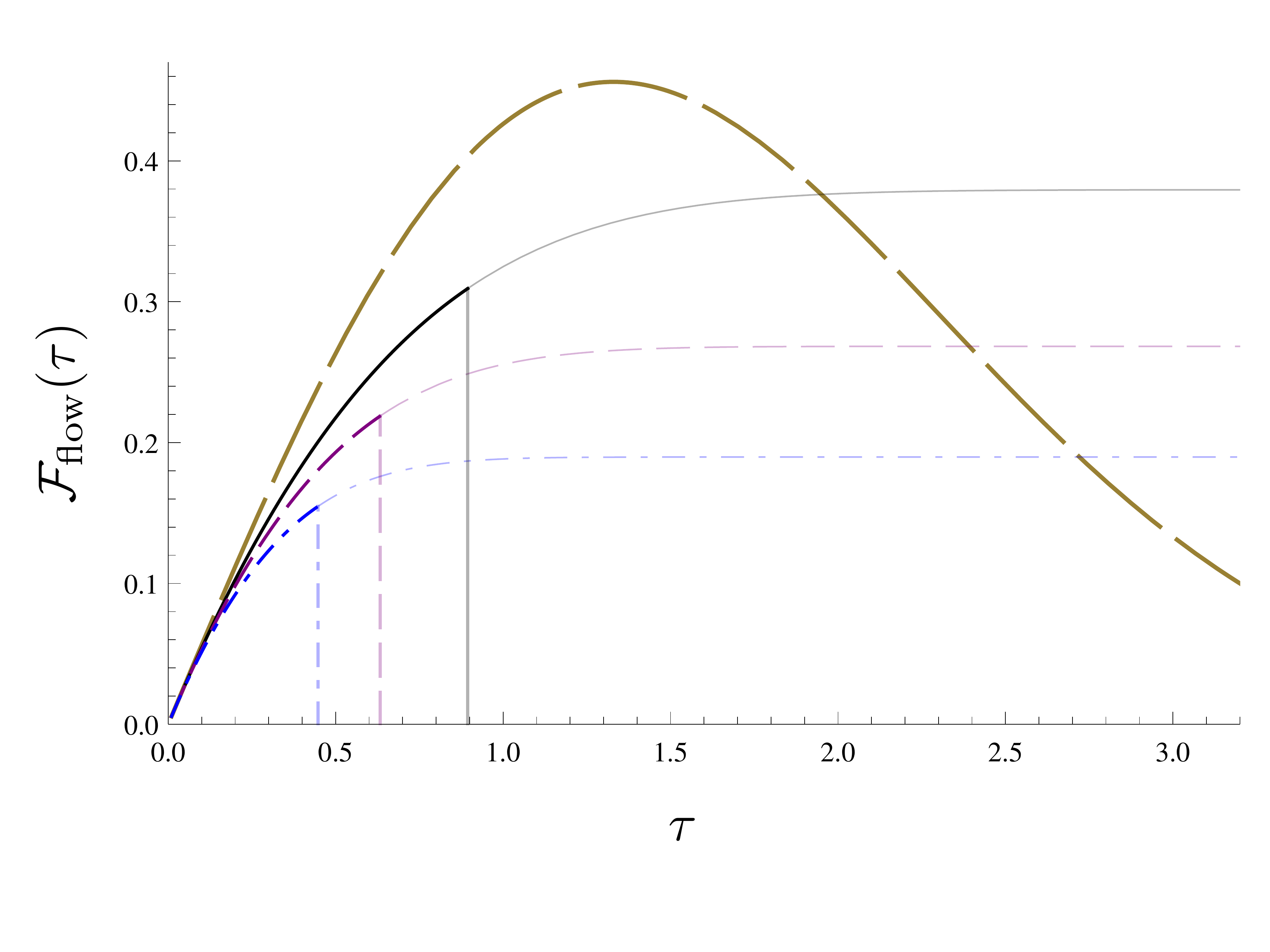}
		\caption{\sl The rate of flow $\F_{\rm flow}(\tau)$
		as a function of $\tau$ for various physical regimes. The
		brown, long--dashed curve represents the function
		in \eqn{Fflow}, which corresponds to $x_c >1$ and englobes both the
		`low energy' regime, and  the `intermediate energy' one, depending upon the
		value of the upper limit $\tau_L=\sqrt{2x_c}$ on $\tau$.
		The other curves correspond to different values $x_c < 1$ 
		(i.e. to various  `high--energy' regimes) and are obtained  according
		to \eqn{Fflowg}: $x_c = 0.4$ (black, solid), 
		 $x_c=0.2$ (purple, dashed), and $x_c=0.1$ (blue, dashed-dotted). 
		 	The thick lines represent the respective curves within their physical
	 range of validity ($\tau < \tau_L$), whereas the
thin curves are their extrapolations at larger times $\tau > \tau_L$.
The vertical lines denote the upper time limit  $\tau_L=\sqrt{2 x_c}$.
		}
		\label{fig:Fflowxc}
\end{figure}

Using the first equality in \eqn{Fx0xc} together with the expression \eqref{Dradex2}
for  $\delta D_{\rm br}(x,\tau)$,  one finds, after simple manipulations,
\beq\label{Fx0g}
\F(x_0,\tau)\,=\,\frac{\abar}{\sqrt{\pi}}\int_{x_0}^\xc %{\rm d} x\
\frac{\rmd x}{\sqrt{x}}\
\gamma\Big(\frac{1}{2},\zeta\Big)\,,\eeq
with $\zeta\equiv \zeta(x_c-x,\tau)$ as defined in \eqn{zeta}. 
We are mostly interested in the
limit $x_0\to 0$ of this result, which represents the energy flux carried by the turbulent flow :
 \beq\label{Fflowg}
\F_{\rm flow}(\tau)\,=\,
\frac{\abar}{\sqrt{\pi}}\int_{0}^\xc %{\rm d} x\
\frac{\rmd x}{\sqrt{x}}\
 \gamma\Big(\frac{1}{2},\zeta\Big)\,.\eeq
As explained in Sect.~\ref{sec:low}, this is the rate at which the energy leaks out of the
spectrum and accumulates into a condensate at $x=0$. It is straightforward
to numerically compute the integral in \eqn{Fflowg} and thus study the flow
as a function of $\tau$ for various values  $x_c\ll 1$. The results 
are shown in Fig.~\ref{fig:Fflowxc}, together with the respective prediction
of the `low--energy' case $x_c >1$, that is, the function $\F_{\rm flow}(\tau)$ in
\eqn{Fflow}. In principle, one should consider these curves only for $\tau$ values within the
physically allowed range, i.e. for  $\tau\le \tau_L=\sqrt{2x_c}$. But in Fig.~\ref{fig:Fflowxc} we also
show them for larger values $\tau> \tau_L$~; this is interesting too, but for a different
physical problem (see below).

By comparing curves which refer to different values of $x_c$, one can better appreciate 
the role of the kinematical constraint $x\le x_c\ll 1$ in slowing down the branching process
and thus reducing the energy flow. The plots in Fig.~\ref{fig:Fflowxc} make clear that,
when lowering $x_c$, one reduces not only the total duration $\tau_L$ of the branching process,
but also the rate for energy loss at any given time $\tau < \tau_L$. This trend
is natural on physical grounds:  by decreasing $x_c$, one limits the phase--space 
for medium--induced radiation to emissions which carry lower and lower fractions of the
total energy of the leading particle. Fig.~\ref{fig:Fflowxc} also shows that the deviation between
curves corresponding to different values of $x_c$ increases with time; for $\tau \sim \tau_L$,
this deviation is seen to be sizable including for the smallest values of $x_c$ under consideration.

It would be interesting to understand the systematics of these plots via analytic studies.
To that aim, one may attempt a small--$\tau$ expansion of the flow in \eqn{Fflowg} based
on the corresponding expansion of $\delta D_{\rm br}(x,\tau)$ in \eqn{Dradexp1}. (This is
tantamount to performing the small--$\zeta$ expansion of the 
function $\gamma(1/2,\zeta)$ in \eqn{Fflowg}.) At leading
order, one should use the dominant contribution to $\delta D_{\rm br}(x,\tau)$, that is,
(minus) the function $D_{\rm rad}^{(1)}(x,\tau)$ in \eqn{D1}. One thus finds
 \beq\label{Fflow0}
\F_{\rm flow}(\tau)\,\simeq\,-
\int_{0}^\xc {\rm d} x\
\frac{\partial \,}{\partial\tau}\,D_{\rm rad}^{(1)}(x,\tau)\,=\,2\abar^2\tau
\int_{0}^\xc %{\rm d} x\
\frac{\rmd x}{\sqrt{x(x_c-x)}}\,=\,2\pi\abar^2\tau\,.
\eeq
This estimate, which is independent of $x_c$, holds only for sufficiently small times, such that
$\zeta(x_c,\tau)=\pi\abar^2\tau^2/x_c\ll 1$, where it describes indeed the common behavior
of all the curves exhibited in Fig.~\ref{fig:F0xc}. But this approximation is unable to capture
the lift in degeneracy with increasing $\tau$. 
One may expect to be able to compute corrections to \eqn{Fflow0} by
using the higher order terms in the expansion \eqref{Dradexp1} of $D_{\rm rad}$,
but this turns out not to be possible: for all the terms in this expansion beyond $D_{\rm rad}^{(1)}$,
the integral over $x$ in \eqn{Fflowg} develops a non--integrable singularity at its upper
endpoint $x_c$.  

%This shows that, unlike the spectrum $\delta D_{\rm br}(x,\tau)$, the energy flux is not
%an analytic function of the `small parameter' $\zeta(x_c,\tau)=\pi\abar^2\tau^2/x_c$.

In the next subsection, we shall exploit the second equality in \eqn{Fx0xc} to deduce
an exact, analytic, result for $\F_{\rm flow}(\tau)$ (see \eqn{Fflowexact}). But for the purposes
of the present discussion, it suffices to consider just
one more term in the small--$\tau$ expansion of $\F_{\rm flow}(\tau)$. This can be obtained
by expanding the exact result in \eqn{Fflowexact} and reads
 \beq\label{Fflow1}
\F_{\rm flow}(\tau)\,\simeq\,2\pi\abar^2\tau\,\left(1-\,\frac{\abar\tau}{\sqrt{x_c}}
\right).
\eeq
As expected, the corrective term above lifts the degeneracy between different values of $x_c$.
The relative importance of this term increases with time and
becomes independent of $x_c$ when $\tau\sim\tau_L$
(since $\tau_L$ itself scales like $\sqrt{x_c}$): ${\abar\tau_L}/{\sqrt{x_c}}=\sqrt{2}\abar$. 
Hence, this correction would be negligible in the formal weak coupling limit, but it is numerically 
important for realistic values of $\abar$ : e.g. $\sqrt{2}\abar\simeq 0.4$ for $\abar=0.3$. And indeed,
the inclusion of this correction greatly improves the accuracy of the small--time expansion,
as it will be shown later,  in Fig.~\ref{fig:F0xc}~:
the limited expansion in \eqn{Fflow1} provides an excellent approximation to the exact 
result for any $\tau\le\tau_L$.

Consider now the behavior of the flow for relatively large times
$\tau \gg \tau_L$, that is, outside of the physical range for jet evolution. This corresponds to
a different physical problem, which is closer to the familiar turbulence set--up --- a steady source
acts for arbitrarily large time and eventually builds up a time--independent energy spectrum ---, 
except that our source has a rather unusual spectrum: rather than being localized near $x_c$
(e.g. $ \mcal{S}(x)=\delta(x-x_c)$), the function
$\mcal{S}_0(x)={\abar}/{\sqrt{x}}$ has a long tail at small $x\le x_c$,  
as expected for {\em radiation}. The associated steady flow at large times can be
obtained as follows: from Sect.~\ref{sec:rad} we recall  that,
when $\pi\abar^2\tau^2/x_c\gg1$, the spectrum reaches
the steady shape in \eqn{Dradas} (see also Fig.~\ref{fig:Drad}).
From that moment on, the energy contained in the spectrum cannot increase anymore. 
For this to be possible, the energy flux associated with branchings must precisely equilibrate 
the rate for energy injection by the source; that is,
the r.h.s. of \eqn{balance} must vanish:
 \beq\label{Fx0as}
\F(x_0,\tau)\,\simeq\,\int_{x_0}^\xc {\rm d} x\,
\mcal{S}_0(x)\,=\,2 {\abar}\big({\sqrt{x_c}}-\sqrt{x_0}\big)
\,.\eeq
As expected, this result is independent of time and fixed by
the source. For $x_0=0$, it yields
 \beq\label{Fflowas}
\F_{\rm flow}(\tau)\,\simeq\,
2 {\abar}{\sqrt{x_c}}\quad\mbox{when}\quad \pi\abar^2\tau^2/x_c\gg1\,,\eeq
which is indeed consistent with both the numerical results in
Fig.~\ref{fig:F0xc} and the large--time asymptotics of \eqn{Fflowg}, as one can easily 
check\footnote{At large times, one has $\zeta\gg 1$ for any $x$, 
hence one can approximate $\gamma(1/2,\zeta)\simeq \sqrt{\pi}$ within \eqn{Fflowg}.}.
%which immediately leads to  \eqn{Fflowas}.)

For comparison, let us also notice the spectrum and flux that would be generated
by a localized source $\mcal{S}(x)=A\delta(x-x_c)$
which acts for $\tau\ge 0$. (This problem has been already considered
in Ref.~\cite{Blaizot:2013hx}.) For generic $\tau$,
the corresponding spectrum coincides (up to a factor of $A$) with the r.h.s. of \eqn{intG}
evaluated at $x_1= x_c$. For large times $\pi\abar^2\tau^2/x_c\gg1$, this reaches the
steady shape 
 \beq\label{Das}
D_{\rm as}(x)\,=\, \frac{A}{2\pi\abar}\,\sqrt\frac{x_c}{x(x_c-x)}\,.
\eeq
In the same limit, the energy flux is both steady and strictly uniform, $\F_{\rm as}(x_0)=A$, 
as in standard turbulence. For $x\ll x_c$, these results are consistent with 
the Kolmogorov--Obukhov relation \eqref{Kolmog}.

 \begin{figure}[t]
	\centering
	\includegraphics[width=0.7\textwidth]{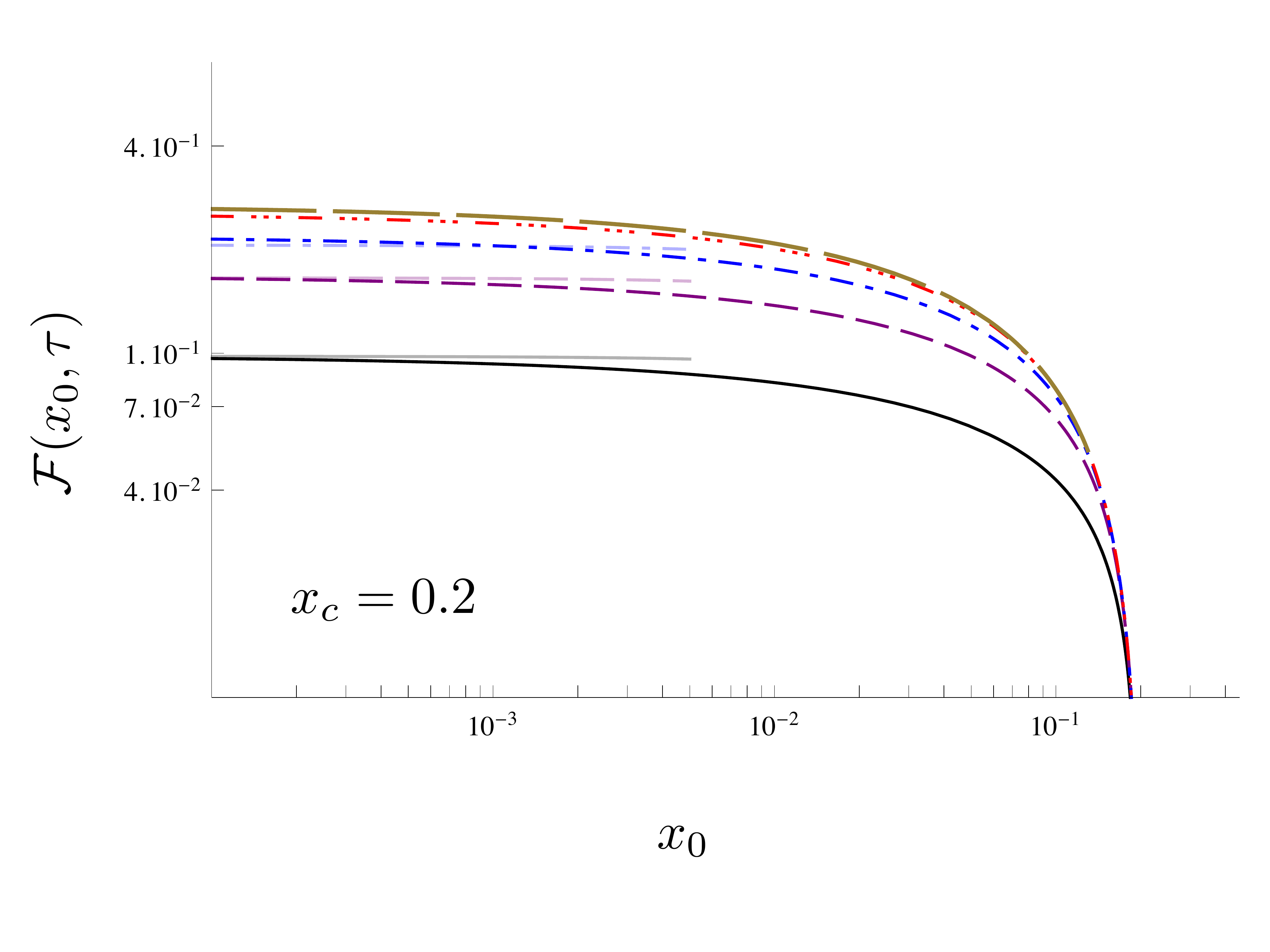}
		\caption{\sl Plot (in log-log scale) of the energy flux $\F(x_0,\tau)$, cf. \eqn{Fx0g}, as a function of $x_0$ for 
		$\abar=0.3$, $x_c=0.2$,
		 and various values of $\tau$ : $\tau=0.2$ (solid, black), $\tau=0.4$ (purple, dashed),
		$\tau=0.63$ (blue, dashed--dotted), $\tau=1$ (red, dashed--triple--dotted).
		The thin curves, shown for $\tau\le\tau_L=0.63$ and $x_0\le 0.005$, 
 represent the  approximation \eqref{Fx00} valid at small $\tau$
		and small $x_0$.
		The enveloping curve (brown, long--dashed) is the limiting curve at large $\tau$, 
		cf. \eqn{Fx0as}.}
		\label{fig:Fsmallxc}
\end{figure}

It is finally interesting to study the $x_0$--dependence of the energy flux in this high--energy case.
This is expressed by \eqn{Fx0g} that we have plotted  in Fig.~\ref{fig:Fsmallxc}
as a function of $x_0$ for different values of $\tau$ and for $x_c=0.2$. 
Good analytic approximations can also be obtained.
For relatively small times $\pi\abar^2\tau^2/x_c\ll 1$, 
and for $x_0$ not too close to $x_c$, it is convenient to rewrite  \eqn{Fx0g} as
\beq\label{Fx0dep}
\F(x_0,\tau)\,=\,\F_{\rm flow}(\tau)\,-\,\frac{\abar}{\sqrt{\pi}}\int_0^{x_0}
\frac{\rmd x}{\sqrt{x}}\
\gamma\Big(\frac{1}{2},\zeta\Big)\,.\eeq
When $\zeta\ll 1$, we can use the Taylor expansion of the function $\gamma(1/2,\zeta)$,
which is rapidly converging. To the same accuracy as in \eqn{Fflow1}, i.e. to
second order in $\abar\tau$, it is enough to use $\gamma(1/2,\zeta)\simeq 2\sqrt{\zeta}$,
which yields 
 \beq\label{Fx00}
\F(x_0,\tau)\,\simeq\,2\pi\abar^2\tau\left\{
1-\,\frac{\abar\tau}{\sqrt{x_c}}
-\frac{2}{\pi}\arcsin\sqrt{\frac{x_0}{x_c}}
\right\}
\,.
\eeq
At larger times 
$\pi\abar^2\tau^2/x_c\gg1$, and also for $x_0$ very close to $x_c$ and any $\tau$,
the flux takes the form in \eqn{Fx0as}. 
Both the numerical results in Fig.~\ref{fig:Fsmallxc}
and the analytic approximations in Eqs.~\eqref{Fx00} and  \eqref{Fx0as}
demonstrate that the flux associated with
branchings is quasi--uniform (i.e. independent of $x_0$) for any $x_0\ll x_c$. 
As already mentioned, this signals a phenomenon
of wave turbulence. Additional evidence in that sense will emerge from the analysis in 
the next subsection. 

\subsection{The energy flux revisited: democratic branchings}
\label{sec:flow2}

In this subsection, we shall present an alternative calculation of the energy flux,
which exploits the second equality in \eqn{Fx0xc}, i.e. the $x$--integral of the branching
term $\mcal{I}[D_{\rm rad}]$. As we shall see, the main virtue of this alternative method
is that it involves the gluon spectrum {\em quasi--locally in $x$}~: in order to compute
the flux $ \F(x_0,\tau)$ at small $x_0\ll x_c$, we need the spectrum $D_{\rm rad}(x,\tau)$
at small $x\ll x_c$ as well. This property has important consequences, of both practical and
conceptual nature. In practice, it will allow us to derive an exact analytic expression for
the rate of flow $\F_{\rm flow}(\tau)=\F(x_0=0,\tau)$ and to establish the analog of the
Kolmogorov--Obhukov relation for the problem at hand.
%Also, it will permit us to obtain the limited, but
%accurate, expansion in \eqn{Fflow1} via purely perturbative arguments. 
%This is useful since the same arguments can be used in relation with the complete splitting 
%kernel in \eqn{Kdef}, for which the exact spectrum is not analytically known. 
 At a conceptual level, the locality of the branching process
in energy (or in $x$) is a fundamental property of a turbulent process  \cite{KST,Nazarenko}. 
This property is quite unusual in the context of a gauge theory, where splittings are generally
very asymmetric due to the `infrared' ($x\to 0$) singularity of bremsstrahlung. Its emergence
in the context of the medium--induced gluon cascade \cite{Blaizot:2013hx,Baier:2000sb,Kurkela:2011ti}
is a non--trivial consequence of coherence phenomena associated with multiple scattering, 
which lead to a profound modification in the splitting rate as compared to 
bremsstrahlung in the vacuum.
%like the  (QCD version of the) LPM
%effect \cite{Baier:1996kr,Baier:1996sk,Zakharov:1996fv,Zakharov:1997uu,Baier:1998kq} .

The integral of the branching term occurring  in  \eqn{Fx0xc} can be decomposed as
\beq
-\int_{x_0}^\xc {\rm d} x\,\mcal{I}[D_{\rm rad}](x,\tau)\,=\,
\int_{x_0}^\xc {\rm d} x\,\mcal{L}(x,\tau)\,+\,\int_{x_0}^\xc {\rm d} x\,\mcal{G}(x,\tau)\,,
\eeq
where the two terms in the r.h.s. are the  respective contributions of the `loss' 
and `gain' term in the rate equation. The `loss' contribution is easily evaluated as
\beq\label{Pvirt}
\int_{x_0}^\xc {\rm d} x\,\mcal{L}(x,\tau)\,=\,
\int_0^1 \rmd z \,z\, {\cal K}(z)  
 \int_{x_0}^{x_c}\rmd x\,\frac{D_{\rm rad}(x,\tau)}{\sqrt{x}}\,.
\eeq
In the `gain' contribution, it is useful to change the integration variable as
$x\to x'\equiv x/z$ :
  \begin{align}\label{Preal}
  \int_{x_0}^\xc {\rm d} x\,\mcal{G}(x,\tau)\,=\,&-
 \int_{x_0}^{x_c}\rmd x\,
 \int \rmd z \,\Theta\Big(z-\frac{x}{x_c}\Big)\, {\cal K}(z)  
  \,\sqrt\frac{z}{x}
  D_{\rm rad}\Big(\frac{x}{z}, \tau\Big)\nn
 =\,&- \int \rmd z \,z\, {\cal K}(z)  \,\Theta\Big(z-\frac{x_0}{x_c}\Big)\, 
 \int_{x_0/z}^{x_c}\rmd x'\,\frac{D_{\rm rad}(x',\tau)}{\sqrt{x'}}\,,
  \end{align}
where, in the second line, the upper limit $x_c$ on $x'$ follows from the condition $z> x/x_c$ ;
also, the last $\Theta$--function, which enforces $z>x_0/x_c$, guarantees that the lower limit $x_0/z$ 
in the integral over $x'$ remains smaller than the upper limit $x_c$.
As usual, the `gain' and `loss' contributions taken separately develop
singularities from the endpoint at $z=1$ of the integral over $z$, but these singularities
cancel in the sum of the two contributions. Hence, the overall result is well defined and reads
  \begin{align}\label{Px0}
  \F(x_0,\tau)\,=\, \abar
 \int_{x_0/x_c}^1 \rmd z \,z\, {\cal K}(z)  
 \int^{x_0/z}_{x_0}\rmd x\,\frac{D_{\rm rad}(x,\tau)}{\sqrt{x}}\,
 +  \abar\int_0^{x_0/x_c} \rmd z \,z\, {\cal K}(z)  
 \int^{x_c}_{x_0}\rmd x\,\frac{D_{\rm rad}(x,\tau)}{\sqrt{x}}
 \,.
  \end{align}
To  better appreciate the physical interpretation of this result, let us return to the individual,
`loss' and `gain', contributions, as shown in \eqn{Pvirt} and respectively \eqn{Preal}.

The interpretation of the `loss' term in \eqn{Pvirt} is quite clear: this is the energy transferred 
per unit time from one parton generation to the next one via the branching of 
any of the `hard' modes with $x_0<x<x_c$. (Recall that ${\cal K}(z)/\sqrt{x}$
represents the splitting rate for the parent mode $x$ into daughter modes $zx$
and $(1-z)x$. Also the factor of $z$ within the first integral can be equivalently replaced
by $z\to [z+(1-z)]/2=1/2$, due to the symmetry property ${\cal K}(z)={\cal K}(1-z)$; hence,
this factor truly accounts for the contribution of {\em both} daughter gluons.)
However, some of these splittings do not contribute to the energy flux at $x_0$~:
this is the case for the splittings with $zx>x_0$ (a condition which can be satisfied only
for $z$ values which are large enough, namely $z>x_0/x_c$), for which the daughter gluons 
are still harder than $x_0$. The contributions of these splittings is therefore subtracted 
by the `gain' term in \eqn{Preal}, which is negative indeed. Accordingly, the net result is 
the sum of two types of contributions, represented by the two terms in the r.h.s. of \eqn{Px0} :
\texttt{(i)} relatively hard splittings with $x_0/x_c< z <1$, but
% (`quasi--local in energy' or `quasi--democratic'),
such the parent gluon $x$ was close enough to $x_0$ (within the 
strip at $x_0<x<x_0/z$), and 
\texttt{(ii)} relatively soft splittings with $z<x_0/x_c$,  %(`non-local in energy' or `highly asymmetric'), 
in which case the parent gluon can be located anywhere between $x_0$ and $x_c$.

The following observations are useful for what follows.
In the limit where $x_0\ll x_c$, the first term in the r.h.s of \eqn{Px0} dominates 
over the second one and controls the rate of flow. This is 
clear from the fact that the second term in  \eqn{Px0} vanishes when $x_0\to 0$, while the first
one preserves a finite value in that limit, as we shall shortly see. 
%(The limit $x_0\to 0$ of the first terms turns out to be
%quite subtle,  since the corresponding integral over $x$ is logarithmic; see below.)
Furthermore, still for $x_0\ll x_c$, the second term is controlled by very asymmetric splittings
($z<x_0/x_c\ll 1$), whereas the first one is rather dominated by {\em quasi--democratic
branchings}, that is, by generic $z$ values in the bulk, which are not specially
close to either the lower limit $z=x_0/x_c\ll 1$, or the upper limit $z=1$, of the $z$--integral.
Indeed, this integral is rapidly convergent both at small $z$, because of the 
factor of $z$ in the integrand, and at $z\to 1$, because the result of the integral over $x$ 
linearly vanishes in that limit. As already mentioned, the prominence  
of `quasi--democratic branchings' 
is an essential condition for the emergence of wave turbulence: e.g.
this permits the existence of fixed--point (KZ) solutions, which
requires fine cancellations between the ({\em a priori} non--local) 
`gain' term and the (always local) `loss' term.

\begin{figure}[t]
	\centering
	\includegraphics[width=0.7\textwidth]{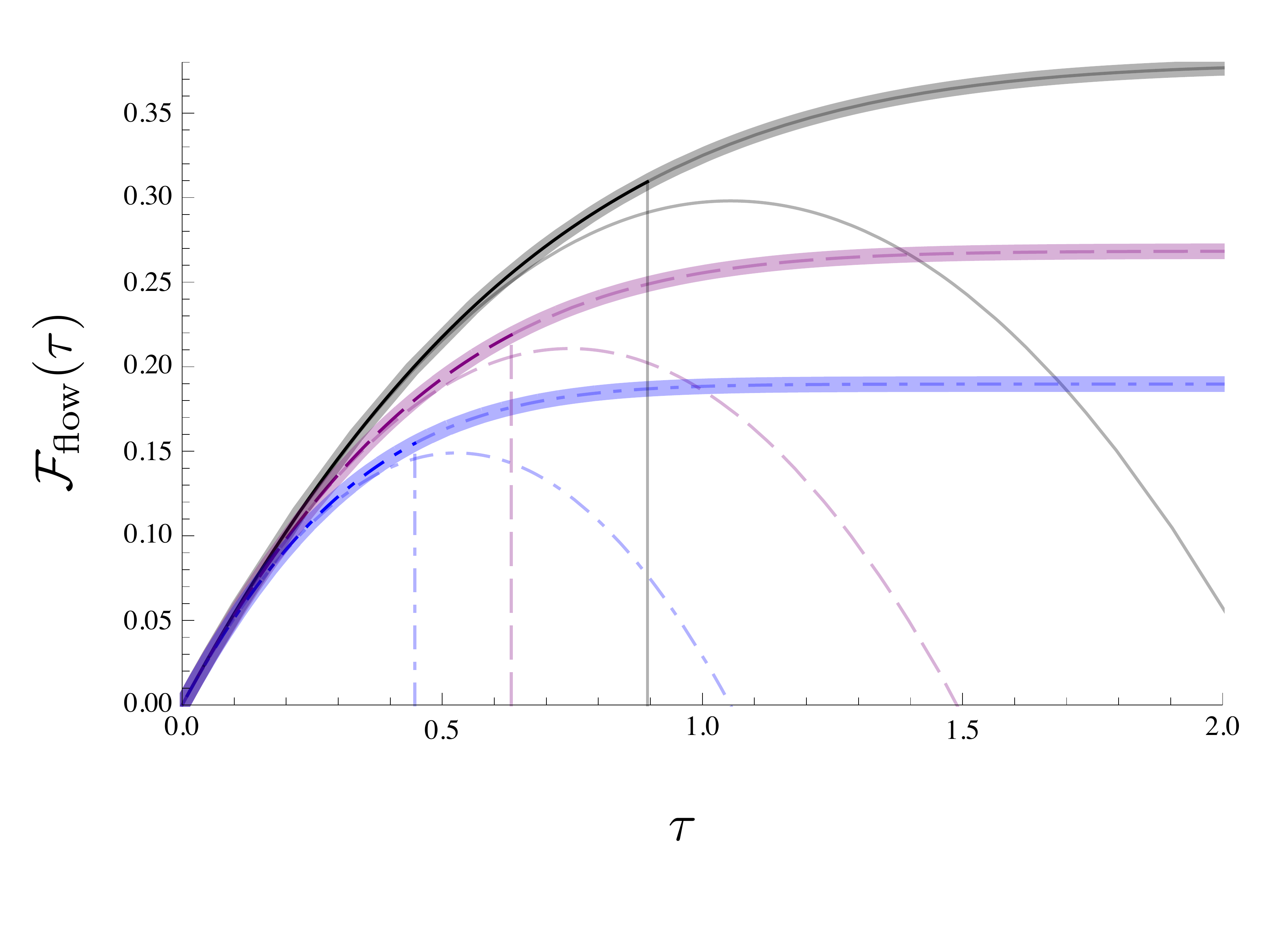}
		\caption{\sl The rate of flow $\F_{\rm flow}(\tau)$
		as a function of $\tau$ in the high--energy regime for various values of
		$x_c$~:  $x_c = 0.4$ (black, solid), 
		 $x_c=0.2$ (purple, dashed), and $x_c=0.1$ (blue, dashed-dotted). 
		 	The thick lines represent the respective curves within their physical
	 range of validity ($\tau < \sqrt{2 x_c}$), as computed by numerical integration in \eqn{Fx0g}.
	 The thin curves following the thick ones are the predictions of \eqn{Fx0g} for larger times,
	  outside the physical range ($\tau > \sqrt{2 x_c}$).
The thin curves deviating from the thick ones correspond to the limited expansion 
in  \eqn{Fflow1}. Finally, the 
very thick (opaque) curves are the new, fully explicit, analytic result in \eqn{Fflowexact}.
The vertical lines denote the physical upper limit on time  $\tau_L=\sqrt{2 x_c}$.		}
		\label{fig:F0xc}
\end{figure}

This locality allows us to construct an exact solution for the energy flux
in the limit $x_0\to 0$ and for the simplified kernel ${\cal K}_0(z)$ (for which the
spectrum is analytically known). When  $x_0\to 0$,
only the first term in \eqn{Px0} survives.
The fact that the respective integral over $z$ is not specially sensitive to its lower
limit $x_0/x_c$ means that the relevant values of $z$ do {\em not} scale like $x_0$ 
when $x_0\to 0$. Hence, the upper limit $x_0/z$ of the integral
over $x$ vanishes when $x_0\to 0$, so like the corresponding lower limit. Accordingly,
this integral is controlled by very small values of $x$, which scale
like $x_0$ and in particular are much smaller than $x_c$. It is then justified to evaluate
this integral using the dominant behavior of the spectrum for $x\ll x_c$,
that is, the KZ spectrum in \eqn{Dradsc}.
With this scaling behavior $\sim 1/\sqrt{x}$, the integral over $x$ is logarithmic and its result is independent of $x_0$.
One thus finds % (recall the notation $\F_{\rm flow}(\tau)=\F(x_0=0,\tau)$)
\beq\label{Fflowexact}
\F_{\rm flow}(\tau) \,=\, 2\pi \abar^2\tau \,\big[1-h(\zeta_0)\big]\,,\qquad
   \zeta_0\,\equiv\,\zeta(x_c,\tau)\,=\,\frac{\pi\abar^2\tau^2}{x_c}\,,
%\left\{ \frac{1}{\sqrt{\pi}}\,\Gamma\Big(\frac{1}{2},\zeta\Big)
 %  +\frac{1-\rme^{-\zeta}}{\sqrt{\pi\zeta}}\right\}\quad\mbox{with}\quad 
 %  \zeta=\frac{\pi\abar^2\tau^2}{x_c}\,,
 \eeq
where the overall factor $2\pi$ has been generated as
\beq\label{v0}
2\pi= \int_{ 0}^1 \rmd z \,z\, {\cal K}_0(z)  \ln\frac{1}{z}\,=\,
 \int_{ 0}^1 \rmd z\, \frac{1}{\sqrt{z} (1-z)^{3/2}}\ln\frac{1}{z}\,.
 \eeq
Using the properties of the function $h(\zeta)$ discussed in Sect.~\ref{sec:rad}, 
one can easily check both the small--$\tau$ expansion of the flow, as anticipated
in \eqn{Fflow1}, and its large--$\tau$ asympotics in \eqn{Fflowas}. 
As a check of  \eqn{Fflowexact}, we display this result in Fig.~\ref{fig:F0xc} (as a function
of $\tau$ for several values of $x_c$) versus the result of the numerical integration in  \eqn{Fx0g}.
One can also see in this figure that the limited expansion \eqref{Fflow1} is indeed a very good
approximation for any $\tau$ in the physical range, as already noticed in Sect.~\ref{sec:Fsmallxc}.
This is understandable since the first correction beyond \eqn{Fflow1} in the small--$\tau$ 
expansion of \eqn{Fflowexact} is exactly vanishing, 
as manifest on \eqn{Dradexp1}. 

By inspection of Eqs. \eqref{Dradsc} and \eqref{Fflowexact}, it is obvious that the spectrum
at small $x$ is proportional to the flow, in the sense of \eqn{Kolmog}. The above construction
of \eqn{Fflowexact} explains the physical origin of this proportionality and also suggests
that it is quite general: it holds for any splitting kernel with the singularity structure shown
in \eqn{Kdef}, since any such a kernel leads to democratic branchings and to a spectrum
which at small $x$ has the shape of the scaling spectrum $D_{\rm sc}(x)= 1/{\sqrt{x}}$. The
time dependence of the spectrum (again at small $x$) depends upon the detailed structure
of the branching kernel (it is generally different for the full kernel ${\cal K}(z)$ and for the 
simplified one ${\cal K}_0(z)$), and also upon the nature of the `source' at large $x$
(it is e.g. different for a source localized at $x_c$, $\mcal{S}(x)=A\delta(x-x_c)$, as opposed
to a radiation source $\mcal{S}_0(x)=\theta(x_c-x){\abar}/{\sqrt{x}}$). But the rate of flow  
$\F_{\rm flow}(\tau)$ has exactly the same time--dependence as the spectrum, 
and the proportionality relation \eqref{Kolmog} universally holds, with a proportionality factor which
is kernel--dependent though:
 \beq\label{Kolmogv}
  D(x,\tau)\,\simeq\,\frac{1}{v\abar}\,
  \frac{\F_{\rm flow}(\tau)}{\sqrt{x}}\qquad\mbox{for}\quad x\,\ll\,x_c\,.\eeq
Here, $\upsilon$ is a pure number, defined by the obvious generalization of \eqn{v0} :
\beq\label{v}
\upsilon\equiv \int_{ 0}^1 \rmd z \,z\, {\cal K}(z)  \,\ln\frac{1}{z}\,=\,
 \int_{ 0}^1 \rmd z \frac{f(z)}{\sqrt{z} (1-z)^{3/2}}\ln\frac{1}{z}\,\simeq\,4.96\,.
 \eeq
On both  \eqn{v} or \eqn{v0}, it is obvious that the respective integral over $z$ is dominated
by generic values in the bulk, as expected for quasi--democratic branchings.
As discussed after \eqn{Eflowtau}, $\upsilon$ has the physical interpretation of the average number
of soft primary gluons with energies $\omega\sim\omega_s(t) = \abar^2 \hat q t^2/2$
that are emitted by the leading particle during a time $t$.

\eqn{Kolmogv} is particularly useful
in a steady situation, where the energy flux is {\em a priori} known, since
determined by the external source. (This is the case in the familiar turbulence problem, 
where the Kolmogorov--Obhukov relation has been originally identified.)
As a simple, yet non--trivial, application of this type, consider the
steady situation reached when the external source $\mcal{S}_0(x)=\theta(x_c-x){\abar}/{\sqrt{x}}$
acts for sufficiently large time $\abar^2\tau^2\gg {x_c}$. 
The corresponding flow is given by \eqn{Fflowas} and then
\eqn{Kolmogv} can be used to deduce the asymptotic spectrum at large times and small $x$~:
\beq\label{Dasv}
  D(x,\tau\to\infty)\,\simeq\,\frac{2}{v}\,
  \sqrt{\frac{x_c}{x}}\qquad\mbox{for}\quad
 % \abar^2\tau^2\,\gg\,{x_c}\quad\mbox{and}\quad
   x\,\ll\,x_c\,.\eeq
This result is interesting in that it represents a non--perturbative prediction associated with the full
kernel, for which exact analytic solutions are not known.
(For the simplified kernel, $v\to 2\pi$ and \eqn{Dasv} reduces to \eqn{Dradas}, as it should.)

 \begin{figure}[t]
	\centering
	\includegraphics[width=0.7\textwidth]{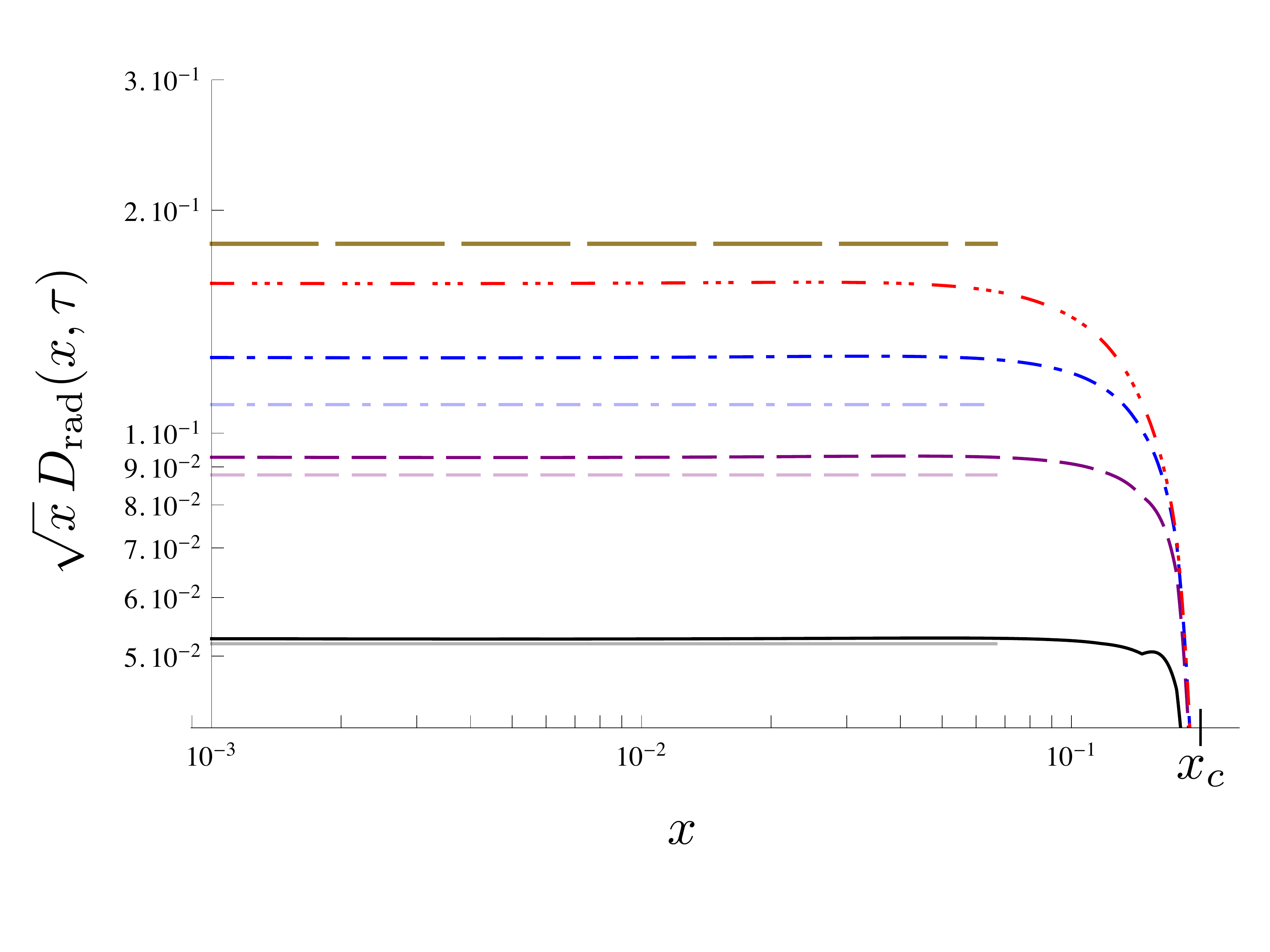}
		\caption{\sl The numerical solution to the rate equation \eqn{Drad0}
        with the full splitting kernel $\mcal{K}$ from \eqn{Kdef}, for $x_c=0.2$ and
         various values of $\tau$ : $\tau=0.2$ (solid, black), $\tau=0.4$ (purple, dashed),
		$\tau=0.63$ (blue, dashed--dotted), $\tau=1$ (red, dashed--triple--dotted).
		The thin curves, shown for $\tau\le\tau_L=0.63$ and $x\le 0.07$, 
 represent the small--$\tau$ and small--$x$ approximation
   in		 \eqn{Dradexp2}.
	The enveloping curve (brown, long--dashed) is the limiting curve at large $\tau$, 
		cf. \eqn{Dasv}.}
		\label{fig:DexactK}
\end{figure}

Still for the full kernel,  \eqn{Kolmogv} can also be used in the reversed way, namely 
to deduce the flow from the 
spectrum in the  small--time regime at $\abar^2\tau^2\ll {x_c}$.
Indeed, in this limit and for $x\ll x_c$, the spectrum can be computed in perturbation
theory, via iterations (see the discussion in Appendix~\ref{sec:PT}).
To second order in $\abar\tau$, the result turns out to be the same as for the simplified 
kernel ${\cal K}_0(z)$, namely (compare to \eqn{Dradexp1})
\beq\label{Dradexp2}
   D_{\rm rad}(x,\tau)\,\simeq\, \frac{\abar\tau}{\sqrt{x}}\left(
   1\,-\,\frac{\abar\tau}{\sqrt{x_c}}   \right)\qquad\mbox{for}\quad
  \abar^2\tau^2\,\ll\,{x_c}\quad\mbox{and}\quad
   x\,\ll\,x_c\,.\eeq
By using this approximation together with \eqn{Kolmogv}, we can obtain the generalization of 
 \eqn{Fflow1}  to the case of the complete kernel:
 \begin{align}\label{Fflownew}
\F_{\rm flow}(\tau) \,=\, \upsilon \abar^2\tau \left(
   1\,-\,\frac{\abar\tau}{\sqrt{x_c}}   \right)\,.
   \end{align}
This result is quite useful, in particular for phenomenology, 
in that it offers a rather accurate estimate for
the energy loss via flow for the case of the physical kernel. 
This will be further discussed in the next section.
In Fig.~\ref{fig:DexactK} we show the numerical solution to the rate equation \eqref{Drad0}
with the full splitting kernel $\mcal{K}$ in \eqn{Kdef}, together with its analytic approximations 
valid at small $x$~:  \eqn{Dradexp2}  at small $\tau$ and \eqn{Dasv}
at large $\tau$. In particular, we have checked that this special number $\upsilon\simeq 4.96$
can be indeed read off the asymptotic behavior of the numerical solution at large time,
in agreement with \eqn{Dasv}.

\section{Physical discussion: energy loss at large angles}
\label{sec:eloss}

In this section, we shall summarize the results obtained in the previous sections and use them
to compute one of the most interesting observables for the phenomenology of di--jet asymmetry
at the LHC: the energy lost by the gluon cascade via soft quanta propagating at large angles.
Specifically, we shall successively consider the following quantities:

\texttt{(i)} {\em the flow energy  ${\cal E}_{\rm flow}(\tau)$}: this is the energy fraction carried away
by the turbulent flow and which formally ends up in a condensate at $x=0$;

\texttt{(ii)} {\em the thermalization energy ${\cal E}_{\rm th}(\tau)$}: this is 
the energy fraction which is carried by quanta with $x < x_{\rm th}\equiv T/E$, which are assumed
to thermalize and hence transmit their energy to the medium.
(As in Sect.~\ref{sec:low}, we assume that the thermalization mechanism
acts as a `perfect sink', i.e. it does not modify the energy flux at $x\ge x_{\rm th}$~;
cf. the discussion at the end of Sect.~\ref{sec:lowflux}.)

\texttt{(iii)} {\em the energy transported at angles larger than a given value $\theta_0$}:
the definition of this quantity requires some additional discussion and is postponed after
the study of the two previous ones.

The {\em flow energy} can be calculated in two alternative ways: as the $\tau$--integral of the respective
flux $\F_{\rm flow}(\tau)$, which is explicitly given by  \eqn{Fflowexact}, or as the $x$--integral of the
change $\delta D_{\rm br}(x,\tau)$ in the spectrum due to branchings, as shown in \eqn{Dradex2}:
\begin{align}\label{Eflowxc} 
  {\cal E}_{\rm flow}(\tau)\,\equiv \,\int_0^{\tau}\rmd\tau'\,\F_{\rm flow}(\tau')\,=\,
  \int_{0}^\xc {\rm d} x\
\delta D_{\rm br}(x,\tau)\,.
 \end{align}
The second representation above relies on the fact that the flow energy is by definition the difference
between the total energy supplied by the source  ${\cal S}_0(x)$ and the radiation energy which 
remains in the spectrum.  Here, we shall use this second representation to numerically compute
${\cal E}_{\rm flow}$, but rely on the first one for analytic estimates. Indeed, we know that 
already the limited expansion of the flow shown in  \eqn{Fflownew} 
is very accurate for any $\tau\le\tau_L$~;  this can be easily integrated over time to give
 \begin{align}\label{Eflowexp} 
  {\cal E}_{\rm flow}(\tau)\,\simeq\,\frac{ \upsilon}{2}\,
  \abar^2\tau^2\,\left(1-\,\frac{2}{3}\frac{\abar\tau}{\sqrt{x_c}}
  \right)\,.
   \end{align}
This estimate holds for the full kernel ${\cal K}(z)$,  but the corresponding result for
the simplified kernel ${\cal K}_0(z)$ is simply obtained by replacing $\upsilon\to 2\pi$ in the prefactor.

 \begin{figure}[t]
	\centering
	\includegraphics[width=0.7\textwidth]{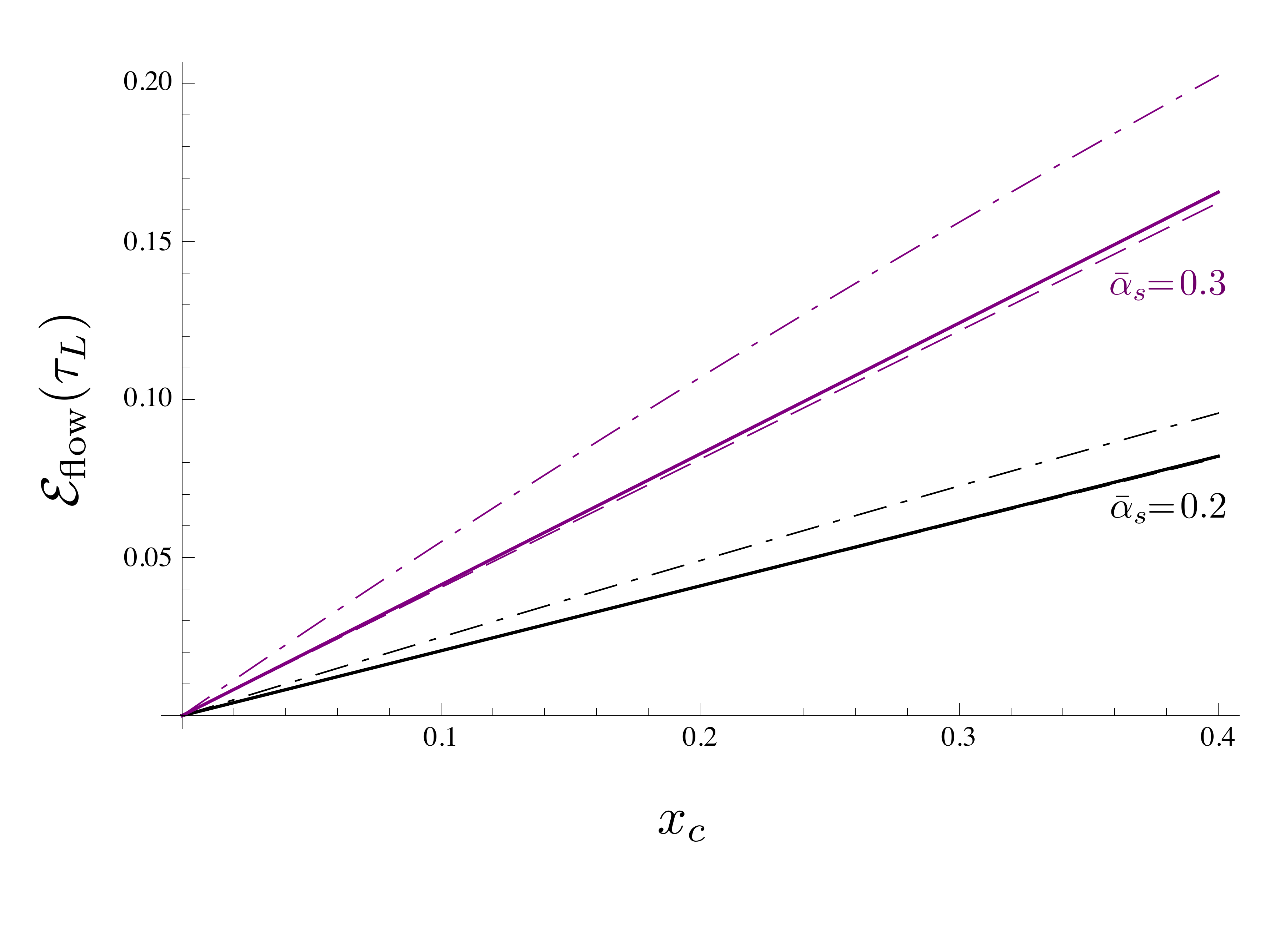}
		\caption{\sl The energy fraction ${\cal E}_{\rm flow}(\tau_L)$
		carried by the turbulent flow, i.e.  \eqn{Eflowxc} with $\tau=\tau_L\equiv\sqrt{2x_c}$,
		plotted as a function of $x_c$ for 
		two values of the coupling constant: $\abar=0.2$ (black) and
		$\abar=0.3$ (purple). Solid lines: the exact result obtained by numerical integration
		in the second equality in \eqn{Eflowxc}.
		Dashed lines: the weak--coupling expansion in
		 		  \eqn{Eflowexp}, that is,  ${\cal E}_{\rm flow}=2\pi\abar^2 x_c(1-2\sqrt{2}\abar/3)$.
		  (For $\abar=0.2$, this approximation can hardly be distinguished from the exact curve.)
		For comparison, we also show, with dashed--dotted lines, the respective predictions
		of the `low--energy case', i.e. \eqn{Eflow} with $\tau=\sqrt{2x_c}$.}
		\label{fig:Esmallxc}
\end{figure}

In Fig.~\ref{fig:Esmallxc} we show the flow energy evaluated at the end of the evolution
($\tau=\tau_L=\sqrt{2x_c}$) as a function of $x_c$ and for two values of $\abar$.
%{\bf [Please remove the LO results from this plot, i.e. the dashed curves.]}
We here compare
the respective exact results, cf. \eqn{Eflowxc}, with the limited expansion in \eqn{Eflowexp} 
(which is seen to be quite accurate) and with the prediction \eqref{Eflow} of the `low--energy case'
which here is extrapolated to $x_c\ll 1$, that is, outside its physical range of validity.
The purpose of this extrapolation is to emphasize that, by ignoring the kinematical constraint
$x\le x_c$, one would significantly overestimate the energy loss via flow.
Remarkably, the plots in  Fig.~\ref{fig:Esmallxc} show that the quantity ${\cal E}_{\rm flow}(\tau_L)$
is a linear function of $x_c$. This property is obvious for the limited expansion in \eqn{Eflowexp},
but is in fact exact within the present effective theory, as we now show. Namely, by using 
 \eqn{Dradex2} for  $\delta D_{\rm br}(x,\tau)$,  we can write
%(recall that $\tau_L=\sqrt{2x_c}$)
\begin{align}\label{Eflowxc2} 
  {\cal E}_{\rm flow}(\tau_L)\,=\, {\abar\tau_L} \int_{0}^\xc \frac{{\rm d} x}{\sqrt{x}}\
   h\bigg(\frac{\pi\abar^2\tau_L^2}{x_c-x}\bigg)\,=\,\sqrt{2}\abar x_c
   \int_{0}^1 \frac{{\rm d} u}{\sqrt{u}}\  h\bigg(\frac{2\pi\abar^2}{1-u}\bigg)\,,
 \end{align}
where the r.h.s. is indeed linear in $x_c$, as anticipated.
This is interesting in that it implies that the energy which is lost via flow, namely (cf. \eqn{Eflowexp}),
\beq\label{DEflow}
\Delta E_{\rm flow}\,\equiv\,{E}\,{\cal E}_{\rm flow}(\tau_L)
\,\simeq\,{\upsilon}\,\bar{\alpha}^2 \omega_c
\,\bigg(1-\,\frac{2\sqrt{2}}{3}{\abar}\bigg)\,,
\eeq
is independent of the energy $E$ of the leading particle and parametrically of order
$\bar{\alpha}^2 \omega_c= \omega_s^2$ (the natural energy scale for multiple branchings).
One should however keep in mind that this conclusion holds only for sufficiently energetic
jets, such that $x_c\ll 1$, or $E\gg\omega_c$. Notice also that the actual value of the
energy loss in \eqn{DEflow} is enhanced by the relatively large numerical factor 
${\upsilon}\big(1-{2\sqrt{2}}{\abar}/3\big)$ ($\simeq 3.5$
for $\abar=0.3$) as compared to its parametric estimate $\bar{\alpha}^2 \omega_c$. This is
mostly due to the factor ${\upsilon}\simeq 4.96$, which we recall is the average number of soft
primary emissions with energies $\omega\sim \omega_s$.

Given the flow energy in  \eqn{Eflowxc}, the {\em thermalization energy} can immediately 
be computed as the sum between
${\cal E}_{\rm flow}(\tau)$ and the energy contained in the small--$x$ bins of 
the spectrum:
 \beq\label{Ethsmallxc}
{\cal E}_{\rm th}(\tau)\,=\, {\cal E}_{\rm flow}(\tau)\,+\,\int_{0}^\xt {\rm d} x \,D_{\rm rad}(x,\tau)\,.
\eeq
In practice, $x_{\rm th}\ll x_c$, hence the above integral can be
estimated by using the dominant behavior of the spectrum for $x\ll x_c$. 
To the same accuracy as in \eqn{Eflowexp}, one finds
 \begin{align}\label{Ethexp} 
  {\cal E}_{\rm th}(\tau)\,\simeq\,\frac{ \upsilon}{2}\,
  \abar^2\tau^2\,\left(1-\,\frac{2}{3}\frac{\abar\tau}{\sqrt{x_c}}
  \right) + 2\abar\tau\sqrt{x_{\rm th}}\,\left(1-\,\frac{\abar\tau}{\sqrt{x_c}}
  \right) 
  \,.
   \end{align} 
We emphasize that this result, which holds for the complete kernel \eqref{Kdef}, 
is fully obtainable from perturbation theory: it only requires
the second iteration to the spectrum in \eqn{Dradexp2}. As manifest in \eqn{Ethexp},
the flow contribution to ${\cal E}_{\rm th}(\tau)$ is formally of higher order in $\abar\tau$,
yet it dominates over the `spectrum' contribution as soon as $x_{\rm th}$ is small enough:
for $\tau=\tau_L$, the flow dominates provided
$x_{\rm th} < x_s=\abar^2x_c$ (or, equivalently, $T< \omega_s$), a condition which is 
well satisfied in practice (see below). 
%This is the same condition as found in the discussion
%of the `low--energy' case in Sect.~\ref{sec:low}. 

\begin{figure}[h]
	\centering
	\includegraphics[width=0.7\textwidth]{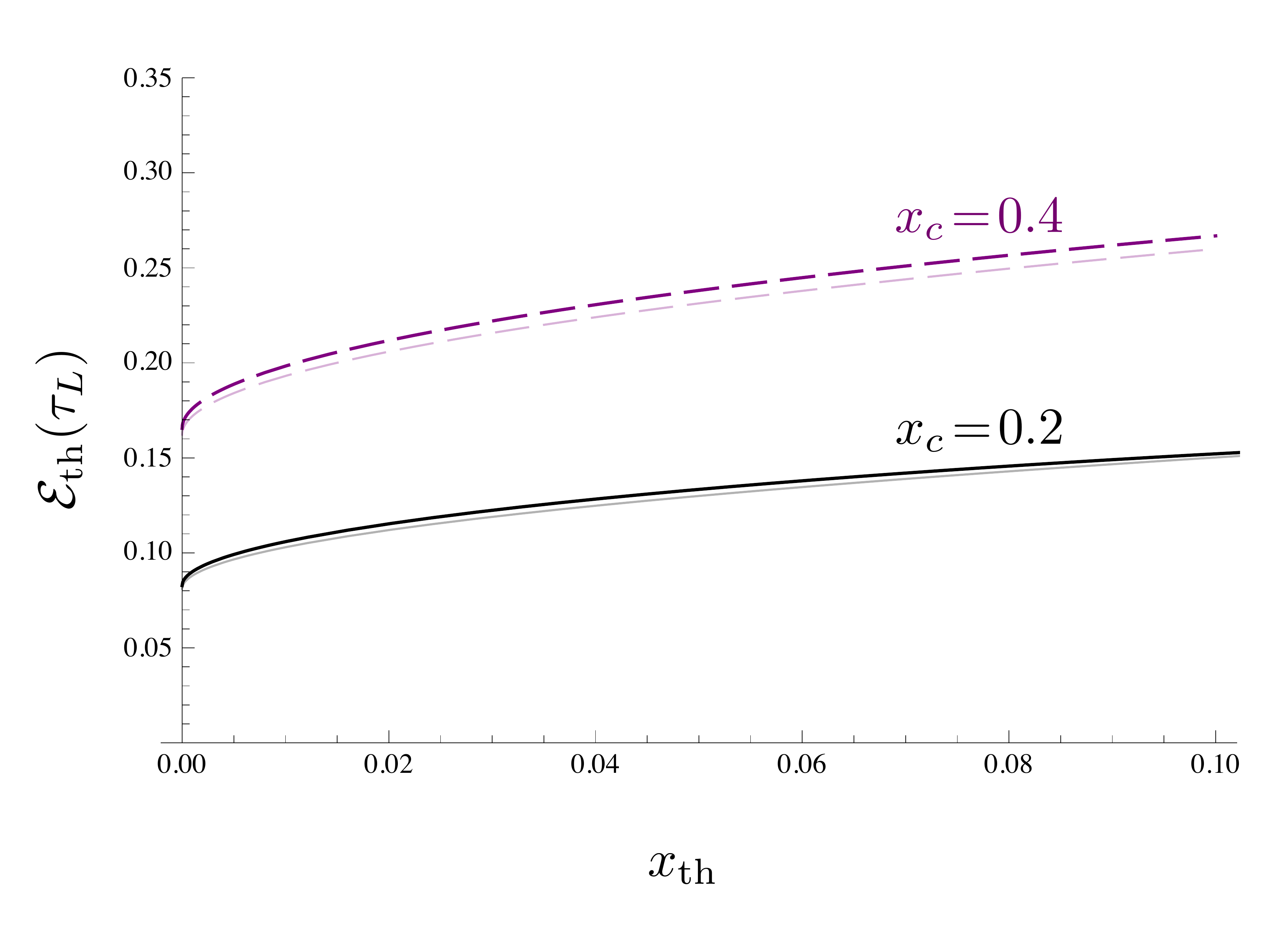}
		\caption{\sl The energy fraction which `thermalizes' ${\cal E}_{\rm th}(\tau_L)$, plotted
		as a function of the thermalization scale $x_{\rm th}$ for two values of $x_c$: 
		$x_c=0.4$ (black, solid) and $x_c=0.2$ (purple, dotted).
		The thick curves are the exact result obtained via  numerical integration in \eqn{Ethsmallxc}.
		The thin, opaque, curves are the respective predictions of the limited expansion in \eqn{Ethexp}
	         with $ \upsilon\to 2\pi$.}
		\label{fig:Ethxth}
\end{figure}

In Fig.~\ref{fig:Ethxth} we plot ${\cal E}_{\rm th}(\tau_L)$ as a function of $x_{\rm th}$ for $x_c=0.2$
and $x_c=0.4$, and for the simplified kernel $\mcal{K}_0$. 
The exact result as obtained via numerical integration in \eqn{Ethsmallxc}
is compared to the limited expansion in \eqn{Ethexp} (where we replace $ \upsilon\to 2\pi$,
of course).

We now turn to the third quantity introduced above, namely the energy fraction
which after a time $\tau$ has been
transported at angles larger than a given value $\theta_0$. We denote this quantity as
${\cal E}(\theta>\theta_0,\tau)$. So far, we have considered only the energy distribution for the gluons in
the cascade, but not also their distribution in transverse momentum $\bk$, or in the polar angle $\theta$
w.r.t. the jet axis (defined as $\sin\theta=k_\perp/\omega$). 
Rather, the $\bk$--distribution  has been explicitly integrated out, as shown
in \eqn{spectrum}, in order to obtain simpler versions for the rate equations.
Yet, it turns out that for qualitative and even semi--quantitative estimates, one can restore
the $\theta$--distribution via the following, simple, argument. 
All the gluons in the cascade which are not too soft
 (namely, those with energy fractions $x\gtrsim x_s=\abar^2x_c$) propagate in the medium
along a distance of order $L$ and hence accumulate via multiple scattering an average transverse
momentum squared $\langle k_\perp^2\rangle\simeq Q_L^2\equiv \hat q L$, which is independent of $x$.
So long as this momentum $Q_L$ is much smaller than the gluon energy $\omega=xE$, one can
estimate the propagation angle according to
\beq\label{thetax}
\theta(x)\,\simeq\,\frac{Q_L}{xE}\,=\,\frac{x_c}{x}\,\theta_c,\qquad\mbox{with}\qquad
\theta_c\,\equiv\,\frac{Q_L}{\omega_c}\,=\,\frac{2}{\sqrt{\hat q L^3}}\,.\eeq
Hence, the interesting quantity ${\cal E}(\theta>\theta_0,\tau)$ can be computed as the energy
fraction ${\cal E}^{\,<}(x_0,\tau)$
carried by the gluons with $x< x_0$, where $x_0\simeq  x_c(\theta_c/\theta_0)$. This is of
course the same as the `thermalization energy' in \eqn{Ethsmallxc} evaluated for $x_{\rm th}=x_0$.
Hence, plotting the following quantity
  \beq\label{E<x0}
{\cal E}^{\,<}(x_0,\tau)\,\equiv\, {\cal E}_{\rm flow}(\tau)\,+\,\int_{0}^{x_0} {\rm d} x \,D_{\rm rad}(x,\tau)\,
\eeq
as a function of $x_c/x_0$ is tantamount to representing the quantity ${\cal E}(\theta>\theta_0,\tau)$ 
as a function of $\theta_0/\theta_c$. This is strictly true so long as the angle $\theta_0$ is not too large, 
namely $\theta_0 \lesssim \theta_c/\abar^2$, in order for the condition $x_0\gtrsim x_s$ to remain satisfied\footnote{\label{foot:7}
The softer gluons with $x\lesssim x_s$ have a shorter lifetime $\Delta t(x) < L$, as shown 
in \eqn{lifetime}. The corresponding  transverse
momentum broadening is estimated as $\langle k_\perp^2\rangle (x) \sim \hat q \Delta t(x)$,
and the relation \eqref{thetax} between the propagation angle $\theta(x)$ and the reference angle
$\theta_c$ gets replaced by (to parametric accuracy) $\theta(x)/\theta_c
\sim \big(1/\sqrt{\abar}\big)(x_c/x)^{3/4}$ \cite{Iancu:2014aza,Kurkela:2014tla,Blaizot:2014ula}.}.
But as we argue now, this is not a serious limitation. Indeed, we have previously explained that,
when $x_0 < x_s$, the r.h.s. of
\eqn{E<x0} is dominated by the first piece, the flow energy, which is independent of $x_0$
(recall the discussion after \eqn{Ethexp}).
% (This follows from the previous discussion of ${\cal E}_{\rm th}(\tau)$ after replacing $x_{\rm th}\to %x_0$~; see e.g. \eqn{Ethexp})  
Hence, for $\theta_0$ larger than $\theta_s\equiv \theta(x_s)\simeq \theta_c/\abar^2$,
 the function ${\cal E}(\theta>\theta_0,\tau)$ is quasi--independent of $x_0$ 
 and approximately equal to ${\cal E}_{\rm flow}(\tau)$.
 An intuitive view of the angles $\theta_c$ and $\theta_s$ in the context of a
typical gluon cascade is provided by Fig.~\ref{fig:cascade}.

As a side remark, we observe that the total energy carried by gluons with energies smaller
than a given scale $\omega_0$, with $\omega_0\le \omega_c$, which is computed as
(below, $x_0\equiv  \omega_0/E\le x_c$)
\beq\label{DeltaE<}
\Delta E^{\,<}(\omega_0)\,=\,E\,{\cal E}^{\,<}(x_0,\tau_L)\,,\eeq
is independent of the energy $E$ of the LP (within the present approximations), but only depends
upon the medium scale $\omega_c$ and upon the energy scale $\omega_0$ of reference.
This follows via manipulations in \eqn{E<x0} which are entirely similar to those in
\eqn{Eflowxc2}.

Returning to  \eqn{E<x0}, we notice that, in practice, it is more convenient to plot the
complementary quantity, namely the energy fraction located at $x$--values {\em larger} than $x_0$, 
 \begin{align}\label{E>x0}
{\cal E}^{\,>}(x_0,\tau)\,\equiv\, 1\,-\,{\cal E}^{\,<}(x_0,\tau)&\,=\,
{\cal E}_{{\rm LP}}(\tau)+ {\cal E}(x_0,x_c,\tau)\nn
&\,=\,
\,1 - 2\abar\tau\sqrt{x_c}\,+\int_{x_0}^\xc {\rm d} x \,D_{\rm rad}(x,\tau)\nn
&\,=\,1 - 2\abar\tau\sqrt{x_0}\,-\int_{x_0}^\xc {\rm d} x \,  \delta D_{\rm br}(x,\tau)\,.
\end{align}
Indeed, this corresponds better to the quantity which is actually measured in the experiments: 
the jet energy $E_J(\theta_0)$ as a function of the jet opening angle $\theta_0$ (i.e. the total energy
in the gluon cascade which propagates along angles $\theta \le \theta_0$). 
As emphasized in the second equality above, this quantity
${\cal E}^{\,>}(x_0,\tau)$ is the sum of the energy fractions carried by the leading particle and
by the modes at $x_0 < x < x_c$.

\begin{figure}[t]
	\centering
	\includegraphics[width=0.7\textwidth]{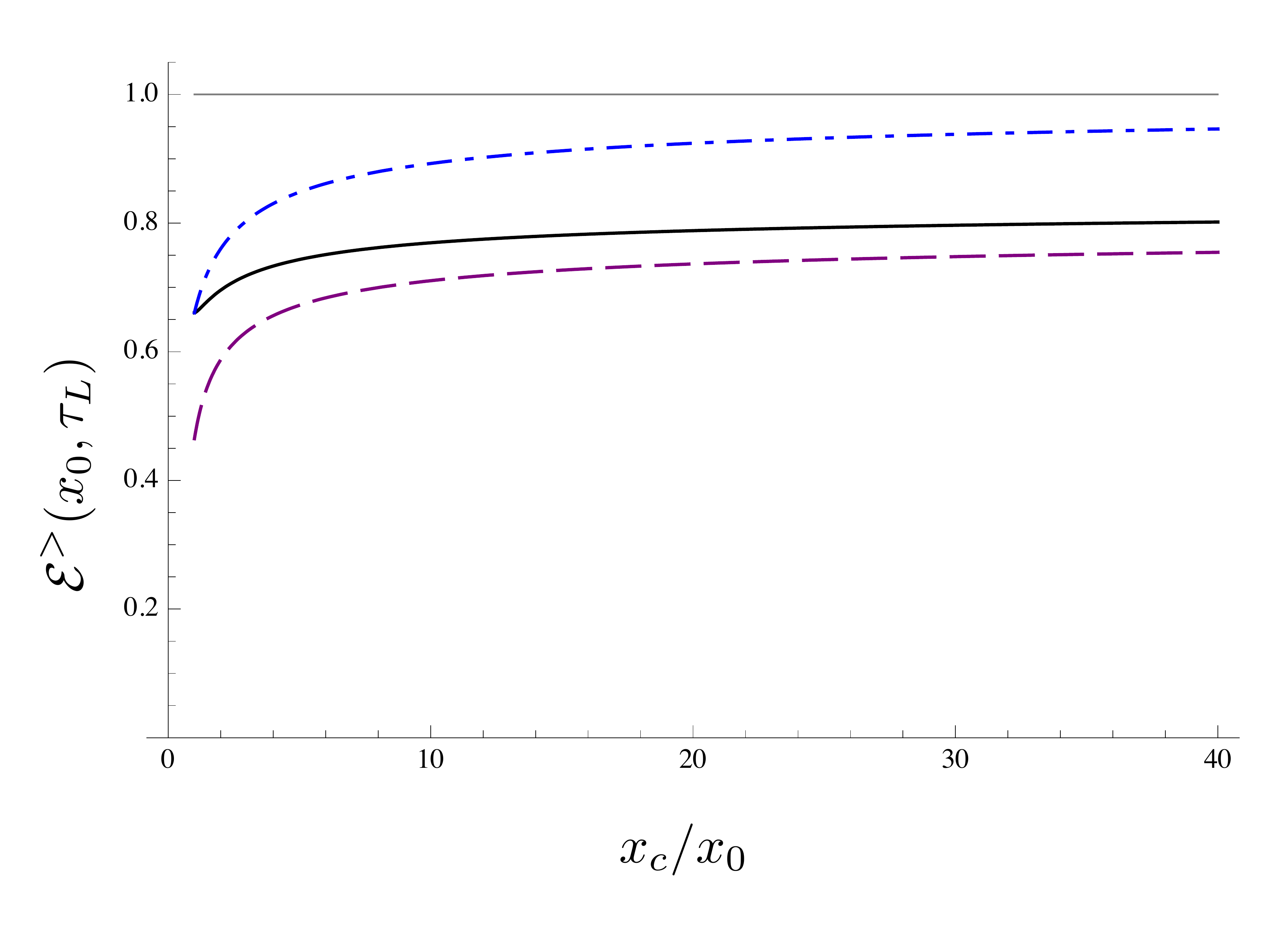}
		\caption{\sl  The energy ${\cal E}^{\,>}(x_0,\tau_L)$ 
		contained in the bins of the spectrum with $x\ge x_0$ at the end of the evolution
	       plotted as a function of $x_c/x_0$ for $x_0\le x_c$,
		$x_c=0.4$, and  $\abar=0.3$.		
		Black, solid, curve:  the full result computed according to \eqn{E>x0}.
		Blue, dotted--dashed, curve: the approximation obtained 
		by neglecting the effects of multiple branchings.
		  Purple, dashed, curve: the respective prediction
		of the low--energy case, \eqn{Dexint}, which is extrapolated to $x_c=0.4$.
		As explained in the text, these curves can also be viewed as representing the
		energy fraction ${\cal E}_J(\theta_0)$ contained within a jet with
		opening angle $\theta_0$ plotted as a function of $\theta_0/\theta_c$.
		}
		\label{fig:Energyxc04}
\end{figure}

In Fig.~\ref{fig:Energyxc04}, the quantity in \eqn{E>x0} is represented as a  function of $x_c/x_0$ for 
$\tau=\tau_L$ and $x_c=0.4$.  One also shows there the single--branching (or BDMPSZ) approximation, 
${\cal E}^{\,>}(x_0,\tau_L)=1 - 2\abar\sqrt{2x_cx_0}$, which is obtained by neglecting the integral of 
$ \delta D_{\rm br}$ in the third line of  \eqn{E>x0}, as well as 
the respective prediction of the low--energy case, \eqn{Dexint}, 
which here is extrapolated outside its physical range.
Two features of these curves are worth emphasizing:

First, the `offset' at large $x_c/x_0$, i.e. the fact that, for the two curves which include the effects
of multiple branchings,
the difference $1-{\cal E}^{\,>}(x_0)={\cal E}^{\,<}(x_0)$ approaches a finite value as $x_c/x_0\to \infty$. 
This non--zero value is, of course, the energy fraction ${\cal E}_{\rm flow}$ 
taken away by the turbulent flow. As also visible
in Fig.~\ref{fig:Energyxc04} (and obvious on physical grounds), this offset is absent if one neglects 
multiple branchings, i.e. if one tries to
describe the energy distribution at large angles on the basis of the BDMPSZ spectrum alone.
For applications to the phenomenology, it is important to notice that the kinematic restriction
to $x< x_c$ (which applies whenever $x_c < 1$) significantly reduces the value of this offset.
This reduction is visible  in both Fig.~\ref{fig:Energyxc04} and Fig.~\ref{fig:Esmallxc}.

Second, as also visible in Fig.~\ref{fig:Energyxc04} (and anticipated after \eqn{E<x0}), 
the variation with $x_c/x_0$ is extremely slow, especially
for the two curves which include the effects of multiple branchings. Physically, this means that, by
increasing the jet opening  angle $\theta_0=(x_c/x_0)\theta_c$, one can recover some of the energy 
that has been transported at large angles, {\em but only very slowly}. This is so because most of this 
energy has been transported, by the turbulent flow, directly at very large angles 
$\theta\gtrsim\theta_{\rm th}$, where it has been lost towards the medium via thermalization.
Here, $\theta_{\rm th}$ is the propagation angle for the very soft quanta with 
$x\sim x_{\rm th}$ and is significantly larger than $\theta_s$
(since $ x_{\rm th}$ is much smaller than $x_s=\abar^2x_c$).  In principle,
this angle $\theta_{\rm th}$ can be estimated within our effective theory
--- to parametric accuracy one finds $\theta_{\rm th}/\theta_c
\sim \big(1/\sqrt{\abar}\big)(x_c/x_{\rm th})^{3/4}$, cf. footnote~\ref{foot:7} ---,
%via the methods discussed in  Refs.~\cite{Iancu:2014aza,Kurkela:2014tla,Blaizot:2014ula}),
but this estimate is probably questionable: the angular distribution of the very soft gluons
with  $x\sim x_{\rm th}$ could be influenced by other effects, like the precise mechanism of 
thermalization, the Bethe--Heitler limit on the medium--induced radiation, or the 
kinematic constraint $k_\perp < \omega$, which are not properly included in
the current formalism. Fortunately though, this theoretical uncertainty is not important
for the angular distribution of the energy loss: 
the relevant curves in Fig.~\ref{fig:Energyxc04} are essentially flat for 
$x_c/x_0\gtrsim 1/\abar^2 \simeq 10$, i.e. for angles $\theta_0\gtrsim\theta_s$.

\comment{
To summarize, the previous analytic estimates and the plots in Figs.~\ref{fig:Ethxth}
and \ref{fig:Energyxc04} demonstrate that the quantity ${\cal E}_{\rm flow}\equiv
{\cal E}_{\rm flow}(\tau_L)$,  i.e. the energy fraction carried  by the turbulent flow, 
provides a good estimate for the energy lost by the gluon cascade 
towards the medium. Via successive branchings, this energy has been transported to
angles which are significantly larger than $ \theta_s = \theta_c/\abar^2\simeq 10\,\theta_c$,
hence it cannot be progressively recovered when increasing the jet opening angle $\theta_0$
beyond $ \theta_s$. }

Let us conclude with a few numerical estimates in view of the phenomenology.
Recent theoretical analyses of the data support an average value for the jet quenching parameter
in the ballpark of $\hat q=1\,{\rm GeV}^2/{\rm fm}$ \cite{Burke:2013yra}. 
%(This estimate is also comparable to the leading order
%estimate at weak coupling for a QGP with temperature  $T=250$~MeV \cite{Baier:1996kr}).
By also choosing an average length $L= 4$~fm for the in--medium path,  one finds
$ \omega_c\simeq 40$~GeV and $\theta_c\simeq 0.05$. This implies that the
characteristic scale for multiple branching is quite hard, $\omega_s=\abar^2\omega_c\simeq
4$~GeV, and in particular significantly harder than the medium `temperature' $T\lesssim 1$~GeV
(the average transverse momentum of the medium constituents).
In the measurements of di--jet asymmetry
at the LHC, one has $E\ge 100$~GeV~; this energy is sufficiently large compared to $\omega_c$
for the `high--energy' regime ($E\gg \omega_c$) to apply.

In this regime, the energy $\Delta E_{\rm flow}$
lost by the gluon cascade via flow is independent of the original energy $E$
(cf. the discussion after \eqn{DEflow}). Using \eqn{DEflow} with $ \omega_c= 40$~GeV
and $\abar=0.3$, one finds
 \begin{eqnarray}\label{DeltaEflow}
{\Delta E_{\rm flow}}\,\simeq\,0.32\,\omega_c\,\simeq\,13\ {\rm GeV}
 \,.\end{eqnarray}
It is also interesting to compute the energy transported at angles larger than $\theta_s=
\theta_c/\abar^2\simeq 0.5$. This is obtained from \eqn{DeltaE<} with $\omega_0\to \omega_s$ and, 
once again, is independent of the energy $E$ of the LP. A good estimate is given by \eqn{Ethexp}
with $x_{\rm th} \to x_s=\abar^2 x_c$, and reads
\beq\label{DeltaE>}
\Delta E(\theta > \theta_s= 0.5)\,=\,E\,{\cal E}^{\,<}(x_s,\tau_L)\,\simeq\,\Delta E_{\rm flow}+2\sqrt{2}
\abar^2\omega_c\big(1-\sqrt{2}\abar\big)\,\simeq\,19\ {\rm GeV}\,.\eeq
The above numbers compare reasonably well with the corresponding experimental results
\cite{Chatrchyan:2011sx,Gulhan:2014}, especially in view of our crude assumptions concerning 
the structure of the medium. 

Consider finally the variation of the jet energy with increasing the jet opening angle $\theta_0$,
i.e. the function $E_J(\theta_0)$.
Our results in Fig.~\ref{fig:Energyxc04} predict that this quantity should be very slowly increasing 
with $\theta_0$. This seems to significantly differ from a
recent analysis of the experimental data in Pb+Pb collisions at the LHC, 
which has reported a considerably steeper angular dependence for $E_J(\theta_0)$ \cite{Gulhan:2014}.
Note however that a similarly steep dependence has also been found in the corresponding 
data for for p+p collisions, and that the {\em difference} between these 
two sets of data looks essentially flat (as a function of $\theta_0$) within the error bars
\cite{Gulhan:2014}.
It looks reasonable to interpret this difference as a measure of the medium effects
in heavy ion collisions. If so, the fact that this difference appears to be slowly varying with  $\theta_0$
(in fact, almost flat) is in good agreement with our predictions in Fig.~\ref{fig:Energyxc04}.

\section*{Acknowledgments}
  We are grateful to Al Mueller for insightful comments on the manuscript.
 We acknowledge useful related discussions with
  Jean-Paul Blaizot and Yacine Mehtar-Tani.
This research is supported by the European Research Council under the Advanced Investigator Grant ERC-AD-267258. Figure 1 has been made with Jaxodraw \cite{Binosi:2003yf}.

\appendix

 \section{Perturbation theory for the rate equation}
\label{sec:PT}

In this Appendix, we shall discuss the perturbative solution to the rate equation with a source,
\eqn{Drad0}, as obtained via successive iterations of the branching term  
$\abar\mcal{I}[D_{\rm rad}]$ in the r.h.s. This is tantamount to an expansion in powers of $\abar\tau$
in which the source term $\mcal{S}_0(x)={\abar}/{\sqrt{x}}$ (including its factor $\abar$) is treated
as a quantity of $\order{1}$. The zeroth order result is
$D_{\rm rad}^{(0)}={\abar\tau}/{\sqrt{x}}$, while the first iteration, as obtained
by evaluating the branching term  $\abar\mcal{I}[D_{\rm rad}]$ with the zeroth order result
and integrating over $\tau$, yields
 \begin{align}\label{IT1}
  D_{\rm rad}^{(1)}(x,\tau)&= \frac{\abar^2 \tau^2}{2}\int \rmd z \,{\cal K}(z)  
  \bigg\{\Theta\Big(z-\frac{x}{x_c}\Big)
   \frac{z}{x}\,-\, \frac{z}{x}
  \bigg\} = \, -\frac{ \abar^2 \tau^2}{2x}\int_0^{x/x_c} \rmd z \, z\,{\cal K}(z)\,. \end{align}
The net result, which is negative, is due to an excess in the phase--space for  
the loss term, at $z <x/x_c$. To simplify the final integral over $z$, we shall restrict
ourselves to the simplified kernel ${\cal K}_0(z)$. In that case, one can easily compute
(say, by changing the integration variable as $z\equiv (u-1)/u$)
\beq\label{intz}
\int_0^{x/x_c} \rmd z \, z\,{\cal K}_0(z)\,=\,\int_0^{x/x_c} \frac{\rmd z }{\sqrt{z}(1-z)^{3/2}}
\,=\,2\sqrt{\frac{x}{x_c-x}}\,.\eeq
When inserted into \eqn{IT1}, this confirms the result  \eqref{D1} for $D_{\rm rad}^{(1)}$.

Note that the small--$x$ limit (in the sense that $x/x_c\ll 1$) of the result in \eqn{intz}
would be the same for the full kernel ${\cal K}(z)$~: indeed, when $z<x/x_c\ll 1$, one can approximate
$f(z)\simeq 1$ in \eqn{Kdef}. This confirms that the limited expansion shown in
\eqn{Dradexp2}  holds for the physical kernel.

Returning to the  simplified kernel ${\cal K}_0(z)$, in which case \eqn{intz} holds
for any $x < x_c$, let us also compute the second iteration, by evaluating the branching term
with the first order correction in \eqn{D1}. One can write
\begin{align}\label{IT2}
  %\frac{\del D_{\rm rad}^{(2)}(x,\tau)}{\del\tau}&=\,\abar
  \mcal{I}[D_{\rm rad}^{(1)}](x,\tau)
  &=\,-\frac{ \abar^2 \tau^2}{x}\int \rmd z \,z\,{\cal K}(z)  
  \bigg\{\Theta\Big(z-\frac{x}{x_c}\Big)\frac{1}{\sqrt{x_c-x/z}}-\frac{1}{\sqrt{x_c-x}}\bigg\}\nn
  &=\frac{2 \abar^2 \tau^2}{\sqrt{x}(x_c-x)}\,-
  \frac{ \abar^2 \tau^2}{x}\int_{x/x_c}^1\rmd z \,z\,{\cal K}(z)  
  \bigg\{\frac{1}{\sqrt{x_c-x/z}}-\frac{1}{\sqrt{x_c-x}}\bigg\}
  \,. \end{align}
Let us denote by $\mcal{J}$ the integral in the second line above. After changing the
integration variable
according to $z=u/(u+1)$, this becomes
 \begin{align}\label{ITu1}
 \mcal{J}&=
\int_{x/x_c}^1\rmd z \,z\,{\cal K}(z)  
  \bigg\{\frac{1}{\sqrt{x_c-x/z}}-\frac{1}{\sqrt{x_c-x}}\bigg\}
   =\,\frac{1}{\sqrt{x_c-x}}\int_{u_0}^\infty\rmd u
   \bigg\{\frac{1}{\sqrt{u-u_0}}-\frac{1}{\sqrt{u}}\bigg\}
  \,, \end{align}
where we denoted $u_0\equiv x/(x_c-x)$. For any finite value of $u_0$, the above integral
over $u$ is well defined and can be evaluated as
  \begin{align}\label{ITu2}
 \mcal{J}&= 
 \,\frac{2}{\sqrt{x_c-x}}\,\lim_{u_M\to\infty}\,\Big\{\sqrt{u_M-u_0} - {\sqrt{u_M}} + {\sqrt{u_0}}\Big\}
\nn
 &= \,\frac{2}{\sqrt{x_c-x}}\,\lim_{u_M\to\infty}\,\bigg\{{\sqrt{u_0}}\,-\,\frac{u_0}{2\sqrt{u_M}}\bigg\}
\,=\,\frac{2\sqrt{x}}{x_c-x}
 \,,\end{align}
where $u_M$ is a sharp upper cutoff on $u$ that has been introduced at intermediate steps
in order to separate the two terms within the braces in the integral in \eqn{ITu1}. When 
inserting the final result from \eqn{ITu2} into the second line of \eqn{IT2}, one finds that
it precisely cancels the other term there, so that the net result of this second iteration is exactly zero:
$ \mcal{I}[D_{\rm rad}^{(1)}]=0$. Accordingly, the perturbative series becomes trivial
(in the sense that all the higher order terms vanish) after the first
iteration, and then the overall result is just the sum of the first two terms:
$D_{\rm rad}= D_{\rm rad}^{(0)}+ D_{\rm rad}^{(1)}$.
This is the result that has been announced towards the end of Sect.~\ref{sec:rad}.

Now, the fact that the function $D_{\rm rad}^{(1)}(x,\tau)$ is an exact fixed point of the branching
term is indeed correct and should not be a surprise: in Sect.~\ref{sec:Fsmallxc}, we have seen that
the very same function of $x$, namely $D_{\rm as}(x)\propto 1/\sqrt{x(x_c-x)}$,
emerges as the exact solution to the rate equation for the case of
a source localized at $x=x_c$ (cf. \eqn{Das}). 
Since the source vanishes at any $x<x_c$, this is tantamount to
saying that $D_{\rm as}(x)$ is an exact fixed point for the branching term: $ \mcal{I}[D_{\rm as}]=0$.
This solution $D_{\rm as}(x)$  becomes divergent when $x\to x_c$, but this is indeed a real
property of that particular problem, because the respective source
$\mcal{S}(x)=A\delta(x-x_c)$ diverges at the end of the spectrum. 

On the other hand, for
the delocalized source $\mcal{S}_0(x)={\abar}/{\sqrt{x}}$, no such a divergence is expected
(as also confirmed by the exact manipulations in Sect.~\ref{sec:rad}),
hence the iterative solution $D_{\rm rad}= D_{\rm rad}^{(0)}+ D_{\rm rad}^{(1)}$ 
cannot be fully right : it fails when
$x\to x_c$. The mathematical reason for this failure can be traced to the subtlety of
the limit $x\to x_c$ in relation with the manipulations in Eqs.~\eqref{ITu1}--\eqref{ITu2}:
clearly, these manipulations become ambiguous when $x\to x_c$, or $u_0\to\infty$,
since this limit $u_0\to\infty$ does not commute with the limit $u_M\to\infty$. 

%%%%%%%%%%%%%%%%%%%%%%%%%%%%%%%%%%
%\bibliographystyle{utcaps}
%\bibliography{../../References/refs}

\providecommand{\href}[2]{#2}\begingroup\raggedright\endgroup

\end{document}